\documentclass{egpubl}
\usepackage{sca2018}

\SpecialIssueSubmission    % uncomment for submission to Computer Graphics Forum, special issue
 \electronicVersion % can be used both for the printed and electronic version

\ifpdf \usepackage[pdftex]{graphicx} \pdfcompresslevel=9
\else \usepackage[dvips]{graphicx} \fi

\PrintedOrElectronic

\usepackage{t1enc,dfadobe}

\usepackage{egweblnk}
\usepackage{cite}
\usepackage{tikz}

\usepackage{algorithm}
\usepackage[noend]{algpseudocode}
\usepackage{graphicx}
\usepackage{caption}
\usepackage{subcaption}
\usepackage{overpic}
\usepackage{color}

\usepackage{amsmath}
\newcommand{\prox}{\mathrm{\mathbf{prox}}}
\newcommand{\x}{\mathbf{x}}
\newcommand{\y}{\mathbf{y}}
\newcommand{\z}{\mathbf{z}}

\newcommand{\norm}[1]{\left\lVert#1\right\rVert}

\newcommand{\half}{\tfrac{1}{2}}
\newcommand{\PiDiv}{\mathrm{\Pi_{\mathrm{DIV}}}}
\newcommand{\PiREAL}{\mathrm{\Pi_{\mathrm{\mathbf{R}_+}}}}
\newcommand{\gv}{\xi} % generic variable (function argument)
\newcommand{\Phy}[1]{\Phi^{(#1)}}
\newcommand{\Phip}[1]{\tilde\Phi^{(#1)}}
\newcommand{\Phiu}[1]{\Delta\Phi^{(#1)}}
\newcommand{\Phic}[1]{\Delta\Phi^{(#1)}_c}
\renewcommand{\u}[1]{\mathbf{u}^{(#1)}}
\newcommand{\up}[1]{\mathbf{\tilde u}^{(#1)}}
\newcommand{\uu}[1]{\Delta\u{#1}}
\renewcommand{\i}[1]{i^{(#1)}}
\newcommand{\ip}[1]{\tilde i^{(#1)}}
\newcommand{\iu}[1]{\Delta i^{(#1)}}
\newcommand{\ic}[1]{\Delta i^{(#1)}_c}

\newcommand{\PhiuNoT}{\Delta\Phi}
\newcommand{\PhicNoT}{\Delta\Phi_c}
\newcommand{\uuNoT}{\Delta\mathbf{u}}

\newcommand{\myrefeq}[1]{Eq.~\eqref{#1}}
\newcommand{\myreffig}[1]{Fig.~\ref{#1}}
\newcommand{\myreftab}[1]{Table~\ref{#1}}
\newcommand{\myrefsec}[1]{Sec.~\ref{#1}}

\newcommand{\myrefalg}[1]{Alg.~\ref{#1}}

\title[Coupled Fluid Density and Motion from Single Views]%
      {Coupled Fluid Density and Motion from Single Views}

\author[M.-L. Eckert, W. Heidrich \& N. Thuerey]
{\parbox{\textwidth}
	{\centering M.-L. Eckert$^{1}$,
	W. Heidrich$^{2}$ 
    and N. Thuerey$^{1}$}
\\
{\parbox{\textwidth}{\centering $^1$Technical University of Munich, Germany\\
       $^2$King Abdullah University of Science and Technology, Saudi Arabia}
}
}

\begin{document}

\maketitle
%-------------------------------------------------------------------------
\begin{abstract}
	We present a novel method to reconstruct a fluid's 3D density and motion based on just a single sequence of images.
This is rendered possible by using powerful physical priors for this strongly under-determined problem. 
More specifically, we propose a novel strategy to infer density updates strongly coupled to previous and current estimates of the flow motion.
Additionally, we employ an accurate discretization and depth-based regularizers to compute stable solutions.
Using only one view for the reconstruction reduces the complexity of the capturing setup drastically and could even allow for online video databases or smart-phone videos as inputs. 
The reconstructed 3D velocity can then be flexibly utilized, e.g., for re-simulation, domain modification or guiding purposes. 
We will demonstrate the capacity of our method with a series of synthetic test cases and the reconstruction of real smoke plumes captured with a Raspberry Pi camera.

%-------------------------------------------------------------------------
%  ACM CCS 2012
%   (see http://www.acm.org/about/class/class/2012)
%The tool at \url{http://dl.acm.org/ccs.cfm} can be used to generate
% CCS codes.


\end{abstract}  

\let\thefootnote\relax\footnotetext{Pre-print. To appear in Computer Graphics Forum (2018).}

\section{Introduction}
% physical / fluid simulation for visual effects
While physical simulations have gained popularity in visual effects, simulations are not easy to control, especially for artists.
As detailed and realistic simulations require high computational power, it is often infeasible to run multiple simulations until both the desired visuals and behavior are reached.
Other approaches for creating visual effects come with their own disadvantages, e.g., compositing video data requires large collections of footage and is difficult to combine with 3D elements.
Since humans are very familiar with fluid phenomena from everyday situations such as pouring milk in coffee or burning candles, the corresponding visual effects have high requirements for accuracy in order to be visually convincing.

% reconstruction instead of simulation
Instead of simulating fluids purely synthetically, our goal is to reconstruct motions of real fluid phenomena in 3D such that the density matches the input image and adheres to the underlying physical models.
Thus, we are solving the inverse problem to fluid simulation.
Reconstructing a real fluid phenomenon directly leads to a plausible
velocity field with a known visual shape, yielding an excellent starting point for, e.g., a guided simulation for visual effects.
While previous fluid capture techniques typically require multiple cameras and a complex setup, 
we focus on reconstructions from a single camera view. 
This allows for the use of a large variety
of existing videos and for the capture of new fluid motions with ease.

% our requirements/contributions
Existing computed tomography methods (e.g.~\cite{Ihrke:tomoFlames}) are designed for inputs with a multitude of cameras and viewing angles.
Their reconstruction quality rapidly drops with the number of views as the corresponding inverse problem becomes under-determined.
Our setting of using just a single view is a challenging scenario, where we aim for plausible but not necessarily accurate reconstructions through the use of physical constraints as priors.
A system with a single camera also has distinct advantages, including reduced costs, a drastically simplified setup, and the elimination of tedious procedures like camera calibration and synchronization.
Without having a reliable, a-priori density reconstruction, estimating flow velocities is significantly more challenging than in previous work~\cite{Gregson:2014}.
Information about the motion of a fluid phenomenon allows us to flexibly re-use the captured data once it is reconstructed. 
It ensures temporal coherence and enables us to edit the reconstruction, to couple it with external physics solvers or to conveniently integrate the reconstructed flow in 3D environments, e.g., to render it from a new viewpoint.
We will demonstrate the capabilities of our method with a variety of complex synthetic flows and with real data from a fog machine video.

In summary, our core technical contributions are
\begin{itemize}
  \item an image formation model for the joint reconstruction of 3D densities
    and flow fields from single-view video sequences,
  \item a new optimization-based formulation of the inverse problem
    with a strong emphasis on temporal coherence of the recovered
    density sequence,
  \item a tailored combination of physics-based as well as geometric
    priors, including new priors to resolve motion uncertainty along
    lines of sight in the single camera view and finally
  \item an efficient implementation based on primal-dual
    optimization to realize an effective single view fluid capture
    system. This system is used for extensive experiments using
    both real and simulated data.
\end{itemize}

\section{Related Work}

{\em Fluid Simulation:}
Forward simulations of fluids have been an extensively studied topic within computer graphics
ever since the introduction of the {\em stable fluids} algorithm \cite{Stam1999}.
Based on this approach for solving the Navier-Stokes equations, 
numerous extensions and improvements have been introduced,
e.g., to retain more natural, swirling motions of fluids \cite{fedkiw2001visual,selle2005vortex},
and to more accurately compute the transport of velocities and other quantities  \cite{Kim05FlowFixer,Selle:2008:USM}.
Additionally, interactions with static and moving objects in the flow often arise
in practical situations. Here, the second order method of Batty et al. \shortcite{Batty2007} is a popular choice.
In our work, we focus on smoke phenomena, i.e., single-phase flows, but	
liquids naturally
play an important role in many special effect applications. For liquids, grid-particle hybrids such as
the particle level-set \cite{Foster01} and the FLIP method \cite{Zhu2005} are
especially popular. 

Many works in graphics also employ physics-inspired methods to increase the apparent
resolution of fluid simulations \cite{Kim:2008:wlt, narain:2008:procTurb, pfaff2010scalable}, which
typically save time compared to a full Navier-Stokes approximation.
For graphics applications, editing and controlling the flow motion is likewise 
a crucial topic \cite{McNamaraAdjointMethod,Pan:2013}. As our algorithm generates dense motion fields, such algorithms could easily
be applied to fine-tune our results.

In the following, we will employ an Eulerian discretization and as such we restrict the discussion
to corresponding algorithms. 
However, for single phase simulations,
vorticity-based Lagrangian methods are popular \cite{angelidis2006controllable,golas2012large}.
Additionally, fully Lagrangian methods are a 
very interesting alternative to grid-based methods for liquid simulations \cite{muller2003particle,IhmsenOSKT14a}.

Our work targets the {\em inverse problem} of reconstructing 3D density and motion from 2D
observations, and in this context the aforementioned Eulerian simulation algorithms play an important role
for our method by providing priors that disambiguate a sparse set of observations.

{\em Density Reconstruction:} 
In order to capture 3D density data from a collection of 2D observations, computed tomography is an established algorithm
originating from medical
applications~\cite{kak1988tomography}.
In computer graphics, the same principle is typically used for visible light instead of X-rays.
E.g., Ihrke et al.~\shortcite{Ihrke:tomoFlames} reconstruct flames from 4-8 input images by solving a least-squares problem describing the relation between pixels and voxels in conjunction with a visual hull. 
Due to the high computational cost of the reconstruction, adaptive algorithms have successfully
been proposed \cite{Ihrke:adaptive}.
Specialized but powerful algorithms have been developed for reconstructing objects for which rotational
symmetry can be safely assumed, such as astronomical structures \cite{lintu2007nebulae,wenger2013fast}.

Closer to our area of application, Gregson et al.~\shortcite{Gregson:2012} developed an efficient, stochastic approach for reconstructing fluid densities from 8-16 input videos. The approach achieves high-quality reconstructions with a grid-less, matrix-free solve and can incorporate a variety of regularizers to improve reconstruction quality.
Okabe et al.~\shortcite{Okabe:2015} used only one to two input videos and augment their tomographic approach by appearance transfer. 
They reconstruct an initial density volume with regular tomography and iteratively improve the density until the reconstruction satisfies additional 
view constraints that make use of up to 180 synthetically rendered views. They also found that re-using a front view for a $90^{\circ}$ angle
significantly improves their reconstructions.

All of the approaches discussed above rely on multiple views, either from complex and carefully calibrated measurement setups or from 
internal calculations. In contrast, we will outline how to incorporate flow physics in order to work with only one single projected image sequence.

While we aim to reconstruct the shape and motion of visible phenomena, {\em Schlieren} imaging has successfully
been applied by Atcheson et al.~\shortcite{Atcheson:gas} to capture hot air flows based on refractive index changes from multiple views.
We will focus on single-phase flows below, but highly interesting methods
have been proposed to capture liquids, e.g., using submerged checkerboard
patterns \cite{Morris:stereo} or structured light to reconstruct diffuse liquid surfaces \cite{wang2009physically}.

{\em Velocity Reconstruction:} 
The reconstruction of the motion of a fluid, also known as velocimetry, is typically more difficult than reconstructing volumetric densities, since motion is 
in most cases only observed indirectly by tracking temporal changes in the densities.
Methods like {\em particle image velocimetry} (PIV) \cite{Elsinga:06} add visible particles to the medium and track their movement. 
PIV methods are widely used, but require carefully chosen particles and complex camera setups with great spatial and temporal resolution. 
They also come with the risk of perturbing the measurement due to the insertion of particles. 
Xiong et al.~\shortcite{Xiong:2017} reconstruct a fluid's motion from one single view by an enhanced PIV approach.
They use so-called {\em Rainbow PIV} to capture different illumination colors depending on the depth of the particle.
However, their setup requires specialized hardware and currently targets motions of a small water tank. 
To the best of our knowledge, our method is the first to propose a generic method for coupled tomographic density updates 
and velocity reconstruction.

A widely used approach for dense motion estimation is optical flow introduced by Horn and Schunck~\shortcite{Horn:OF}.
It provides an excellent basis for flow velocity calculations.
Corpetti et al.~\shortcite{Corpetti:2002} use optical flow with an additional divergence and curl smoothness prior to regularize 2D cloud motions.
The approach by Chen et al.~\shortcite{chen2016dense} is also fully 2D and incorporates a skeleton reconstruction
step to improve the image space motion estimates.
Gregson et al.~\shortcite{Gregson:2014} adapted an optical flow algorithm to reconstruct fluid motions in 3D.
They solve a convex optimization problem for optical flow while constraining the motion to be divergence-free. While this method can reconstruct volumetric motions with high accuracy, the algorithm requires full volumetric density reconstructions, e.g., using computed tomography as input, which are often very difficult or impossible to acquire in visual effects settings. 
We will employ a similar incompressibility constraint in our optimization, but a key difference is that our approach works with a single 2D video instead of a series of volumetric inputs.
The appearance transfer approach mentioned above \cite{Okabe:2015} also proposed a motion reconstruction, which, however, is based on a projected, 2D  optical flow and as such only yields a rough motion estimate.
In these methods, incompressibility is a crucial constraint for flow motions, but it is difficult to enforce in practice.

{\em Convex Optimization:}
In this paper, 
we solve a joint inverse problem for reconstructing both fluid density and motion field over time
by optimizing a single objective function. 
Algorithmically, our approach
heavily relies on the framework of {\em proximal
  operators}~\cite{parikh2013proximal} and especially the Primal-Dual
algorithm proposed by Chambolle and Pock~\shortcite{ChambollePD}. 
This framework has recently become very popular for solving inverse
problems in graphics, including projective dynamics
simulations~\cite{Narain2016admm}, fluid control~\cite{Inglis:2017},
and domain-specific languages~\cite{heide2016proximal} for a host of imaging problems. 
Similar convex optimization algorithms were also used
in the works discussed above~\cite{Gregson:2014,Xiong:2017}.

% --- DO NOT DELETE ---
% Local Variables:
% mode: latex
% mode: flyspell
% mode: TeX-PDF
% End:

\section{Method}
\begin{figure*}[t]
	\centering
	\begin{overpic}[trim={0.4cm 1.8cm 9.cm 1.4cm},clip,width=0.123\linewidth]{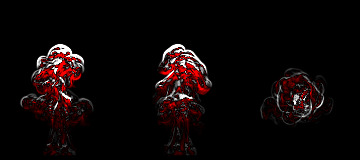}
		\put(3,60){\small \color{white}{$a)$}}\end{overpic}\hspace{0em}%
	\includegraphics[trim={4.7cm 1.8cm 4.7cm 1.4cm},clip,width=0.123\linewidth]{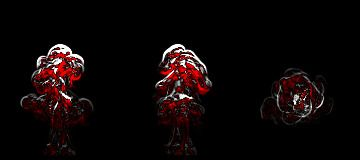}\hfill
	\begin{overpic}[trim={0.4cm 1.8cm 9.cm 1.4cm},clip,width=0.123\linewidth]{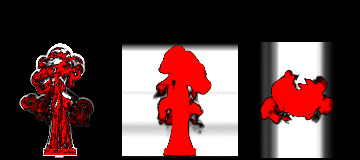}
		\put(3,60){\small \color{white}{$b)$}}\end{overpic}\hspace{0em}%
	\includegraphics[trim={4.7cm 1.8cm 4.7cm 1.4cm},clip,width=0.123\linewidth]{images/volVelNaive/0/volUcalc_30}\hfill
	\begin{overpic}[trim={0.4cm 1.8cm 9.cm 1.4cm},clip,width=0.123\linewidth]{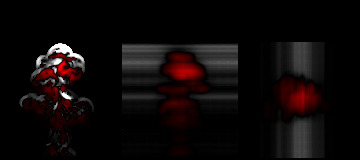}
		\put(3,60){\small \color{white}{$c)$}}\end{overpic}\hspace{0em}%
	\includegraphics[trim={4.7cm 1.8cm 4.7cm 1.4cm},clip,width=0.123\linewidth]{images/volVelNaive/1/volUcalc_30}\hfill
	\begin{overpic}[trim={0.4cm 1.8cm 9.cm 1.4cm},clip,width=0.123\linewidth]{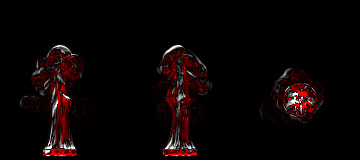}
		\put(3,60){\small \color{white}{$d)$}}\end{overpic}\hspace{0em}%
	\includegraphics[trim={4.7cm 1.8cm 4.7cm 1.4cm},clip,width=0.123\linewidth]{images/volVelNaive/2/volUAdv_60}
	\caption{Front and side views of density updates with different methods. F.l.t.r.: a) ground truth, b) estimated density change for pure tomographic reconstructions, c) differences in images space mapped into the density with tomography, d) our computed update. 
	As shown in the right half of each image, our method yields plausible and natural density updates
	that are slightly too small, but match the ground truth solution well.
	In contrast, tomography approaches with a single view lead to suboptimal and unnatural updates (b and c).}
	\label{fig:volvelNaive}
\end{figure*}

We estimate both density $\Phi$ and velocity $\mathbf{u}$ for a single time step $t$ based on a given state for time $t-1$.
The estimation considers the transport equation governing fluid motion in conjunction with a linear tomographic image formation model that describes an observed image $i$ as line integrals of a scalar density volume $\Phi$.
Since we only use one single 2D view as input, this image formation model is highly under-de\-ter\-mined, which necessitates the use of physical priors to arrive at plausible solutions.
Therefore, we employ a physics-based prediction scheme to add additional constraints to our optimization and couple our density estimation tightly to the velocity estimation. We do not modify the density explicitly.
Apart from a density inflow source estimated from $i$, our densities only move based on the reconstructed velocity.
In the following, we first describe our estimation algorithm at a higher level and then discuss the problem formulation for our density and velocity updates, and how we solve it in depth. 
Calculating the velocity update $\uu{t}$ is the core step of our algorithm. 
In order to consider constraints on the density change induced by the velocity update, we use a temporary variable for the density update $\Phiu{t}$.
\subsection{Density and Velocity Estimation Algorithm}
Having the density estimate $\Phy{t-1}$ and the velocity estimate $\u{t-1}$ of the last time step $t-1$, we predict both for the current time step $t$ and therefore use them to regularize the same two quantities at time step $t$.
The prediction is based on physical priors, i.e., the density and velocity transport over time steps and the incompressibility of the velocity.
Incompressibility is ensured by projecting the velocities onto the space of divergence-free velocities, called $\PiDiv$.

We advect velocity $\u{t-1}$ with itself and ensure its incompressibility to create a velocity guess $\up{t}$.
The density guess $\Phip{t}$ is created by advecting the density $\Phy{t-1}$ forward with the velocity guess $\up{t}$.
Therefore, we only need to solve for a \textbf{velocity update $\uu{t}$} that corrects both density and velocity guesses at each time step.
The density update $\Phiu{t}$ induced by the velocity update must match $\iu{t}$, which is the difference between the input image $\i{t}$ and the 2D projection $\ip{t}$ of the density guess.

The full density $\Phy{t}$ is then the result of advecting the density guess $\Phip{t}$ with the velocity update $\uu{t}$. 
The full velocity $\u{t}$ is the sum of the velocity update $\uu{t}$ and the divergence-free, with the velocity update advected velocity guess $\up{t}$.
~\myrefalg{alg:predUpd} describes our full estimation procedure including prediction, update and alignment steps.

\begin{algorithm}
	\caption{Density and Velocity Estimation}
	\label{alg:predUpd}
	\begin{algorithmic}[1]
		\Procedure{combinedEstimation}{$\Phy{t-1}$, $\u{t-1}$, $\i{t}$}
		\setlength{\parskip}{0.5ex}
		\State $\up{t}$=$\PiDiv$(advect($\u{t-1},\u{t-1}$)) // predict div-free vel
		\State $\Phip{t}$=advect($\Phy{t-1},\up{t}$) // predict density
		\State $\uu{t}$=calculateUpdate($\Phip{t}$, $\iu{t}$) // \myrefsec{sec:optstep}
		\State $\Phy{t}$=advect($\Phip{t},\uu{t}$) // move predicted den forward
		\State $\u{t}$=$\PiDiv$(advect($\up{t},\uu{t}$))+$\uu{t}$ // update vel
		\State $\Phy{t}, \u{t}$=applyInflow($\Phy{t}, \u{t}$)
		\EndProcedure
	\end{algorithmic}
\end{algorithm}
Note that the first two steps (line 2, 3)
are similar to previous work~\cite{Gregson:2014}, while the subsequent
steps differ. A key difference is that we do not assume
a given sequence of volumetric density, but instead infer them together
with the motion (line 4).
In addition, the density $\Phy{t}$ is not directly modified outside the inflow region (line 5).
In the inflow region, we apply the estimated smoke source (line 7). 
Within this region, we prescribe densities based on the input images and optionally also an inflow velocity. 
Details of the inflow estimation will be given in \myrefsec{sec:inflow}.
Another smaller difference from previous work is that we employ
a velocity alignment step~\cite{Thuerey2017,rtliquids2017}. As explained 
in these works, the additive combination of Eulerian flow fields
always requires an alignment of the individual fields.

\subsection{Density and Velocity Updates}
Existing methods for fluid imaging directly update the density $\Phi$
in each time step using tomographic reconstructions from the input
video sequence(s). 
This works well for systems with large numbers of cameras, 
but as the number of projections decreases, the tomography
problem becomes more and more ill-posed.
While temporal priors can be employed in order to 
alleviate this problem to a certain degree, 
these priors do not suffice for
a single-view setting, as we will illustrate below.
Instead, we tightly couple the density update $\Phiu{t}$ and the velocity update $\uu{t}$ through the transport equation.
In this way, the density $\Phy{t}$ is forced to adhere to the input images $\iu{t}$ solely based on the reconstructed motion 
 $\uu{t}$.
The density update $\Phiu{t}$ is an intermediate variable only used for constraining the density change induced by the velocity update.

We illustrate the inherent problems of traditional tomographic methods
by means of a synthetic example with ground truth in~\myreffig{fig:volvelNaive}.
Here, we show density updates (white being positive, red negative) from front and side views
for a simulated smoke plume.
The ground truth density change is naturally distributed with most positive values on the top since the smoke plume is rising upwards. 
While the input view of the traditional tomography density update looks very accurate, the side view reveals its implausible nature due to the equal distribution of density along each line of sight. 
\myreffig{fig:volvelNaive}b) and c) present two variants, the former computing
the update after two tomographic reconstructions, while the latter computes the update in the image space
and then maps this difference into the density volume.
Since the density changes in both cases do not represent any natural fluid motion, there is no divergence-free velocity that could induce or match these observed updates. 
Therefore, such volumetric reconstructions
from existing tomography approaches do not yield a valid basis for capturing the motions of flow phenomena.

We propose targeting the density change $\Phiu{t}$ induced by $\uu{t}$, instead of 
separately reconstructing both $\Phy{t}$ and the flow motion.
The result of such a coupled density change estimation is shown in~\myreffig{fig:volvelNaive}d).
The computed density change looks natural and is induced by advection with $\uu{t}$.
Our formulation not only improves temporal coherence but also reduces the undesired density expansion along the depth direction,
and it helps to eliminate ray artifacts of regular tomographic reconstructions.
This coupling is necessary to compensate for the inherent under-de\-ter\-mi\-na\-tion of the single-view reconstruction.

Starting with an optical flow problem $\frac{\partial\Phi}{\partial t}+\nabla\cdot(\Phi\mathbf{u})=0$, 
we formulate the linearized version of our combined and constrained convex optimization problem for density and velocity updates as follows:
\begin{equation}
\begin{aligned}
& \underset{\Phiu{t},\uu{t}}{\text{minimize}}
& & f(\Phiu{t},\uu{t}) = \norm{\frac{\partial\Phip{t}}{\partial t}+\nabla\Phip{t}\cdot\uu{t}}^2\\
& \text{subject to}
& & \Phiu{t}+\Phip{t}\geq 0, \; P\:\Phiu{t}-\iu{t}=0,\\ &&&\nabla \cdot (\uu{t})=0,
\end{aligned}
\label{eq:problem_statement}
\end{equation}
where we approximate the temporal derivative of the density guess $\frac{\partial\Phip{t}}{\partial t}$ with the density update $\Phiu{t}$
and $f(\Phiu{t},\uu{t})$ is our convex objective function, namely the brightness constancy assumption of optical flow or transport equation governing a fluid's motion. 
While the density change can be negative, the resulting actual density $\Phy{t}$ naturally may not contain any negative densities.
Thus, additional constraints are non-negativity of the total density $\Phy{t}$, incompressibility of the velocity update $\uu{t}$, and our linear image formation model, where $P$ is the matrix that projects the 3D density update $\Phiu{t}$ into the 2D image difference $\iu{t}$.

The key element of our reconstruction approach is the combination of the image formation model and the transport equation, where both density and velocity are unknowns.
Based on this, we achieve a robust reconstruction using only a single view without any additional depth information, as
will be demonstrated in more detail below.
In order to further restrict our solution, we add the following smoothness and kinetic energy (Tikhonov) regularizers to both the density and the velocity~\cite{wedel:2011:sceneFlow}.
In terms of a generic variable $\gv$, these regularizers are given by:
\begin{equation}
\begin{aligned}
	&E_{\text{smooth}}(\gv)&=&\half\norm{\nabla\gv}^2,
	&E_{\text{kinetic}}(\gv)&=&\half\norm{\gv}^2
\end{aligned}
\label{eq:smoothKinRegs}
\end{equation}

\subsection{Optimization Step} \label{sec:optstep}
In order to compute our density and velocity updates as described in \myrefeq{eq:problem_statement}, we make use of the fast primal-dual method (PD) for convex optimization introduced by Chambolle and Pock \shortcite{ChambollePD}.
The PD method is an iterative divide-and-conquer approach based on proximal operators.
Instead of solving for the whole complex optimization problem at once, subproblems are solved separately. 
Therefore, the problem is split into manageable components where proximal operators act as efficient solvers for each subproblem.
Iterative variable updates ensure that the solution converges to the optimal value of the problem in~\myrefeq{eq:problem_statement}. 
The simplified PD updates are given by
\begin{equation}
\begin{aligned}
\x^{k+1} &:= \x^k + \sigma \y^k - \sigma \, \prox_{f, \sigma}(\tfrac{1}{\sigma}\x^k + \y^k) \\
\z^{k+1} &:= \prox_{g, 1/\tau}(\z^k - \tau \x^{k+1}) \\
\y^{k+1} &:= \z^{k+1} + \theta(\z^{k+1}-\z^k),
\end{aligned}
\label{eq:PD_updates}
\end{equation}
where $\{\sigma,\tau,\theta\}$ are parameters that affect convergence, $k$ is the iteration number, $\z$ is the solution of~\myrefeq{eq:problem_statement}, $\x, \y$ are helper variables and $\prox$ are the proximal operators for each subproblem, see~\cite{Inglis:2017}.

We split our optimization problem from~\myrefeq{eq:problem_statement} 
into three different proximal operators, which we will explain in terms of a generic variable $\gv$.
The actual proximal operator calls are given in \myrefalg{alg:calcUpdate}.
By default, we assume that each variable contains concatenated velocity and density,
i.e., $\gv = (\gv_{\PhiuNoT},\gv_{\uuNoT})^T$. When a proximal operator acts on only 
either velocity or density, this will be indicated by a corresponding subscript.

First, $\prox_{f, \sigma}(\gv)$ targets the transport equation where both density $\Phiu{t}$ and velocity $\uu{t}$ update are unknowns. The objective function $f(\Phiu{t},\uu{t})$ is solved in a least-squares sense. The proximal operator for such a quadratic problem~\cite{Boyd:2011} is
\begin{equation}
	\prox_{f, \sigma}(\gv) =\; (\sigma I+A)^{-1}(\sigma\gv-b),
	\label{eq:proxF}
\end{equation}
where 
$A =\begin{bmatrix}I&(\nabla\Phip{t})^T\\(\nabla\Phip{t})&(\nabla\Phip{t})(\nabla\Phip{t})^T\end{bmatrix}$ and $b=0$.

Next, $\prox_{g, 1/\tau}(\gv)$ is split into two separate proximal operators for density and velocity update, which we point out by an additional subscript for the proximal operator.
The remaining constraint concerning the velocity update $\uu{t}$ is the incompressibility:
\begin{equation}
\prox_{g, 1/\tau, \;\uuNoT}(\gv_{\uuNoT}) = \PiDiv(\gv_{\uuNoT}).
\label{eq:proxG_u}
\end{equation}
Making a velocity field $\uu{t}$ divergence-free is an orthogonal projection onto the space of
divergence-free velocity fields $C_{\mathrm{DIV}}$. Here we employ the pressure solver
that is a typical component of an Eulerian fluid simulator.

The second part of $\prox_{g, 1/\tau}(\gv)$ concerns density constraints.
The density update must comply with both the image formation and the non-negativity constraint.
In order to fulfill both constraints on $\Phiu{t}$ at once, we introduce a second, separate PD loop obtaining a correction $\Phic{t}$ of $\Phiu{t}$.
Ensuring the non-negativity of the density is an orthogonal projection onto the set of non-negative real numbers $\mathbf{R}_+$.
We realize this projection efficiently by setting values in $\Phic{t}$ to the maximal value allowed such that $\Phic{t}+\Phiu{t}+\Phip{t}$ is non-negative.
The correction $\Phic{t}$ itself is allowed to be negative.
To match $\Phic{t}$ with the 2D input image $\ic{t}=\i{t}-\ip{t}-\iu{t}$, a second least-squares minimization problem is solved due to the sparse and ill-conditioned nature of the image formation matrix system. 
We apply the same proximal operator scheme for quadratic problems as in~\myrefeq{eq:proxF}.
\begin{equation}
\begin{aligned}
\prox_{g, 1/\tau, \;\PhiuNoT}(\gv_{\PhiuNoT}) = \gv_{\PhiuNoT}+
\underset{\Phic{t}}{\text{arg min}} \norm{P\:\Phic{t}-\ic{t}}^2\\
\text{subject to } \;
\Phic{t}+\Phiu{t}+\Phip{t}\geq 0,
\end{aligned}
\label{eq:proxG_Phi}
\end{equation}
where the two proximal operators of the second PD loop are defined as
\begin{align}
&\prox_{f2, \sigma_2, \PhicNoT}(\gv_{\PhicNoT}) &= &\;(\sigma_2 I+P^TP)^{-1}(\sigma_2\gv_{\PhicNoT}+P^T\ic{t}) \label{eq:proxG_phi_proxF}\\[2pt]
&\prox_{g2, \frac{1}{\tau_2}, \PhicNoT}(\gv_{\PhicNoT}) &= &\;\PiREAL(\gv_{\PhicNoT}). \label{eq:proxG_phi_proxG}
\end{align}

The respective regularizing terms from~\myrefeq{eq:smoothKinRegs} are added to both least squares matrices, i.e. $A$ and $P^TP$.
We use a constant set of PD parameters that we found to converge robustly for all our scenarios:  $(\sigma_{\PhiuNoT},$$\tau_{\PhiuNoT},$$\theta_{\PhiuNoT},$$\sigma_{\uuNoT},$ $\tau_{\uuNoT},$$\theta_{\uuNoT},$$\sigma_{\PhiuNoT 2},$$\tau_{\PhiuNoT 2},$$\theta_{\PhiuNoT 2}) = (10,0.01,1,0.1,5,1,0.01,100,1)$. These parameters are specific to our problem formulation and 
control the speed of convergence for each of the two PD loops.

The optimization for the density and velocity update is summarized in~\myrefalg{alg:calcUpdate}.
As before, variables without subscript, such as $\x, \y, \z$, contain both $\PhiuNoT$ and $\uuNoT$,
while subscripts denote the two separate parts.
\begin{algorithm}
	\caption{Coupled Density and Velocity Update}
	\label{alg:calcUpdate}
	\begin{algorithmic}[1]
		\Procedure{calculateUpdate}{$\Phip{t}$, $\iu{t}$}
		\setlength{\parskip}{1ex}
		\State $\x^{k+1} = \x^k + \sigma \y^k - \sigma \, \prox_{f, \sigma}(\tfrac{1}{\sigma}\x^k + \y^k)$
		\State $\z^{k+1}_{\uuNoT} = \prox_{g, 1/\tau, \uuNoT}(\z^k_{\uuNoT} - \tau \x^{k+1}_{\uuNoT})$
		\State $\z^{k+1}_{\PhiuNoT} = \prox_{g, 1/\tau, \PhiuNoT}(\z^k_{\PhiuNoT} - \tau \x^{k+1}_{\PhiuNoT})$
		\State $\y^{k+1} = \z^{k+1} + \theta(\z^{k+1}-\z^k)$
		\EndProcedure
	\end{algorithmic}
\end{algorithm}
We can also perform this solve from coarse to fine spatial scales in order to realize a multi-scale scheme typically used for optical flow to enable the reconstruction of large displacements, see~\cite{meinhardt2013horn}. Quantities from a coarse scale are advected forward;
the finer solve computes the corresponding velocity update.

\subsection{Ray Sampling and Camera}
To obtain a mapping from the density volume to the image space, we cast a ray for each pixel through the voxelized density and save the corresponding weights in a projection matrix $P$. 
We use piece-wise linear basis functions for interpolation and a visual hull to reduce the size of our equation system. 
As real-world cameras are perspective pinhole cameras, we use perspective cameras in our final reconstruction method.
Orthographic cameras could be used under the assumption of having a large distance to the fluid.
\begin{figure}[t!]
	\centering
	\begin{overpic}[trim={0.1cm 0cm 8.5cm 0.cm},clip,width=0.165\linewidth]{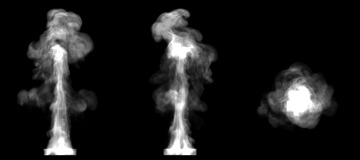}
	\put(4,85){\small \color{white}{$a)$}}\end{overpic}\hspace{0em}%
	\includegraphics[trim={4.3cm 0cm 4.3cm 0.cm},clip,width=0.165\linewidth]{images/noRegs/input/density_104_high}\hfill
	\begin{overpic}[trim={0.1cm 0cm 8.5cm 0.cm},clip,width=0.165\linewidth]{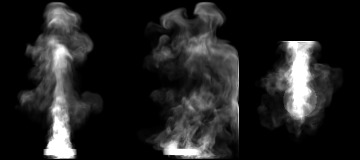}
	\put(4,85){\small \color{white}{$b)$}}\end{overpic}\hspace{0em}%
	\includegraphics[trim={4.3cm 0cm 4.3cm 0.cm},clip,width=0.165\linewidth]{images/noRegs/persp/volRecon_104_high}\hfill
	\begin{overpic}[trim={0.1cm 0cm 8.5cm 0.cm},clip,width=0.165\linewidth]{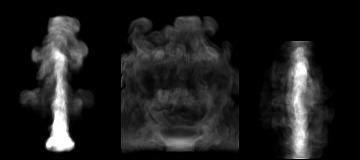}
	\put(4,85){\small \color{white}{$c)$}}\end{overpic}\hspace{0em}%
	\includegraphics[trim={4.3cm 0cm 4.3cm 0.cm},clip,width=0.165\linewidth]{images/noRegs/ortho/volRecon_104_high}
\caption{a) Input, b) perspective, and c) orthographic reconstructions without depth regularizer at $t=104$. Each image shows a front (left) and side view (right). The side views of b) and c) highlight the undesirable depth expansion and the C-shaped motion for the perspective camera view in b).}
\label{fig:noZRegs}
\end{figure}

In~\myreffig{fig:noZRegs}, we show reconstruction results with our method as explained so far for both perspective and orthographic cameras.
The input is a synthetic smoke plume as shown in \ref{fig:noZRegs}a).
It is noticeable that the reconstructed density expands into depth.
While orthographic cameras in \ref{fig:noZRegs}c) reconstruct symmetric density volumes in depth, we observe a C-shaped density for perspective reconstructions in the side view, see \ref{fig:noZRegs}b).
This curved shape is caused by motions being reconstructed as orthogonal to the viewing rays.
While the density update merely acts along each viewing ray, optical flow reconstructs velocities where density gradients are seen, i.e., orthogonal to the viewing direction. 
For perspective cameras, this causes the motion to tilt towards or away from the viewer near the top and bottom of an image.
This distortion is a result of the strongly under-determined single view case. Additional views
could prevent this distortion, as illustrated in \myreffig{fig:frontAs} and~\myreffig{fig:compareTomoOF}.
However, in order to arrive at a working single-view reconstruction algorithm,
we introduce the following depth regularization terms.

\subsection{Depth Regularization}
In order to gain control over the motion in depth, we constrain the velocities in depth more than in other directions where its direct effect is visible in the 2D images. 
First, we increase the weight of the kinetic energy penalty, i.e., kinetic energy regularizer (\myrefeq{eq:smoothKinRegs}), in the depth direction.
The z-velocity component is therefore more strongly regularized than the x- and y-components.

To avoid a drift of the densities to the front or back of the domain, we introduce a second regularizer that constrains the sum of depth velocities in one voxel row to be zero inside the density volume.
We minimize the following energy in \myrefeq{eq:sumRegs} where $S$ is a matrix that sums up the z-velocities along one voxel row in depth:
\begin{equation}
\begin{aligned}
& & E_{\text{sum}}(\uuNoT) = \half\norm{S\uuNoT}^2.
\end{aligned}
\label{eq:sumRegs}
\end{equation}
For perspective reconstructions, we use an additional adaptive z-Tik\-ho\-nov regularizer where we have higher regularizing weights at the bottom and at the top compared to the center of the domain.
The formula for the adaptive z-Tik\-ho\-nov weight is $\lambda_{\text{adapt tiko}}=1+10*(|(Y / 2.) - 1. - j|/(Y / 2.))^2$, where Y is the domain size in y-direction and j is the current voxel's height.
In summary, we have the following weighted regularizing energies to minimize:
\begin{equation}
\begin{aligned}
E_{\text{reg}}(\PhiuNoT,\uuNoT) &&=\; &\alpha_{\PhiuNoT} E_{\text{smooth}}(\PhiuNoT) + \alpha_{\uuNoT} E_{\text{smooth}}(\uuNoT) \\
&&+\; &\beta_{\PhiuNoT}E_{\text{kinetic}}(\PhiuNoT) + \beta_{\uuNoT}E_{\text{kinetic}}(\uuNoT) \\
&&+\; &\lambda_{\text{sum}}E_{\text{sum}}(\uuNoT)
+ \lambda_{\text{tiko}}E_{\text{kinetic}}((\uuNoT\text).z) .
\end{aligned}
\label{eq:allRegs}
\end{equation}

\subsection{Discretization}
In the standard Horn-Schunck optical flow and in previous approaches~\cite{Gregson:2014}, the brightness constancy assumption has been discretized on a collocated velocity grid. 
Pressure solvers on a collocated grid typically introduce checkerboard artifacts if the velocity field is not smooth enough.
The divergence of the velocity is best eliminated on a staggered Marker-and-Cell (MAC) grid~\cite{harlow1965}. 
Using a collocated grid for the optical flow part and a staggered grid for the pressure solver would introduce smoothing and divergence errors from interpolation.
Thus, we use the staggered grid as velocity discretization for all parts of our algorithm.

\subsection{Source Estimation} \label{sec:inflow}
Our algorithm reconstructs density and velocity in 3D with the only input being a sequence of 2D target images.
In order to start the reconstruction, we estimate a smoke source $\Phy{0}$ from the first 2D image at time step $t_0$. 
The source region is constructed by projecting the 2D image into 3D space and by limiting it in depth, and if necessary in other spatial directions to meet a target shape. 
For our results, we typically constrain the source shape to not be larger than a cylinder or a box.
Note that we do not use the ground truth source for synthetic reconstructions 
in order to make them representative of real world settings.
We add the estimated source each time step to the density to model a continuous production of smoke. 
For real world cases where smoke enters the view with an initial velocity, we also prescribe
a manually chosen velocity for the source region.

\section{Results} \label{sec:results}
\begin{figure*}[th!]
	\centering
	\begin{tikzpicture}
	\draw (0,0) node[inner sep=0] {\includegraphics[trim={2.6cm 1cm 2.6cm 6.5cm},clip,width=0.165\linewidth]{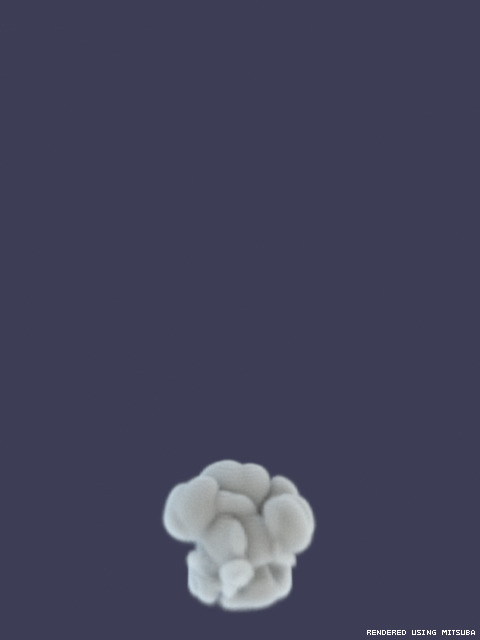}\hspace{0.0cm}%
	\includegraphics[trim={2.6cm 1cm 2.6cm 6.5cm},clip,width=0.165\linewidth]{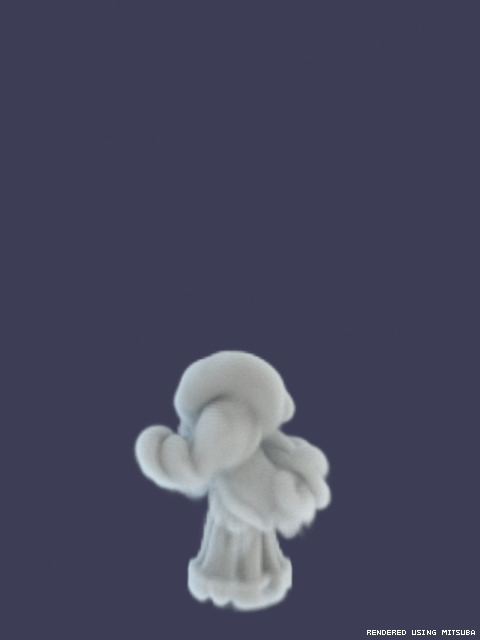}\hspace{0.0cm}%
	\includegraphics[trim={2.6cm 1cm 2.6cm 6.5cm},clip,width=0.165\linewidth]{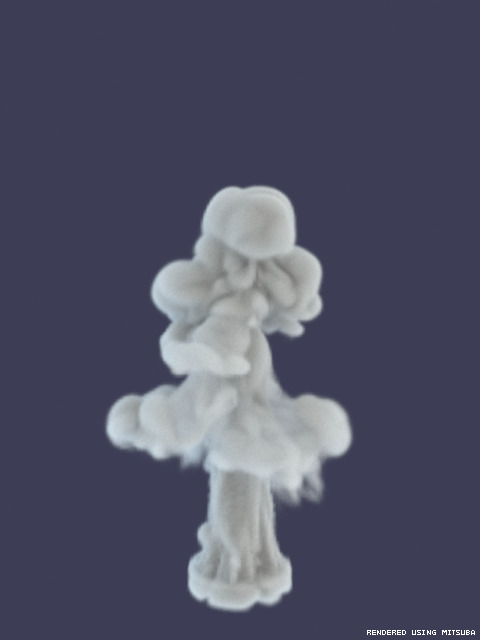}};
	\small \color{white} \draw (-3.4, 1.6) node {$a)$ input, front};
	\end{tikzpicture}
	\hfill
	\begin{tikzpicture}
	\draw (0,0) node[inner sep=0] {\includegraphics[trim={2.6cm 1cm 2.6cm 6.5cm},clip,width=0.165\linewidth]{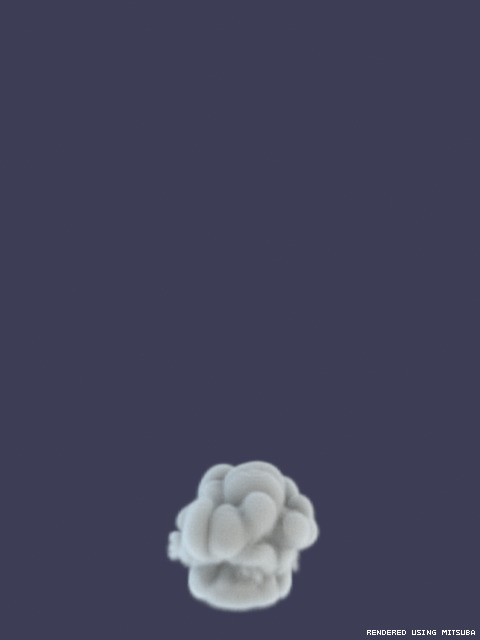}\hspace{0.0cm}%
	\includegraphics[trim={2.6cm 1cm 2.6cm 6.5cm},clip,width=0.165\linewidth]{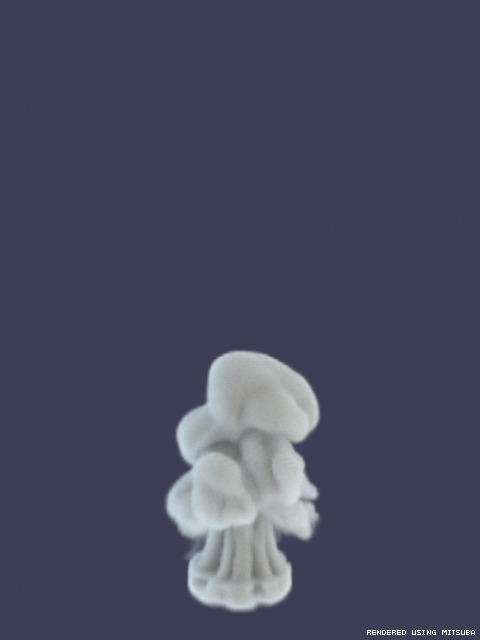}\hspace{0.0cm}%
	\includegraphics[trim={2.6cm 1cm 2.6cm 6.5cm},clip,width=0.165\linewidth]{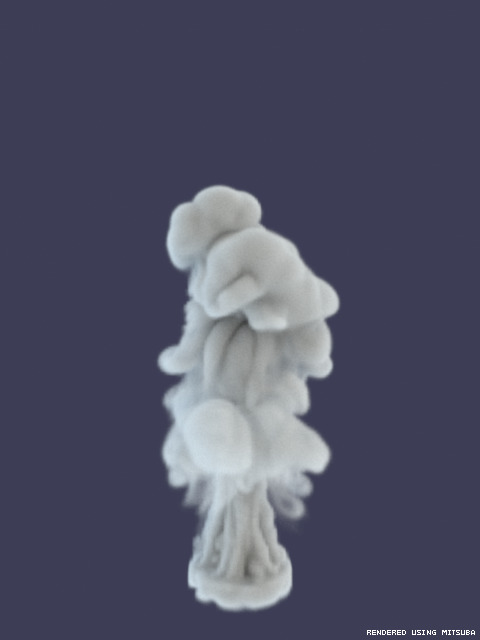}};
	\small \color{white} \draw (-3.4, 1.6) node {$b)$ input, side};
	\end{tikzpicture}\\
	\vspace{0.1mm}
	\begin{tikzpicture}
	\draw (0,0) node[inner sep=0] {\includegraphics[trim={2.6cm 1cm 2.6cm 6.5cm},clip,width=0.165\linewidth]{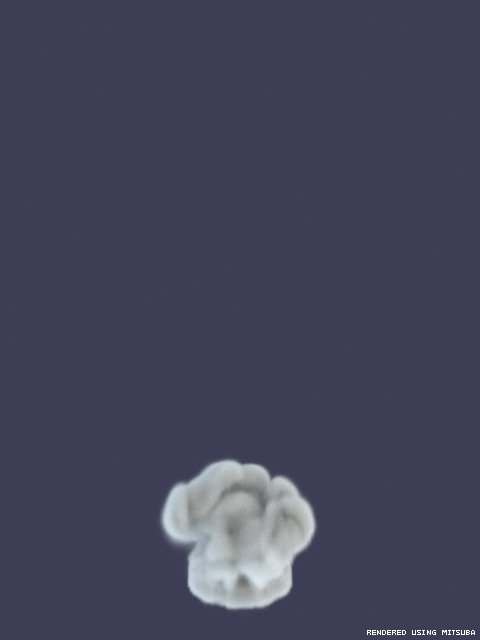}\hspace{0.0cm}%
	\includegraphics[trim={2.6cm 1cm 2.6cm 6.5cm},clip,width=0.165\linewidth]{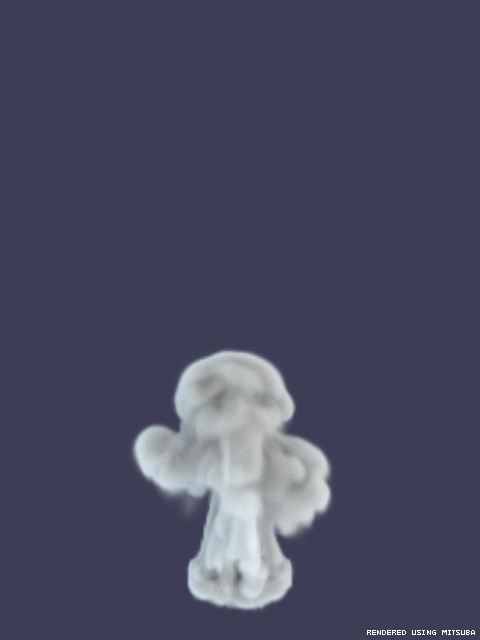}\hspace{0.0cm}%
	\includegraphics[trim={2.6cm 1cm 2.6cm 6.5cm},clip,width=0.165\linewidth]{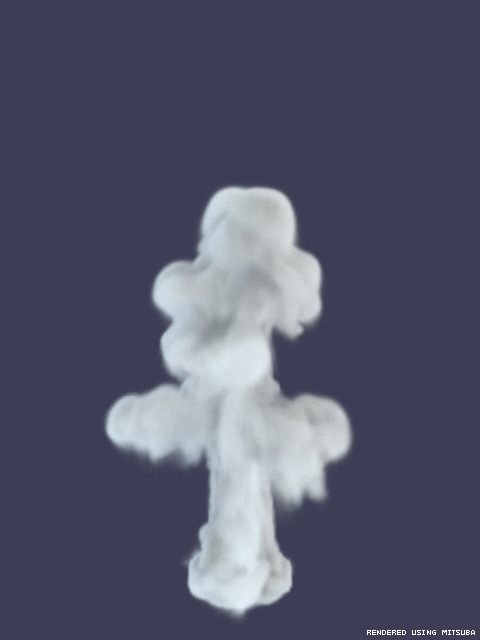}};
	\small \color{white} \draw (-2.88, 1.6) node {$c)$ reconstruction, front};
	\end{tikzpicture}
	\hfill
	\begin{tikzpicture}
	\draw (0,0) node[inner sep=0] {\includegraphics[trim={2.6cm 1cm 2.6cm 6.5cm},clip,width=0.165\linewidth]{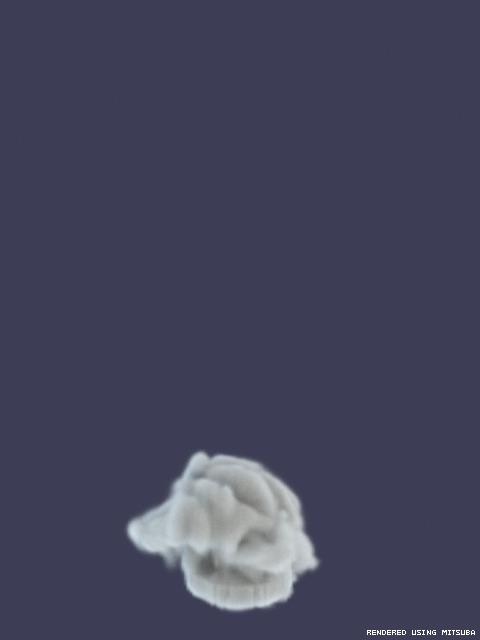}\hspace{0.0cm}%
	\includegraphics[trim={2.6cm 1cm 2.6cm 6.5cm},clip,width=0.165\linewidth]{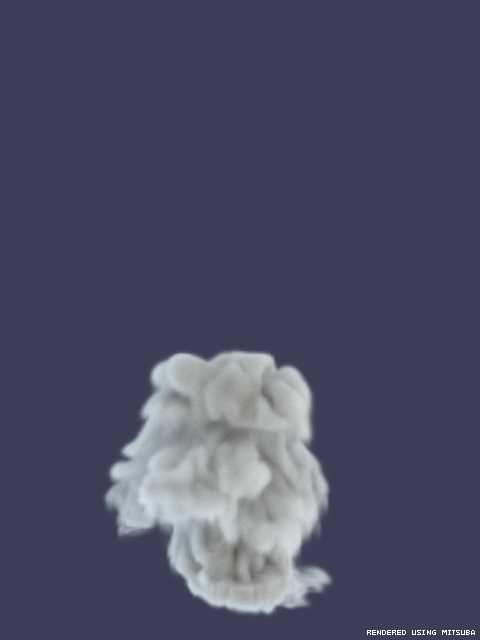}\hspace{0.0cm}%
	\includegraphics[trim={2.6cm 1cm 2.6cm 6.5cm},clip,width=0.165\linewidth]{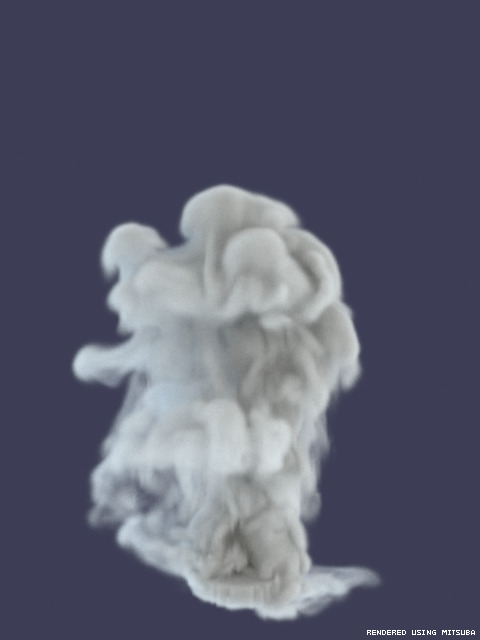}};
	\small \color{white} \draw (-2.88, 1.6) node {$d)$ reconstruction, side};
	\end{tikzpicture}\\
	\vspace{1.5mm}
	\begin{tikzpicture}
	\draw (0,0) node[inner sep=0] {\includegraphics[trim={0cm 2.0cm 0cm 0cm},clip,width=0.165\linewidth]{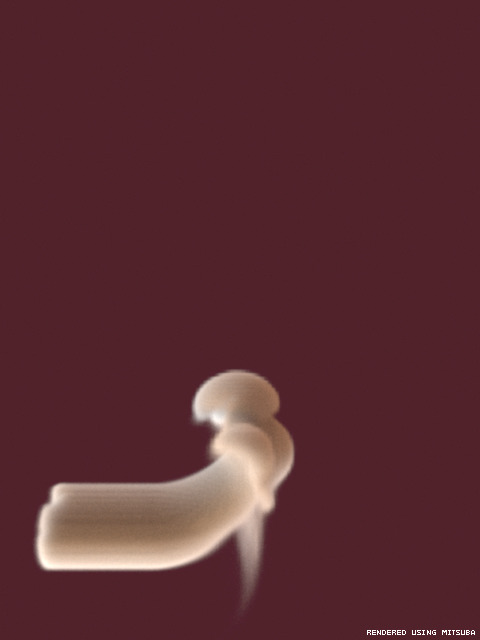}\hspace{0.0cm}%
	\includegraphics[trim={0cm 2.0cm 0cm 0cm},clip,width=0.165\linewidth]{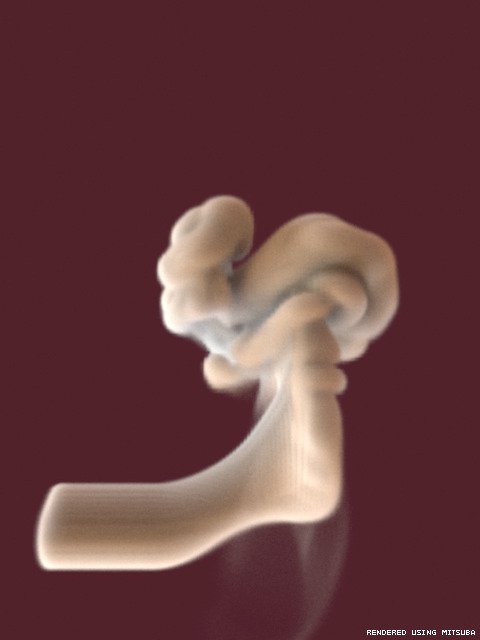}\hspace{0.0cm}%
	\includegraphics[trim={0cm 2.0cm 0cm 0cm},clip,width=0.165\linewidth]{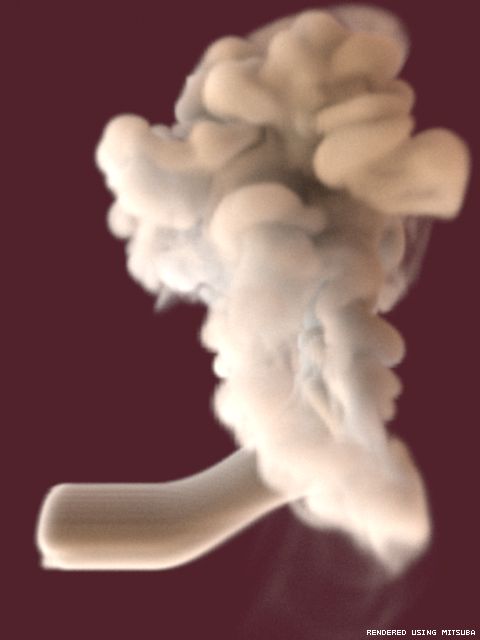}};
	\small \color{white} \draw (-3.4, 1.5) node {$e)$ input, front};
	\end{tikzpicture}
	\hfill
	\begin{tikzpicture}
	\draw (0,0) node[inner sep=0] {\includegraphics[trim={0cm 2.0cm 0cm 0cm},clip,width=0.165\linewidth]{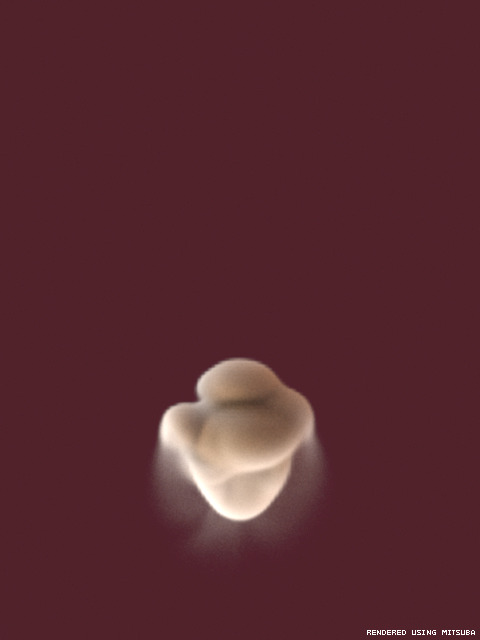}\hspace{0.0cm}%
	\includegraphics[trim={0cm 2.0cm 0cm 0cm},clip,width=0.165\linewidth]{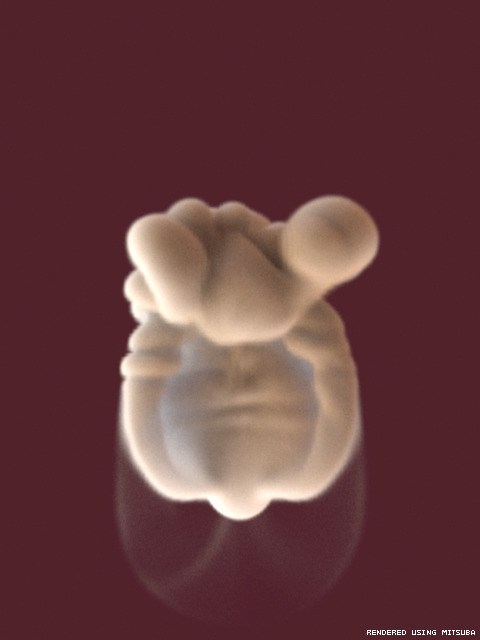}\hspace{0.0cm}%
	\includegraphics[trim={0cm 2.0cm 0cm 0cm},clip,width=0.165\linewidth]{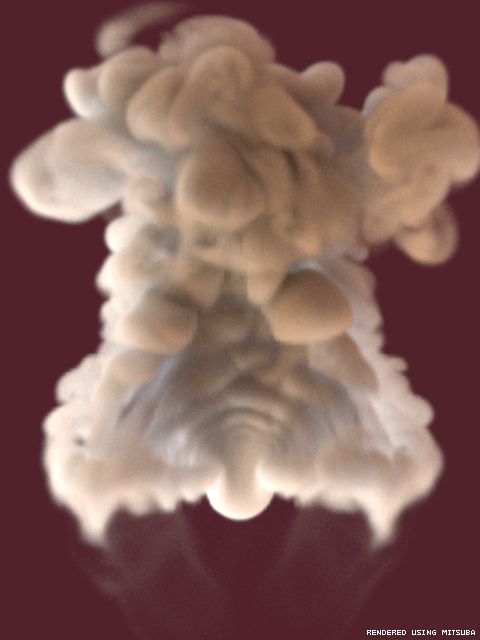}};
	\small \color{white} \draw (-3.4, 1.5) node {$f)$ input, side};
	\end{tikzpicture}\\
	\vspace{0.1mm}
	\begin{tikzpicture}
	\draw (0,0) node[inner sep=0] {	\includegraphics[trim={0cm 2.0cm 0cm 0cm},clip,width=0.165\linewidth]{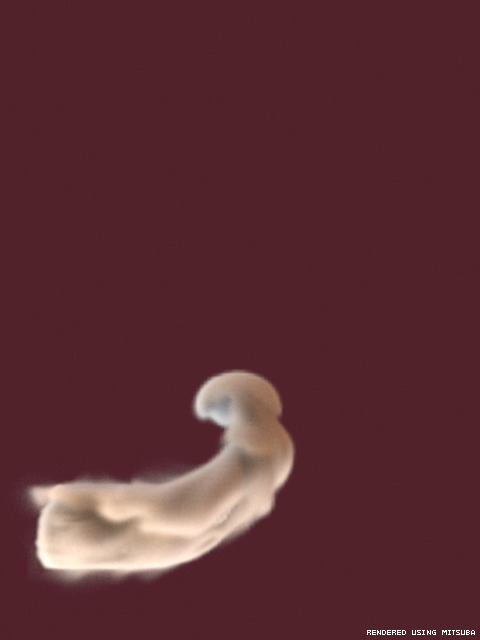}\hspace{0.0cm}%
	\includegraphics[trim={0cm 2.0cm 0cm 0cm},clip,width=0.165\linewidth]{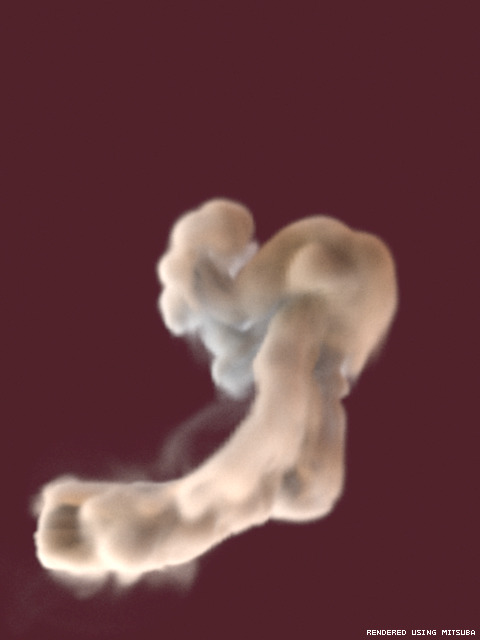}\hspace{0.0cm}%
	\includegraphics[trim={0cm 2.0cm 0cm 0cm},clip,width=0.165\linewidth]{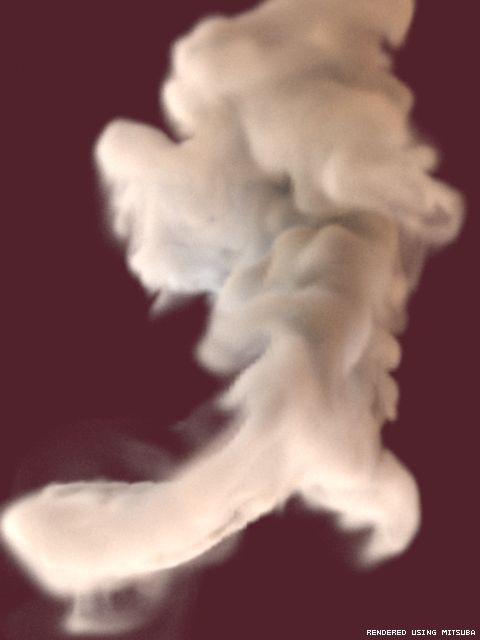}};
	\small \color{white} \draw (-2.88, 1.5) node {$g)$ reconstruction, front};
	\end{tikzpicture}\hfill
	\begin{tikzpicture}
	\draw (0,0) node[inner sep=0] {\includegraphics[trim={0cm 2.0cm 0cm 0cm},clip,width=0.165\linewidth]{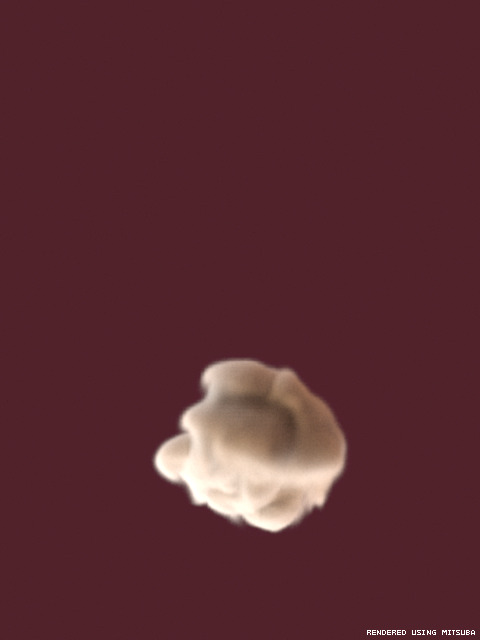}\hspace{0.0cm}%
	\includegraphics[trim={0cm 2.0cm 0cm 0cm},clip,width=0.165\linewidth]{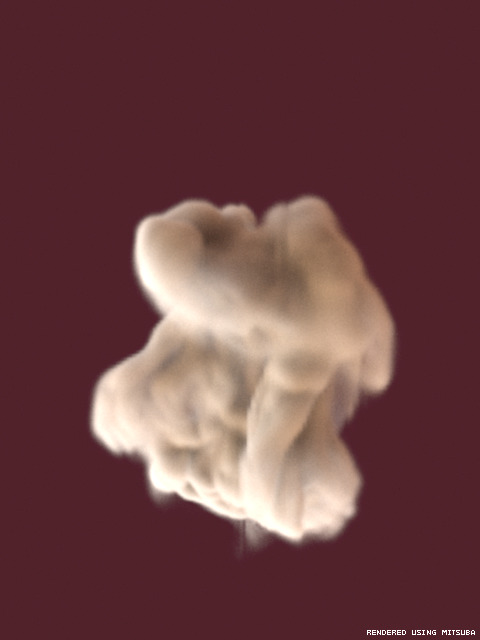}\hspace{0.0cm}%
	\includegraphics[trim={0cm 2.0cm 0cm 0cm},clip,width=0.165\linewidth]{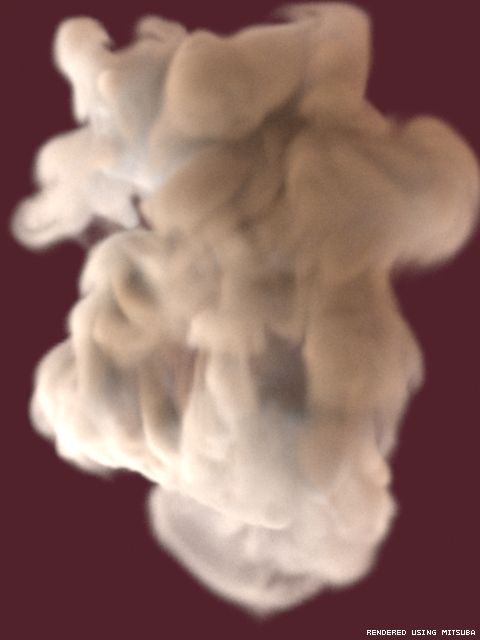}};
	\small \color{white} \draw (-2.88, 1.5) node {$h)$ reconstruction, side};
	\end{tikzpicture}\\
	\vspace{1.5mm}
	\begin{tikzpicture}
	\draw (0,0) node[inner sep=0] {\includegraphics[trim={2cm 1cm 2cm 8.4cm},clip,width=0.165\linewidth]{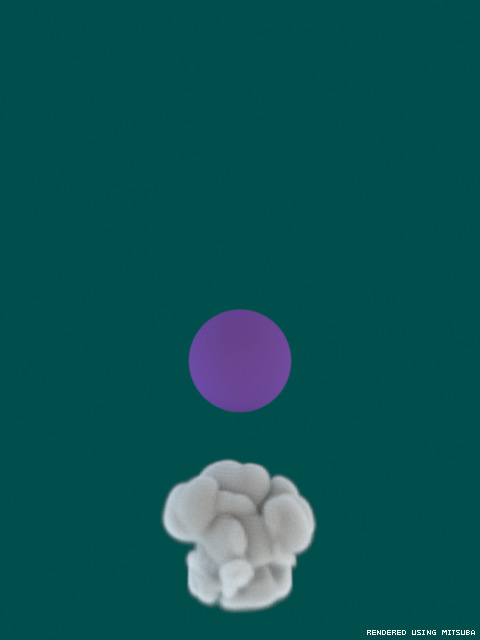}\hspace{0.0cm}%
	\includegraphics[trim={2cm 1cm 2cm 8.4cm},clip,width=0.165\linewidth]{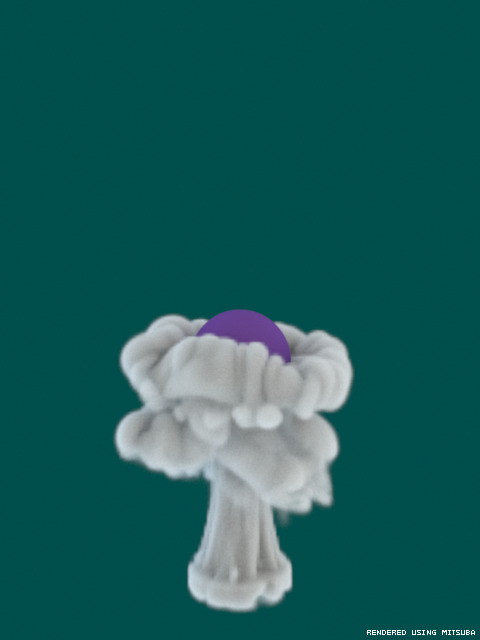}\hspace{0.0cm}%
	\includegraphics[trim={2cm 1cm 2cm 8.4cm},clip,width=0.165\linewidth]{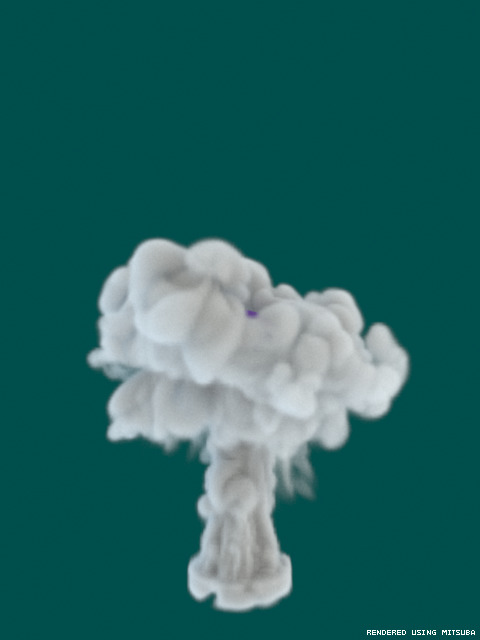}};
	\small \color{white} \draw (-3.4, 1.2) node {$i)$ input, front};
	\end{tikzpicture}\hfill
	\begin{tikzpicture}
	\draw (0,0) node[inner sep=0] {\includegraphics[trim={2cm 1cm 2cm 8.4cm},clip,width=0.165\linewidth]{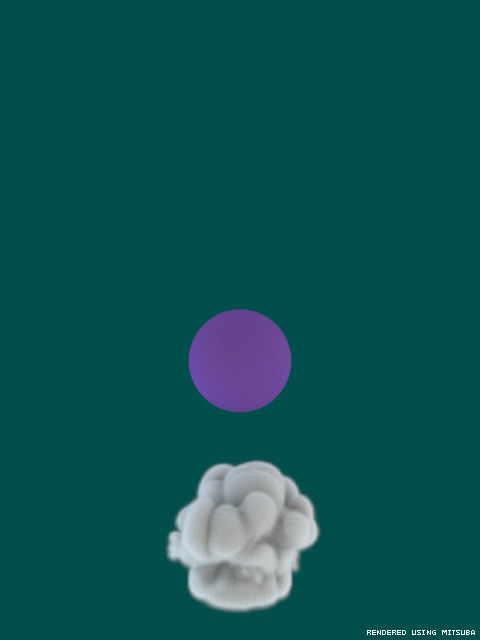}\hspace{0.0cm}%
	\includegraphics[trim={2cm 1cm 2cm 8.4cm},clip,width=0.165\linewidth]{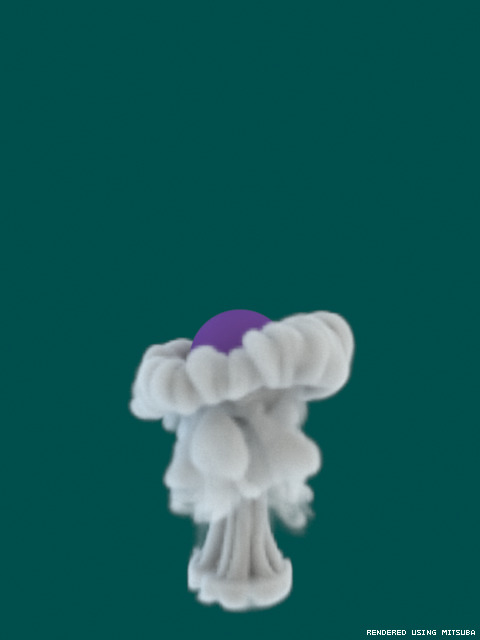}\hspace{0.0cm}%
	\includegraphics[trim={2cm 1cm 2cm 8.4cm},clip,width=0.165\linewidth]{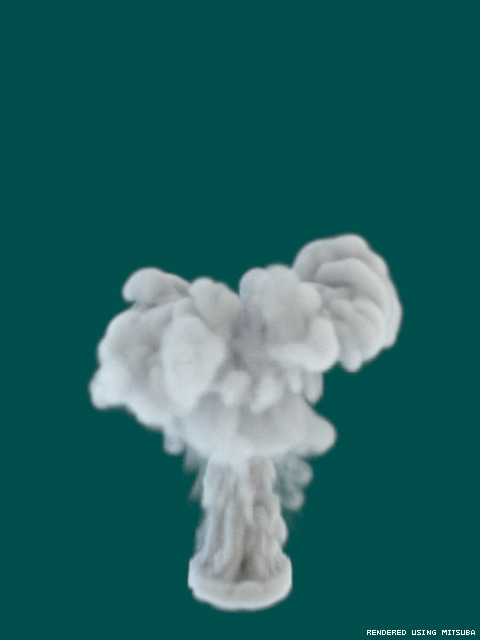}};
	\small \color{white} \draw (-3.4, 1.2) node {$j)$ input, side};
	\end{tikzpicture}\\
	\vspace{0.1mm}
	\begin{tikzpicture}
	\draw (0,0) node[inner sep=0] {\includegraphics[trim={2cm 1cm 2cm 8.4cm},clip,width=0.165\linewidth]{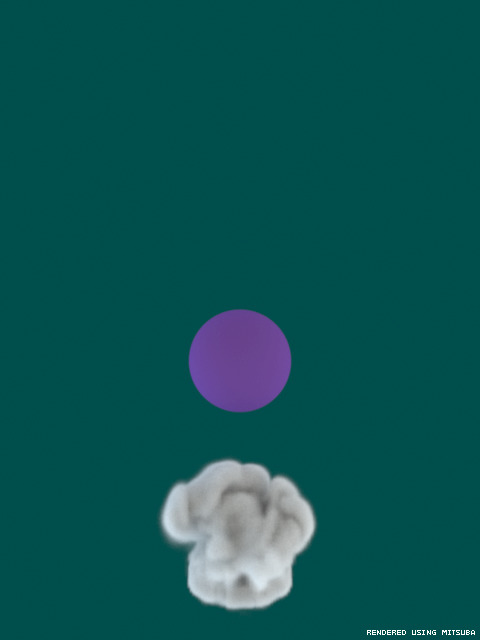}\hspace{0.0cm}%
	\includegraphics[trim={2cm 1cm 2cm 8.4cm},clip,width=0.165\linewidth]{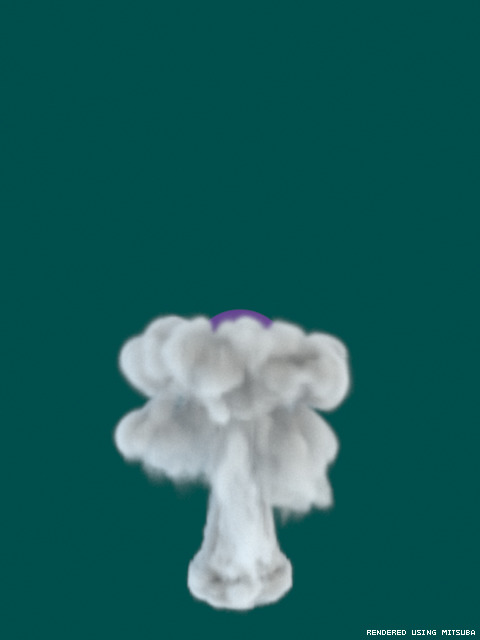}\hspace{0.0cm}%
	\includegraphics[trim={2cm 1cm 2cm 8.4cm},clip,width=0.165\linewidth]{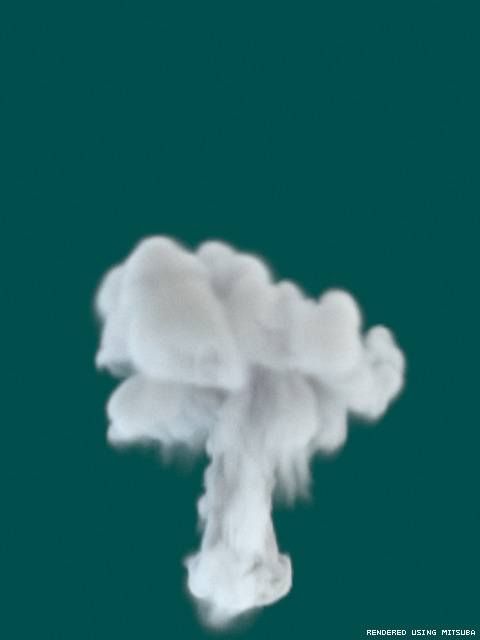}};
	\small \color{white} \draw (-2.88, 1.2) node {$k)$ reconstruction, front};
	\end{tikzpicture}
	\hfill
	\begin{tikzpicture}
	\draw (0,0) node[inner sep=0] {\includegraphics[trim={2cm 1cm 2cm 8.4cm},clip,width=0.165\linewidth]{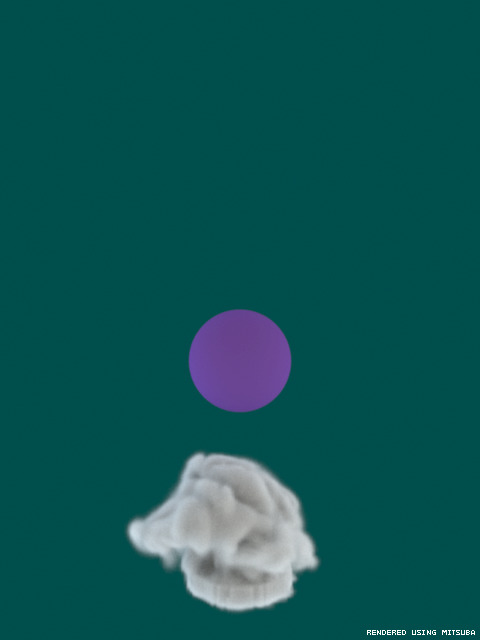}\hspace{0.0cm}%
	\includegraphics[trim={2cm 1cm 2cm 8.4cm},clip,width=0.165\linewidth]{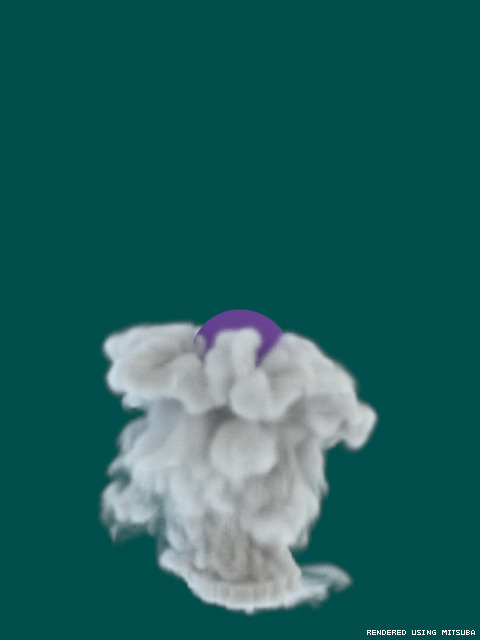}\hspace{0.0cm}%
	\includegraphics[trim={2cm 1cm 2cm 8.4cm},clip,width=0.165\linewidth]{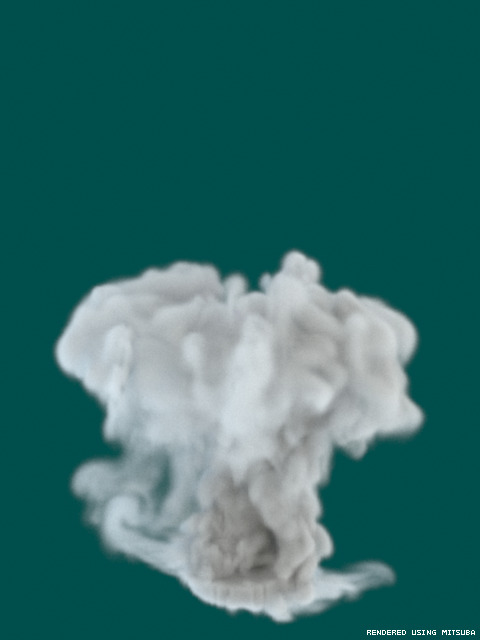} };
	\small \color{white} \draw (-2.88, 1.2) node {$l)$ reconstruction, side};
	\end{tikzpicture}
	\caption{Front and side view of input a,b) and reconstruction c,d) of a plume at $t=30,60,96$, a jet stream e) - h) at $t=30,60,96$, and a plume with sphere i) - l) at $t=30,80,112$.}	\label{fig:jetobs_mitsuba}
\end{figure*}
In order to evaluate our method, we reconstruct synthetic flows for which we have ground truth density and velocity information.
These reconstructions do not use a velocity inflow.
Later on, we also compare our results to previous work and demonstrate
reconstructions of filmed, real-world smoke clouds.

\subsection{Evaluation with Synthetic Inputs}
First, we simulate a rising smoke plume and raycast the density volume at each time step to create a sequence of 2D images that we use as single input to our reconstruction algorithm.  
The image resolution is $480\times640$ while the 3D domain size is $120\times160\times120$.
Our choice of parameters is almost constant for all reconstructions, except for real input data, which requires higher smoothness. 
For our synthetic results, we use the smoothness weights $\alpha_{\mathbf{u}}=1e-1$ and $\alpha_\Phi=1e-3$, the Tikhonov regularizer weight 
$\beta_{\PhiuNoT}=\beta_{\uuNoT}=1e-4$ and the depth-regularizing weights $\lambda_{\text{tiko}}=1e-3,\lambda_{\text{sum}}=10$.

\textbf{Plume with Single View:}
The input fluid simulation and perspective reconstructions are shown in \myreffig{fig:jetobs_mitsuba} a) - d) for front and side views.
Note that the side view is the most challenging viewing angle for single view reconstructions as it is orthogonal to the input view.
The input view in \myreffig{fig:jetobs_mitsuba} a) is matched very well, see c). 
The side views in b) and d) show that the reconstructed density expands in depth.
However, considering that we have no explicit constraints for this motion, our method
still reconstructs a very natural flow behavior.

\begin{figure}[h!]
	\centering
	\begin{tikzpicture}
	\draw (0,0) node[inner sep=0] {
	\includegraphics[trim={4cm 6cm 4cm 6cm},clip,width=0.164\linewidth]{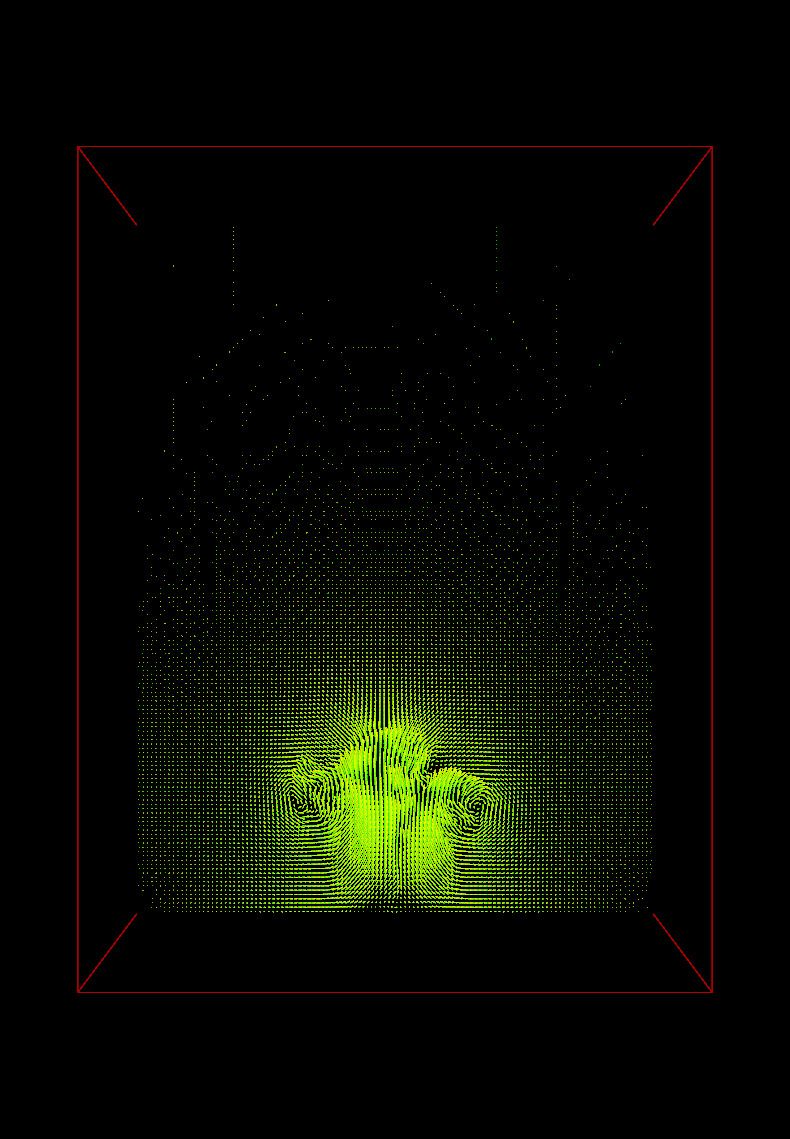}\hspace{0.0cm}%
	\includegraphics[trim={4cm 6cm 4cm 6cm},clip,width=0.164\linewidth]{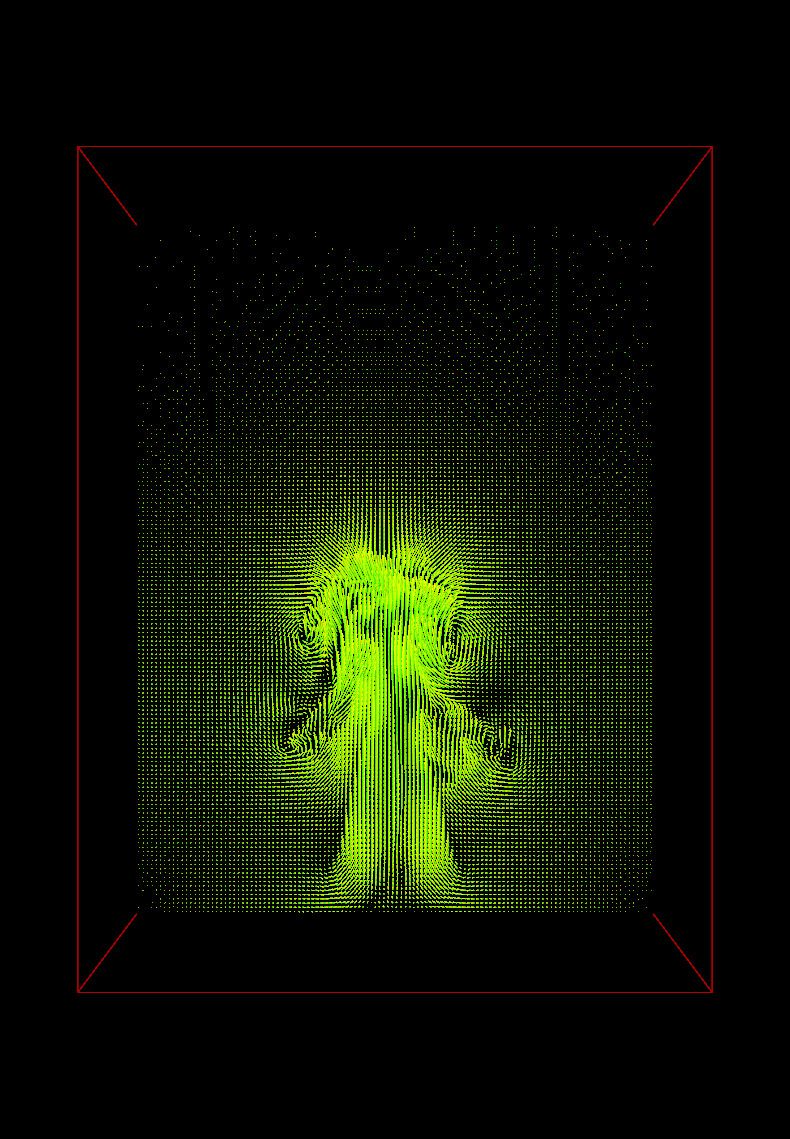}\hspace{0.0cm}%
	\includegraphics[trim={4cm 6cm 4cm 6cm},clip,width=0.164\linewidth]{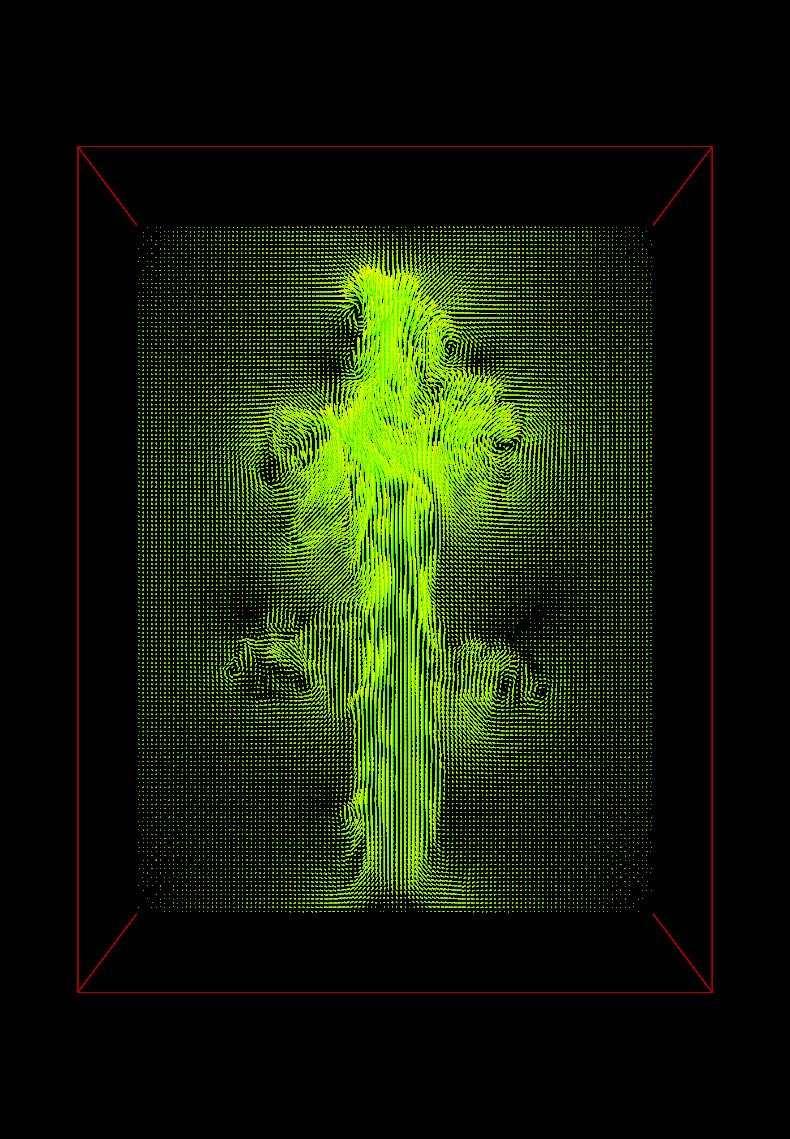}};
	\small \color{white} \draw (-1.15, 0.75) node {$a)$ input, front};
	\end{tikzpicture}
	\hfill
	\begin{tikzpicture}
	\draw (0,0) node[inner sep=0] {
	\includegraphics[trim={4cm 6cm 4cm 6cm},clip,width=0.164\linewidth]{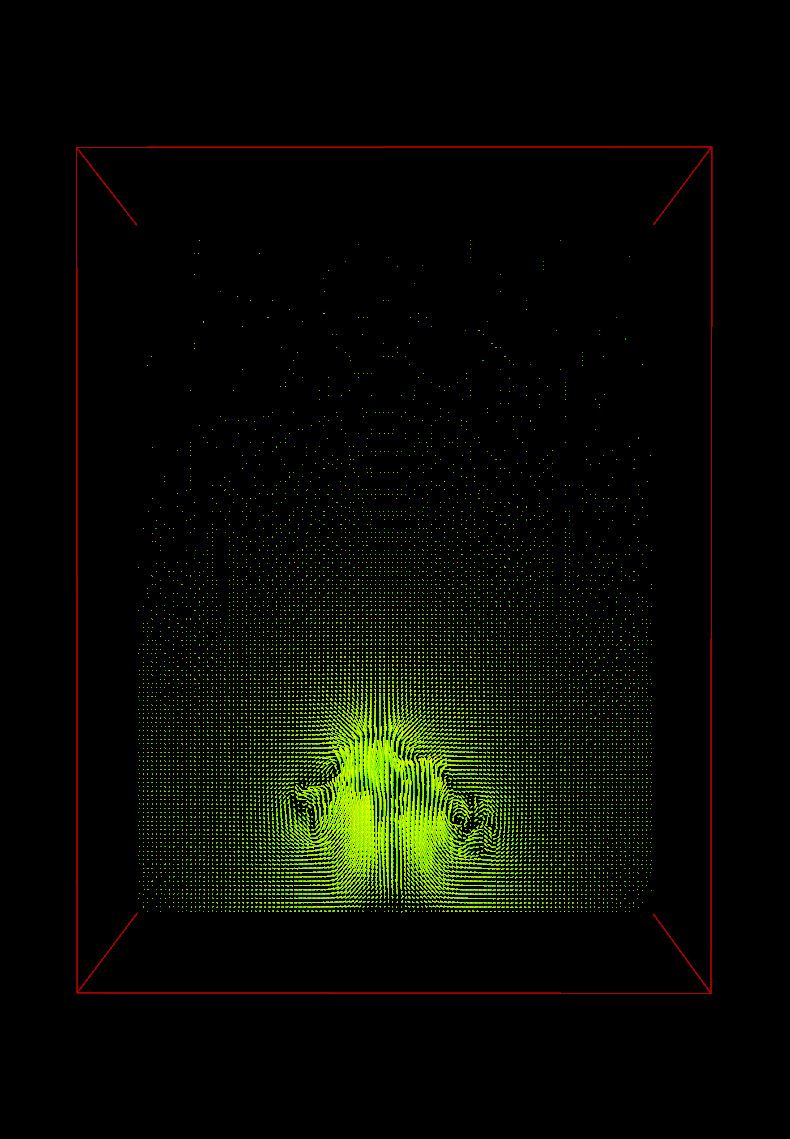}\hspace{0.0cm}%
	\includegraphics[trim={4cm 6cm 4cm 6cm},clip,width=0.164\linewidth]{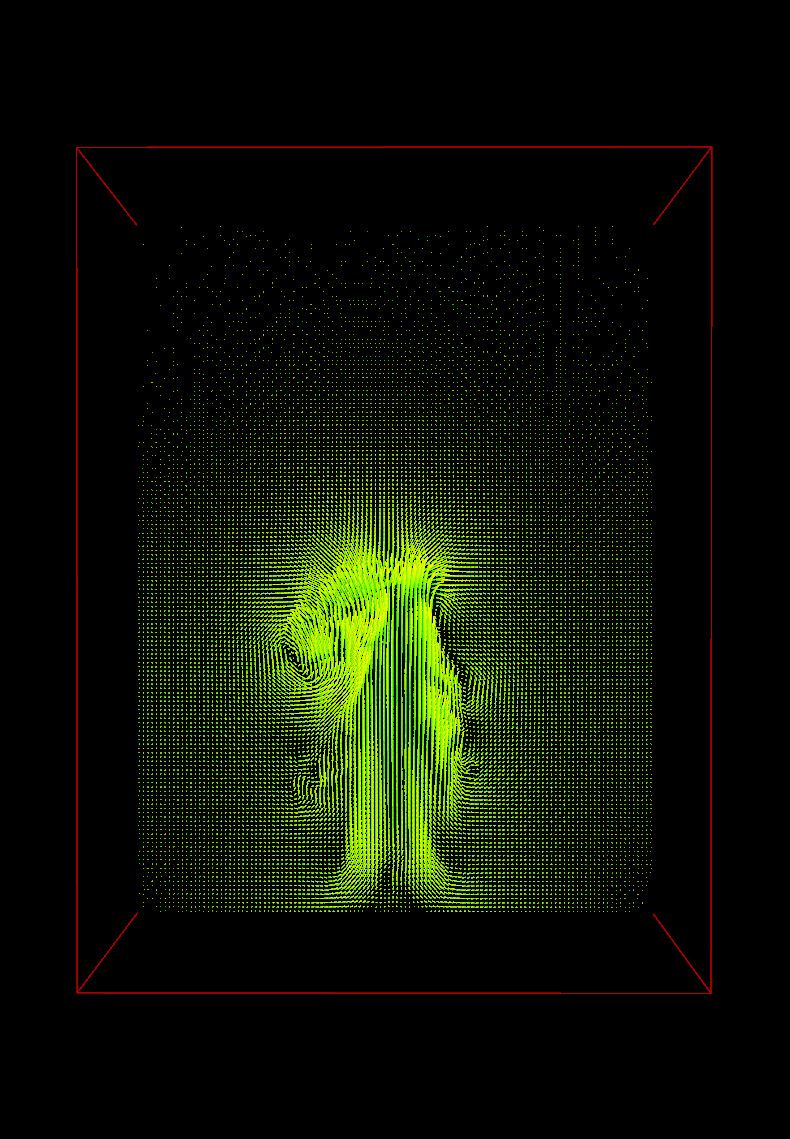}\hspace{0.0cm}%
	\includegraphics[trim={4cm 6cm 4cm 6cm},clip,width=0.164\linewidth]{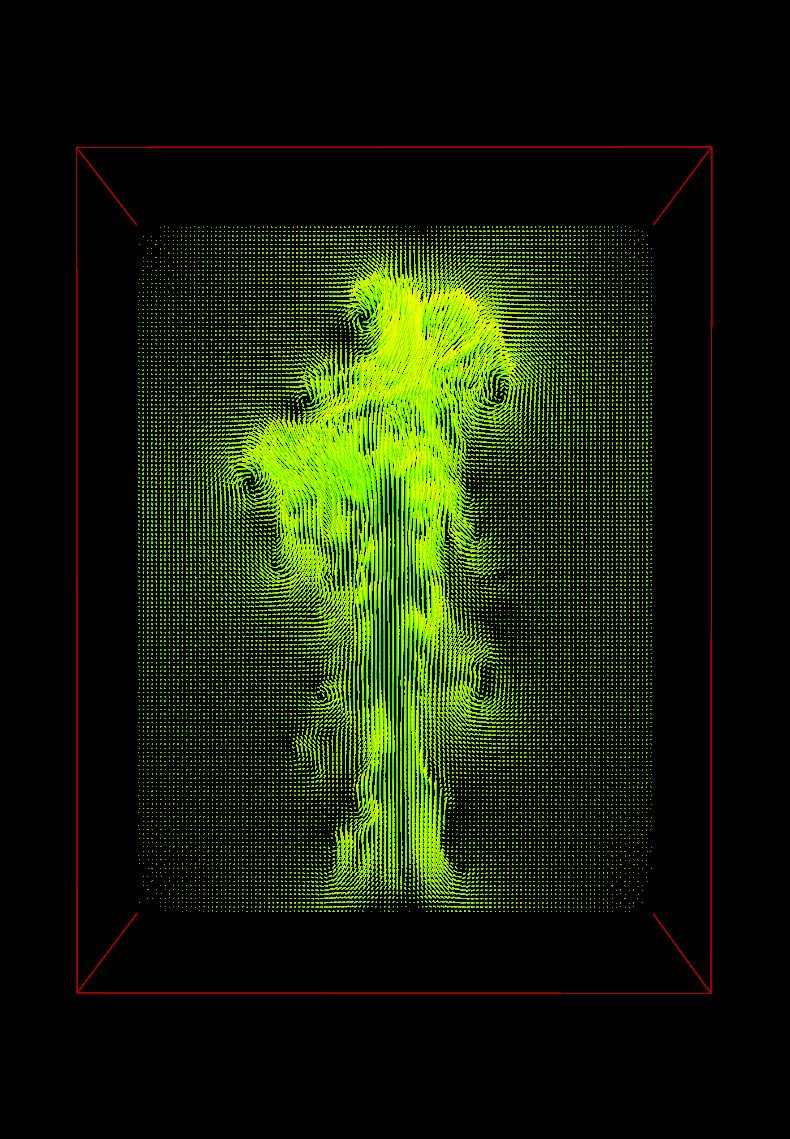}};
	\small \color{white} \draw (-1.15, 0.75) node {$b)$ input, side};
	\end{tikzpicture}\\
	\vspace{0.5mm}
	\begin{tikzpicture}
	\draw (0,0) node[inner sep=0] {
	\includegraphics[trim={4cm 6cm 4cm 6cm},clip,width=0.164\linewidth]{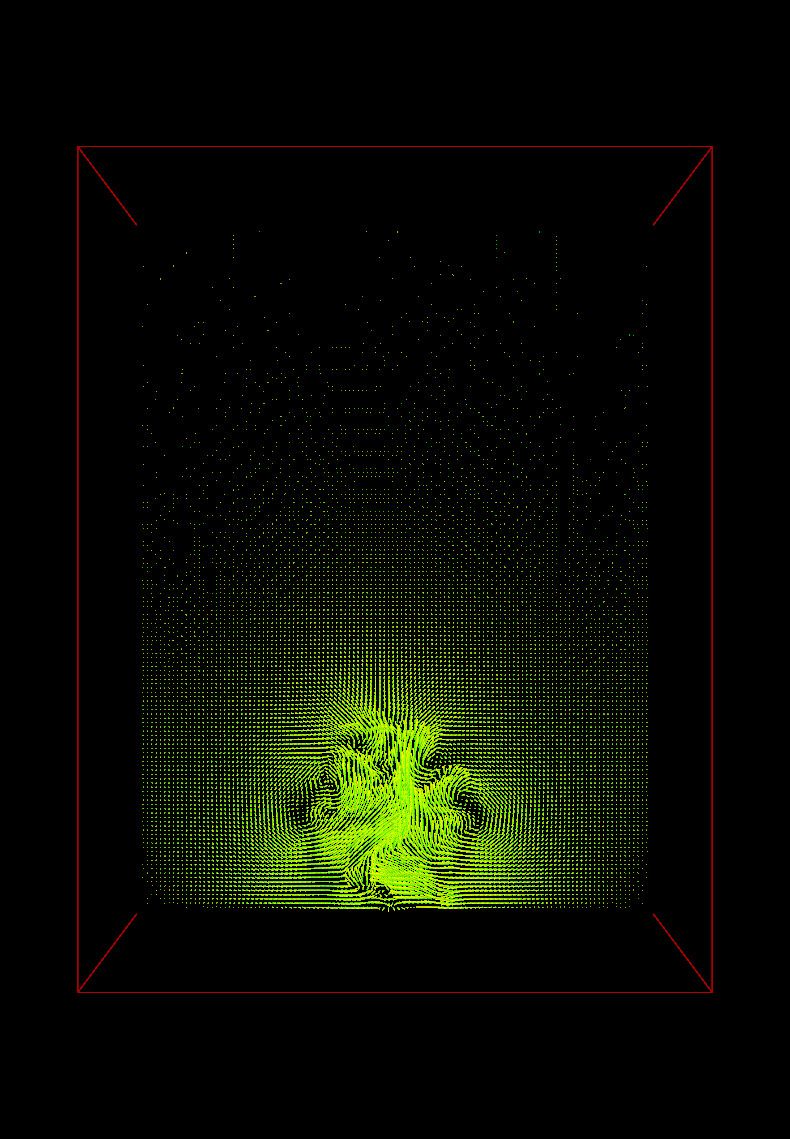}\hspace{0.0cm}%
	\includegraphics[trim={4cm 6cm 4cm 6cm},clip,width=0.164\linewidth]{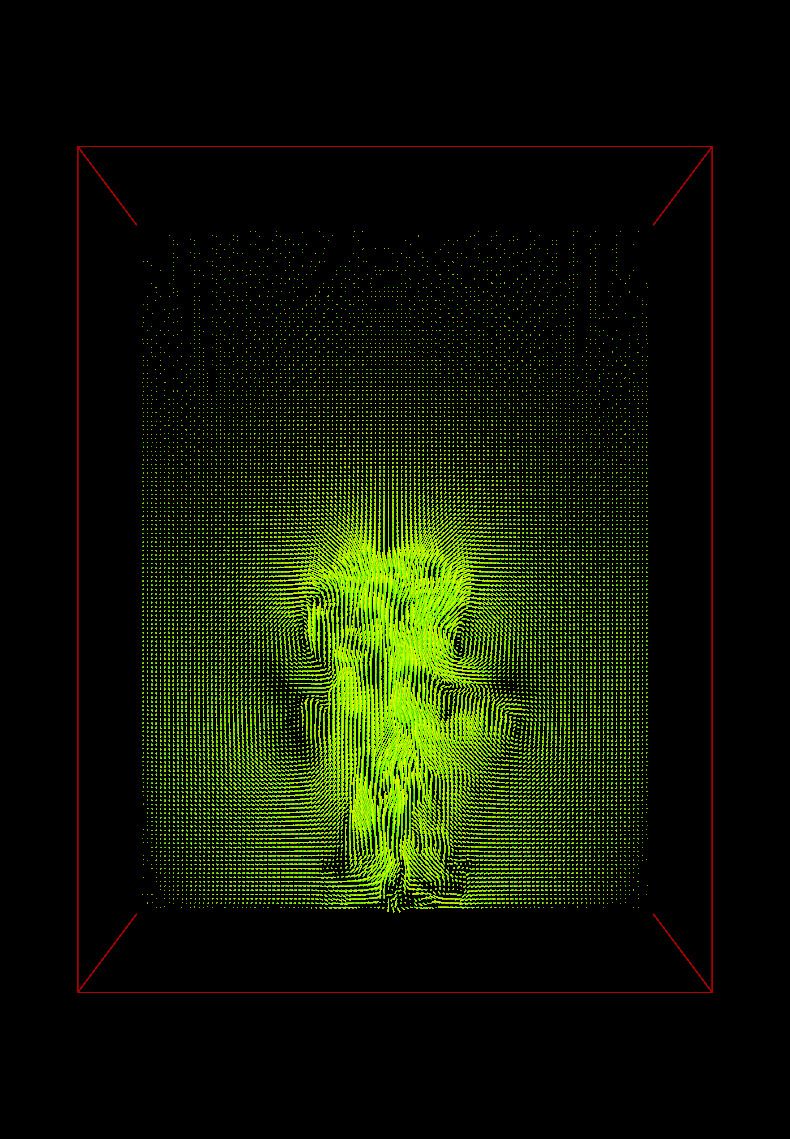}\hspace{0.0cm}%
	\includegraphics[trim={4cm 6cm 4cm 6cm},clip,width=0.164\linewidth]{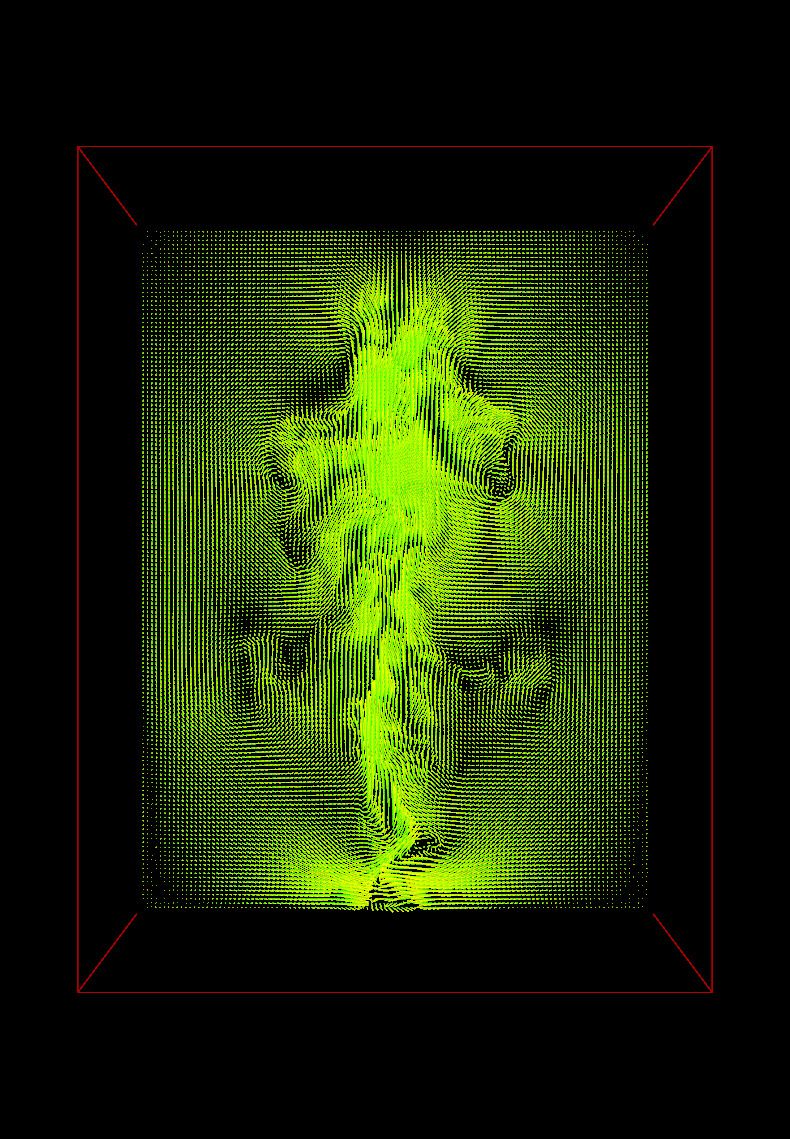}};
	\small \color{white} \draw (-0.85, 0.75) node {$c)$ ours single, front};  
	\end{tikzpicture}
	\hfill
	\begin{tikzpicture}
	\draw (0,0) node[inner sep=0] {
	\includegraphics[trim={4cm 6cm 4cm 6cm},clip,width=0.164\linewidth]{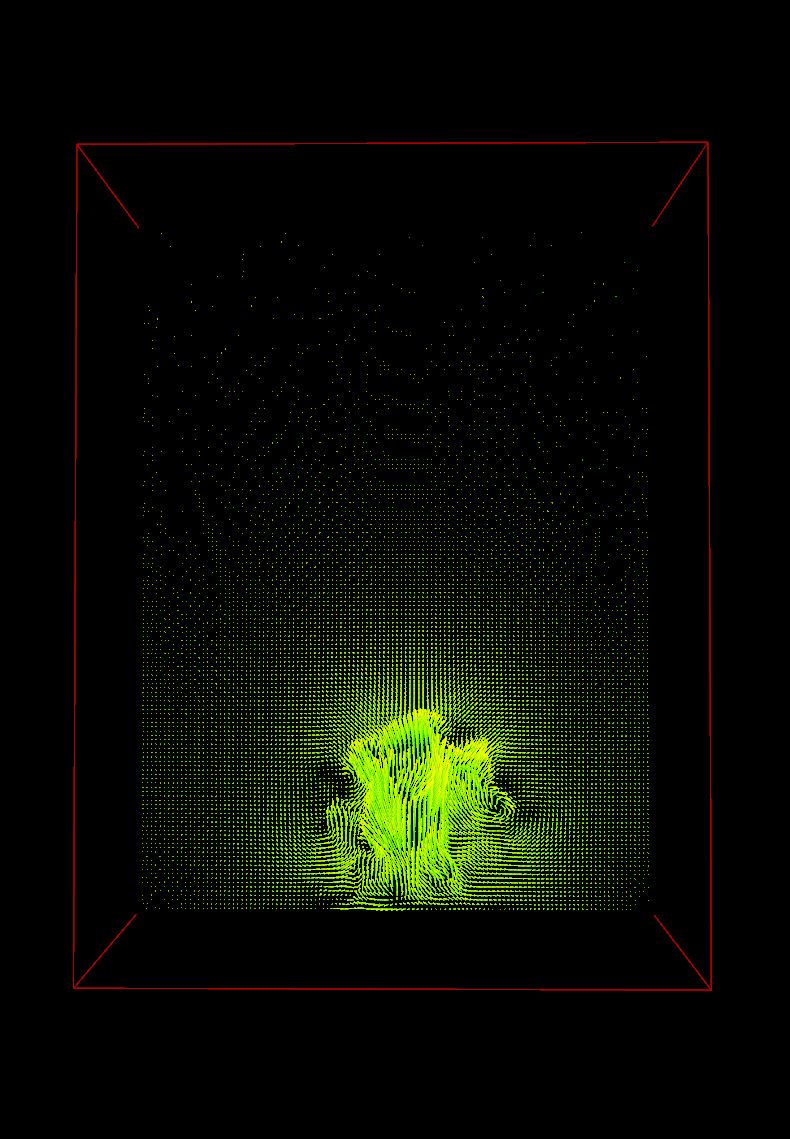}\hspace{0.0cm}%  
	\includegraphics[trim={4cm 6cm 4cm 6cm},clip,width=0.164\linewidth]{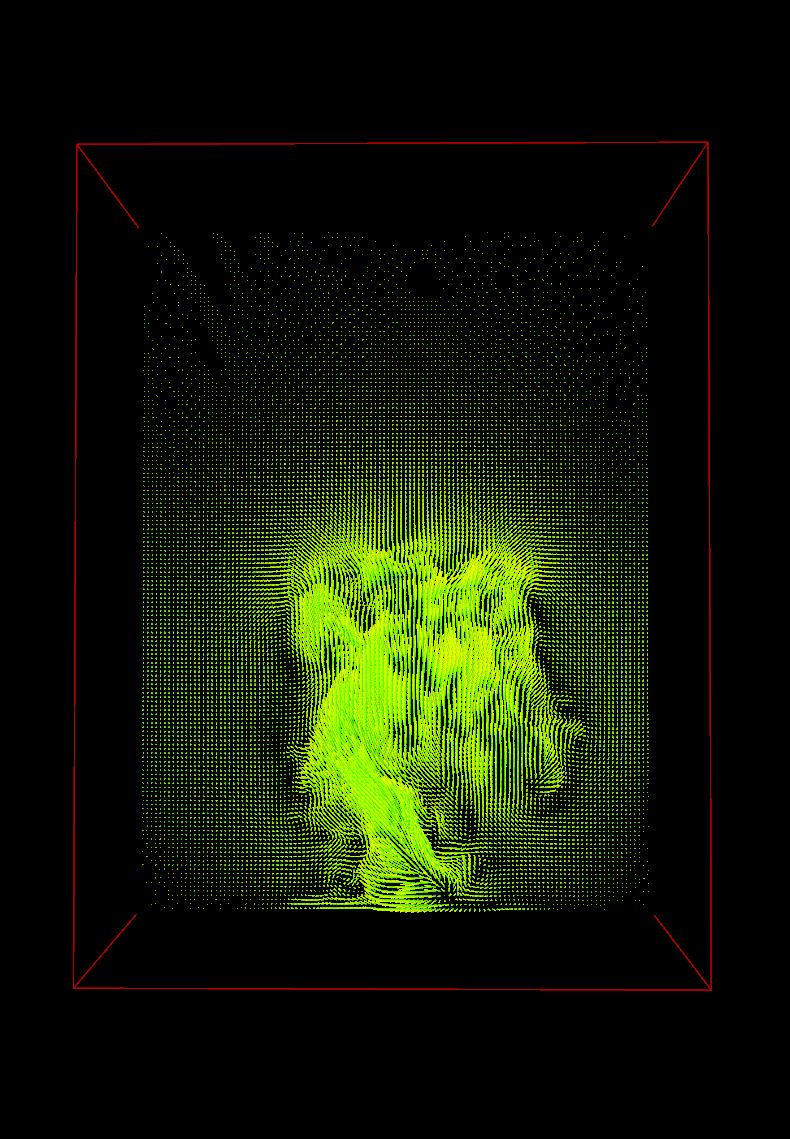}\hspace{0.0cm}% 
	\includegraphics[trim={4cm 6cm 4cm 6cm},clip,width=0.164\linewidth]{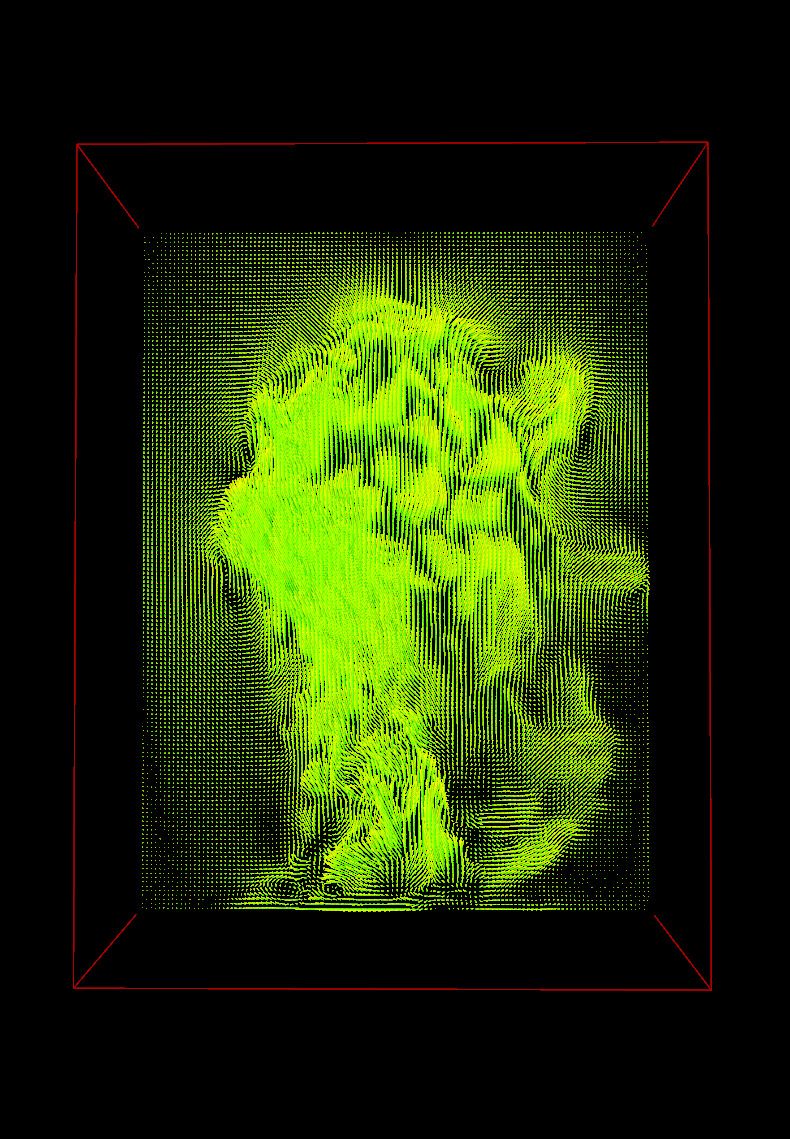}};
	\small \color{white} \draw (-0.85, 0.75) node {$d)$ ours single, side};
	\end{tikzpicture}\\
	\vspace{0.5mm}
	\begin{tikzpicture}
	\draw (0,0) node[inner sep=0] {
	\includegraphics[trim={4cm 6cm 4cm 6cm},clip,width=0.164\linewidth]{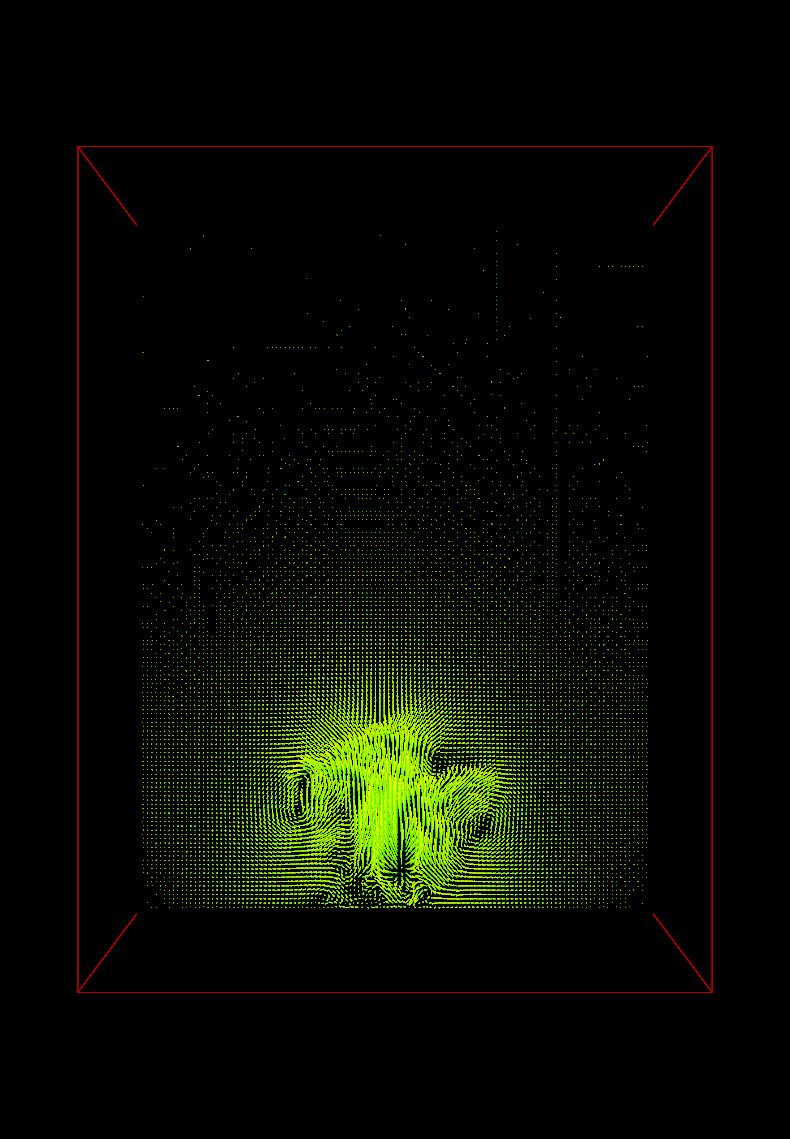}\hspace{0.0cm}% 
	\includegraphics[trim={4cm 6cm 4cm 6cm},clip,width=0.164\linewidth]{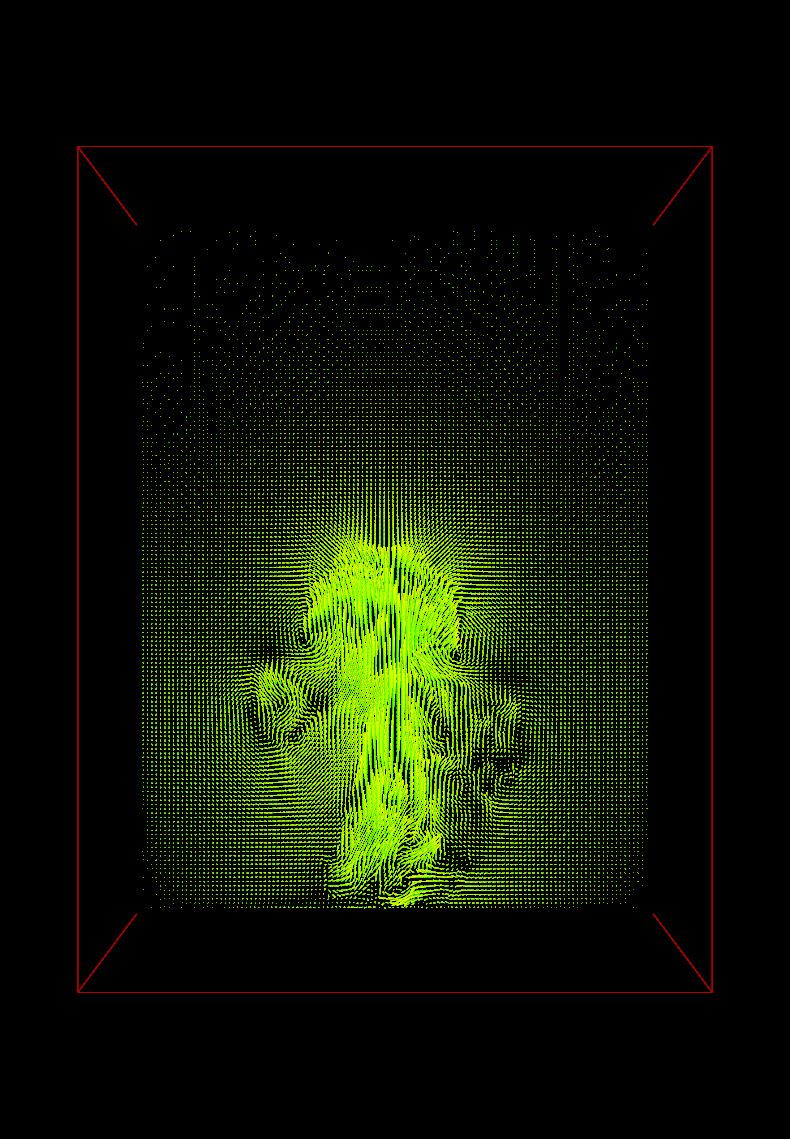}\hspace{0.0cm}%
	\includegraphics[trim={4cm 6cm 4cm 6cm},clip,width=0.164\linewidth]{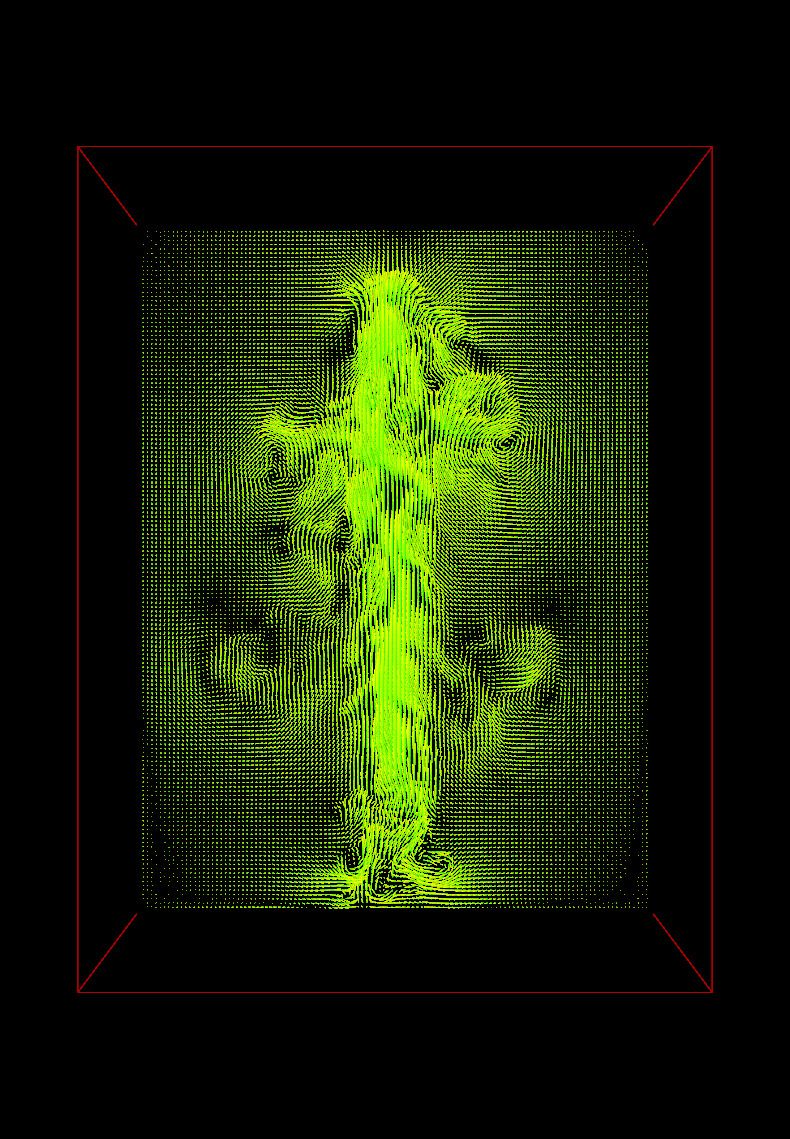}};
	\small \color{white} \draw (-0.8, 0.75) node {$e)$ ours double, front}; 
	\end{tikzpicture}
	\hfill
	\begin{tikzpicture}
	\draw (0,0) node[inner sep=0] {
	\includegraphics[trim={4cm 6cm 4cm 6cm},clip,width=0.164\linewidth]{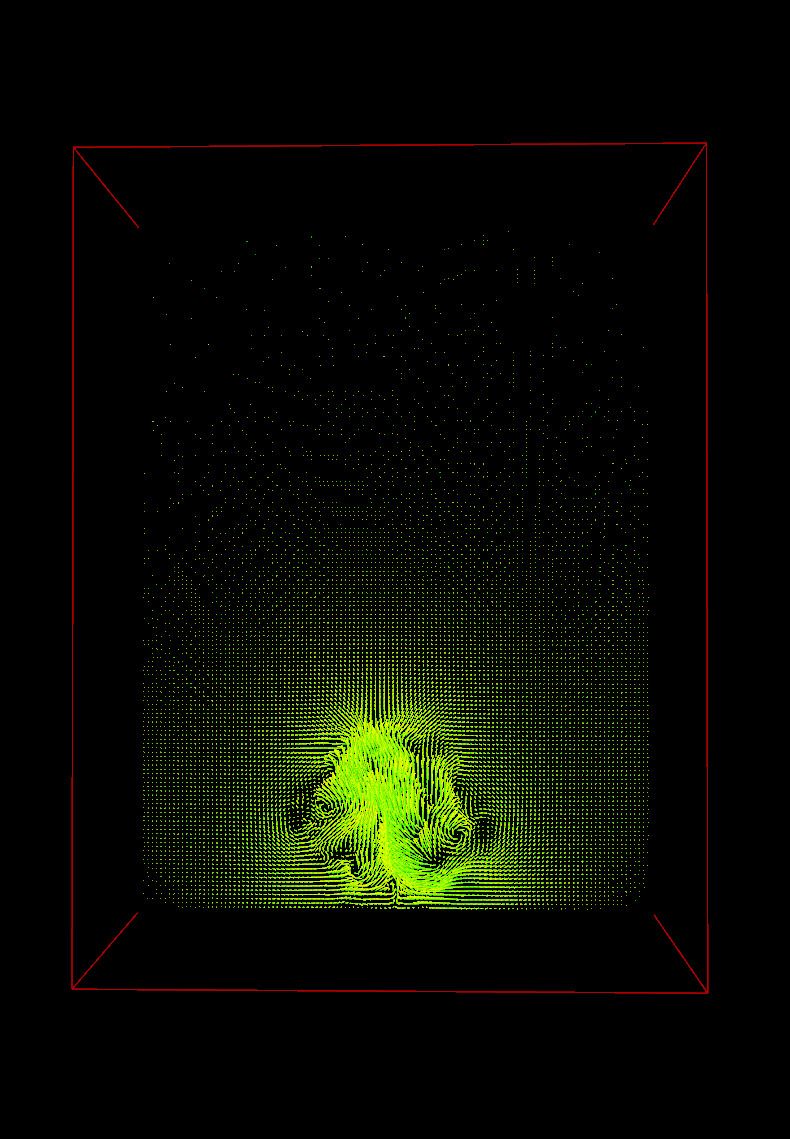}\hspace{0.0cm}% 
	\includegraphics[trim={4cm 6cm 4cm 6cm},clip,width=0.164\linewidth]{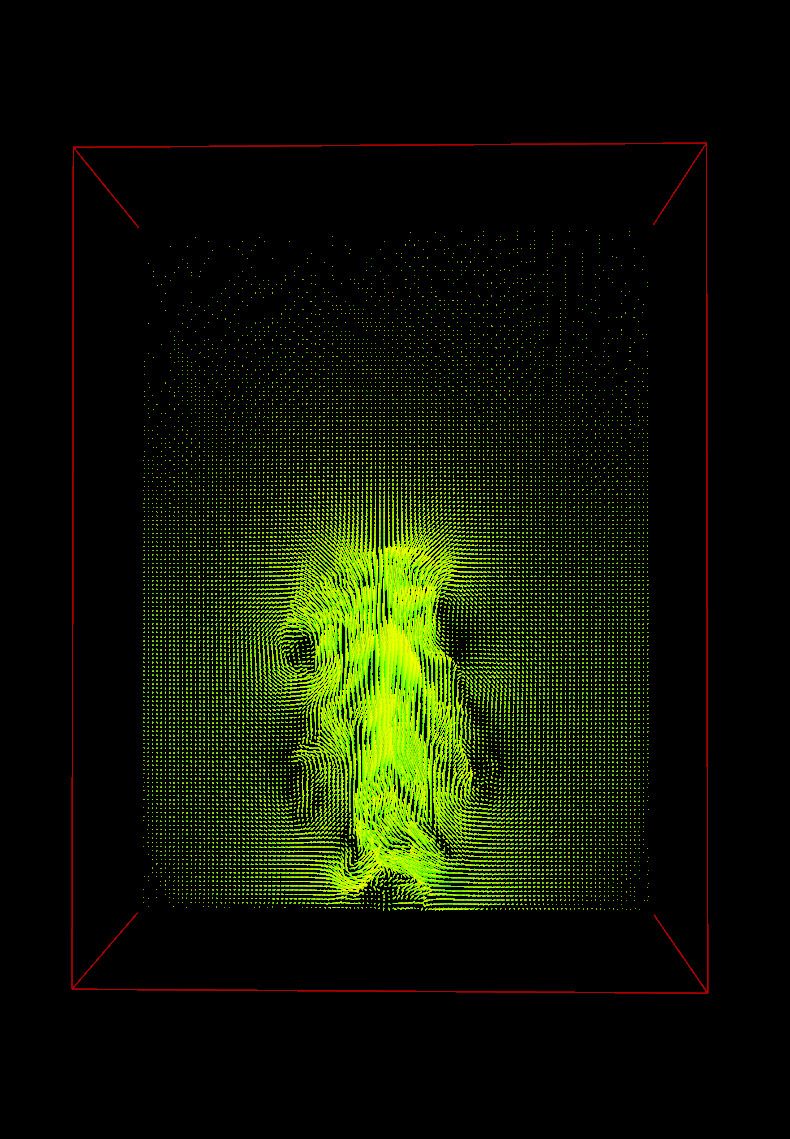}\hspace{0.0cm}% 
	\includegraphics[trim={4cm 6cm 4cm 6cm},clip,width=0.164\linewidth]{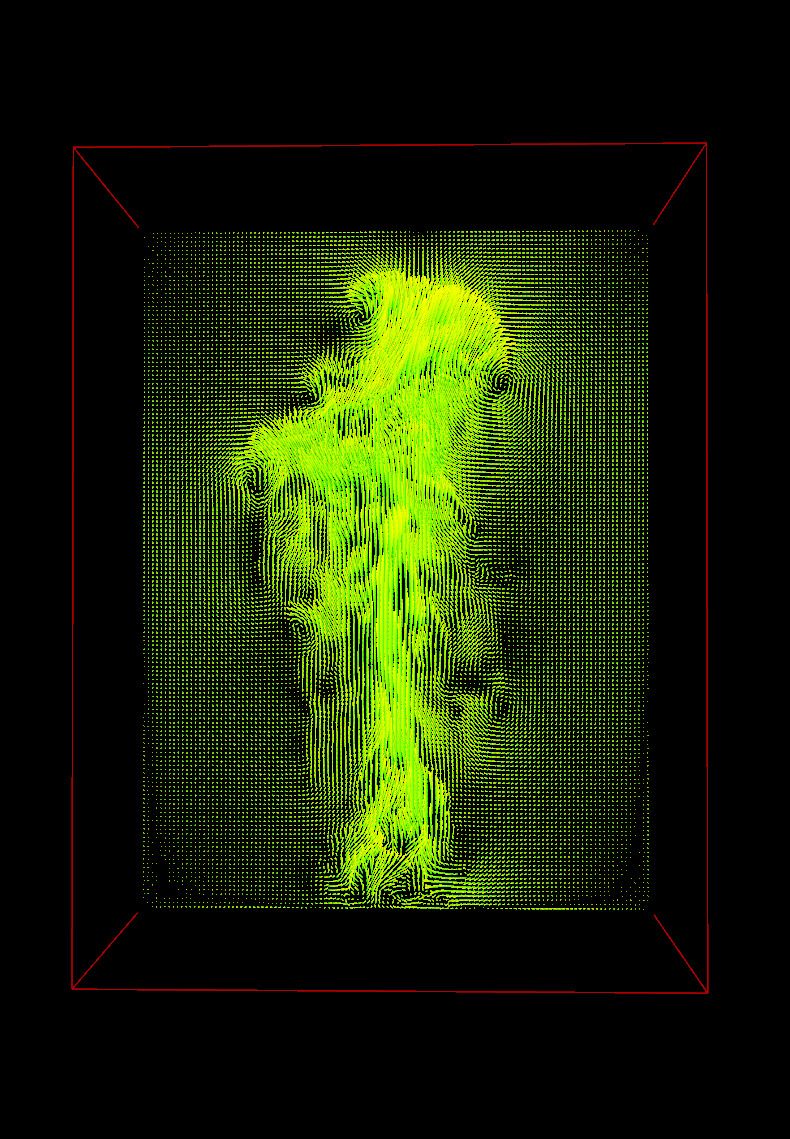}};
	\small \color{white} \draw (-0.8, 0.75) node {$f)$ ours double, side};
	\end{tikzpicture}\\
	\vspace{0.5mm}
	\begin{tikzpicture}
	\draw (0,0) node[inner sep=0] {
	\includegraphics[trim={4cm 6cm 4cm 6cm},clip,width=0.164\linewidth]{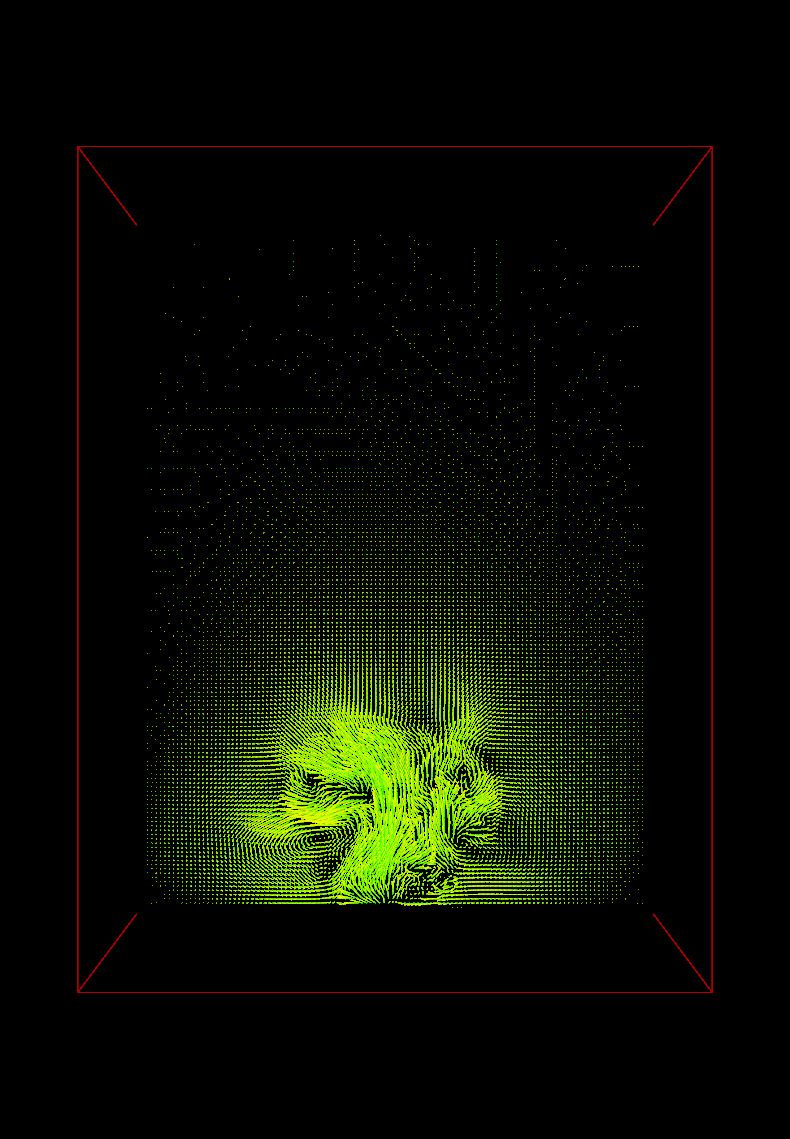}\hspace{0.0cm}% 
	\includegraphics[trim={4cm 6cm 4cm 6cm},clip,width=0.164\linewidth]{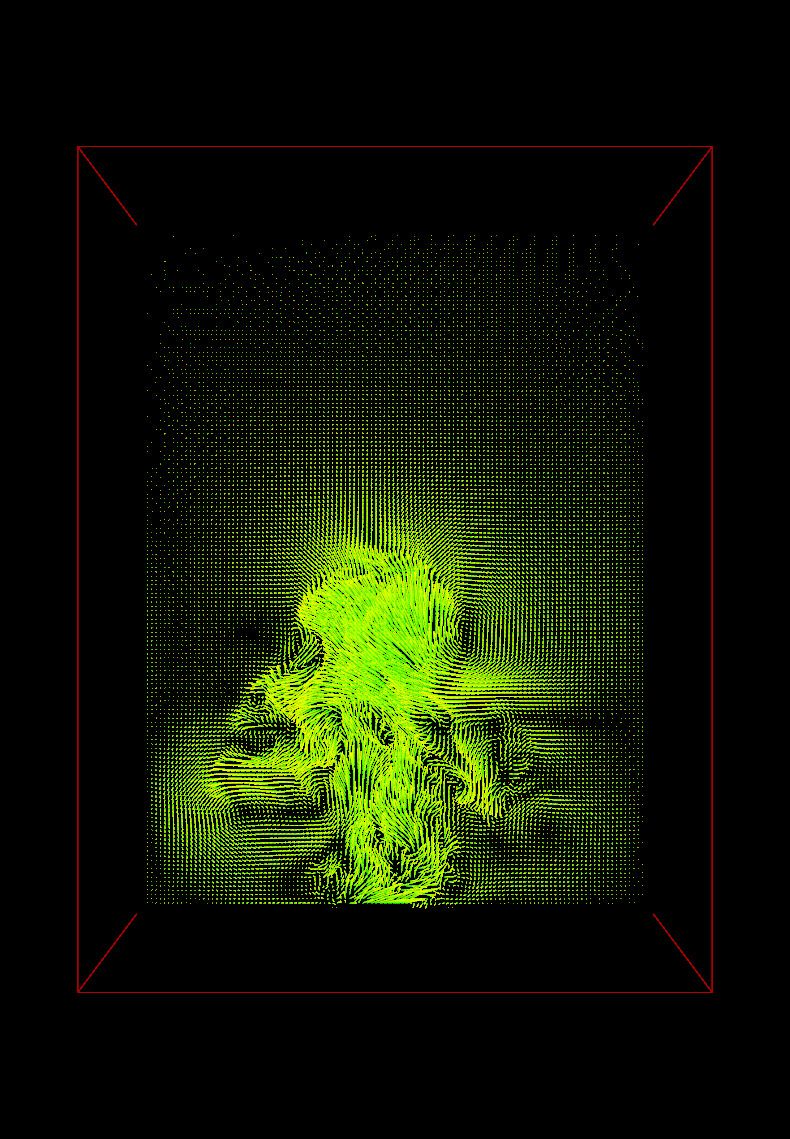}\hspace{0.0cm}% 
	\includegraphics[trim={4cm 6cm 4cm 6cm},clip,width=0.164\linewidth]{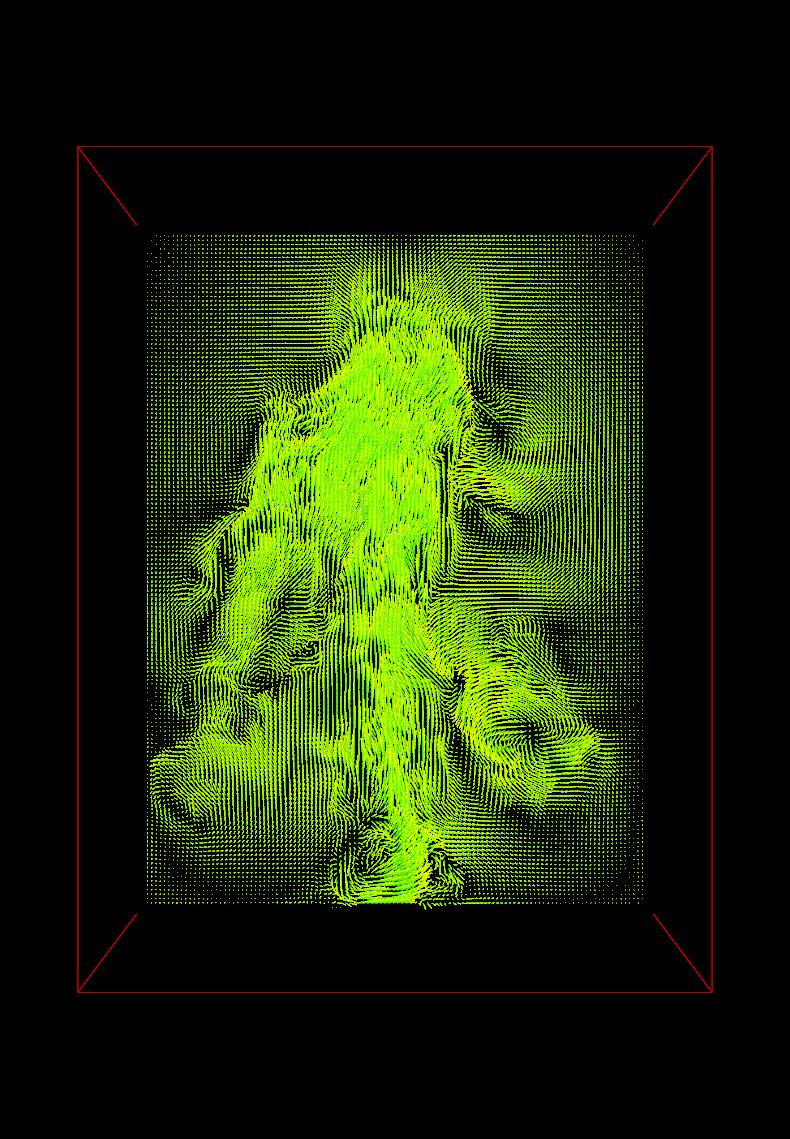}};
	\small \color{white} \draw (-0.65, 0.75) node {$g)$ Gregson et al., front}; 
	\end{tikzpicture}
	\hfill
	\begin{tikzpicture}
	\draw (0,0) node[inner sep=0] {
	\includegraphics[trim={4cm 6cm 4cm 6cm},clip,width=0.164\linewidth]{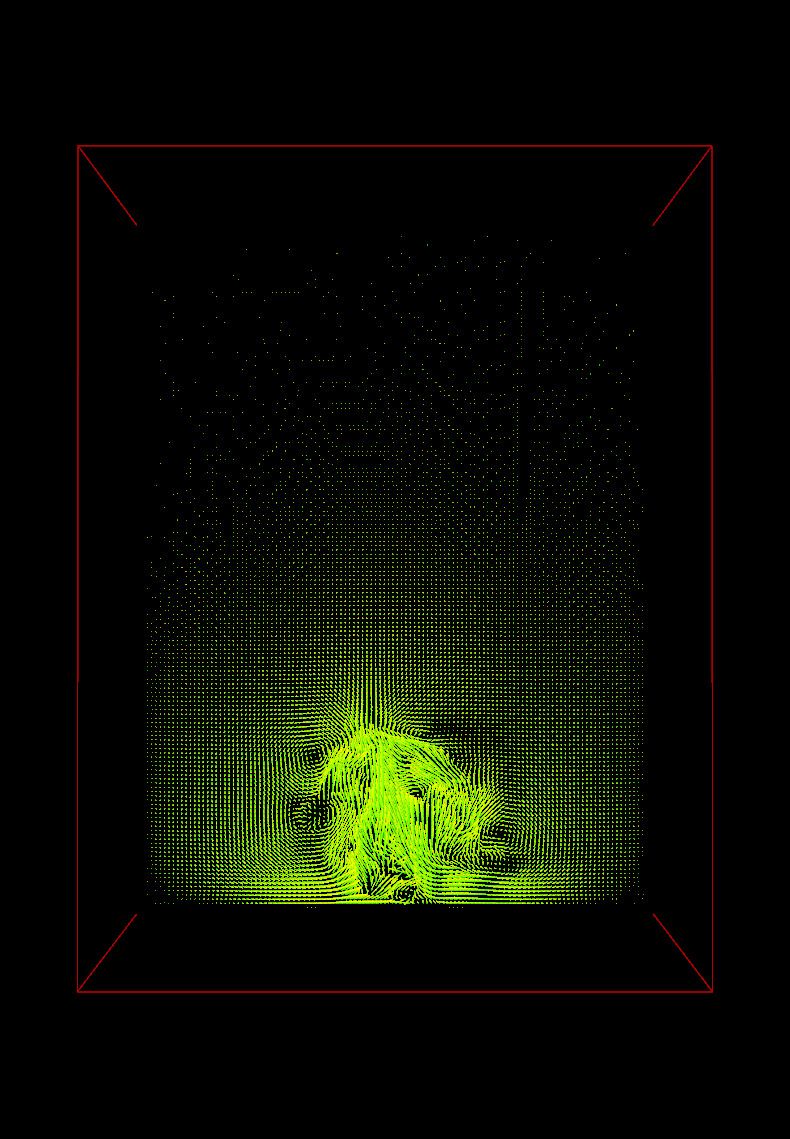}\hspace{0.0cm}% 
	\includegraphics[trim={4cm 6cm 4cm 6cm},clip,width=0.164\linewidth]{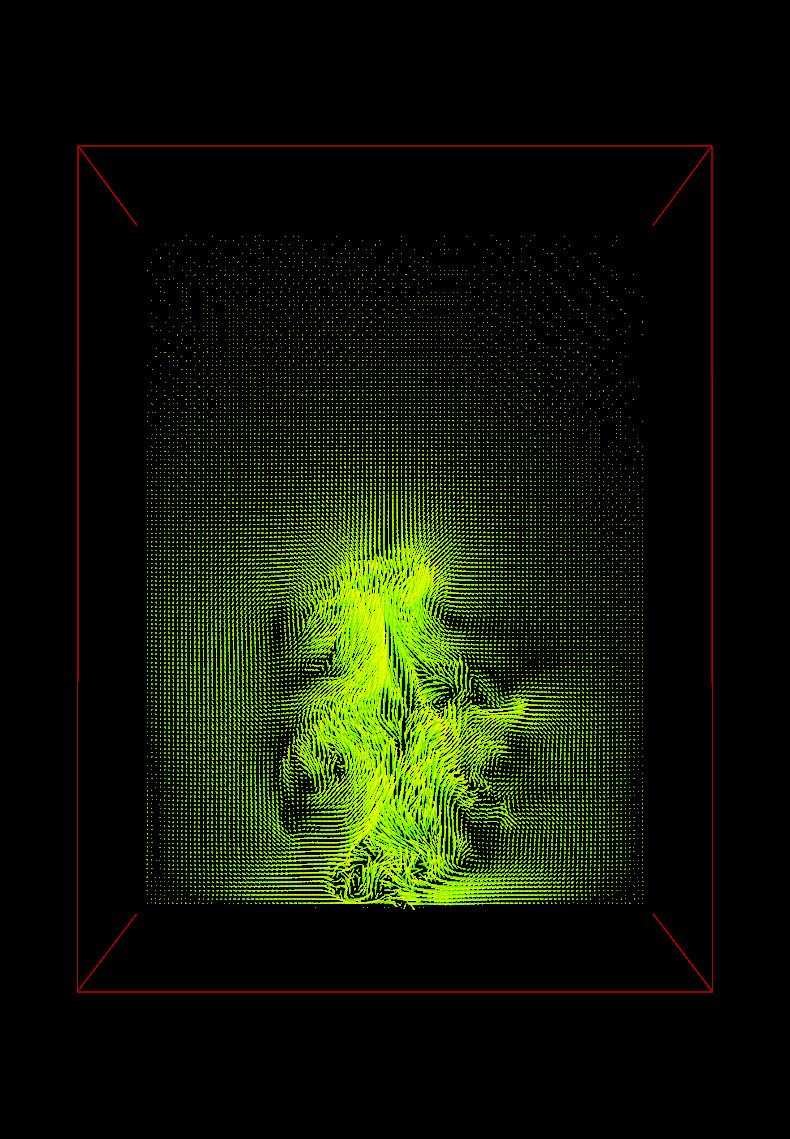}\hspace{0.0cm}% 
	\includegraphics[trim={4cm 6cm 4cm 6cm},clip,width=0.164\linewidth]{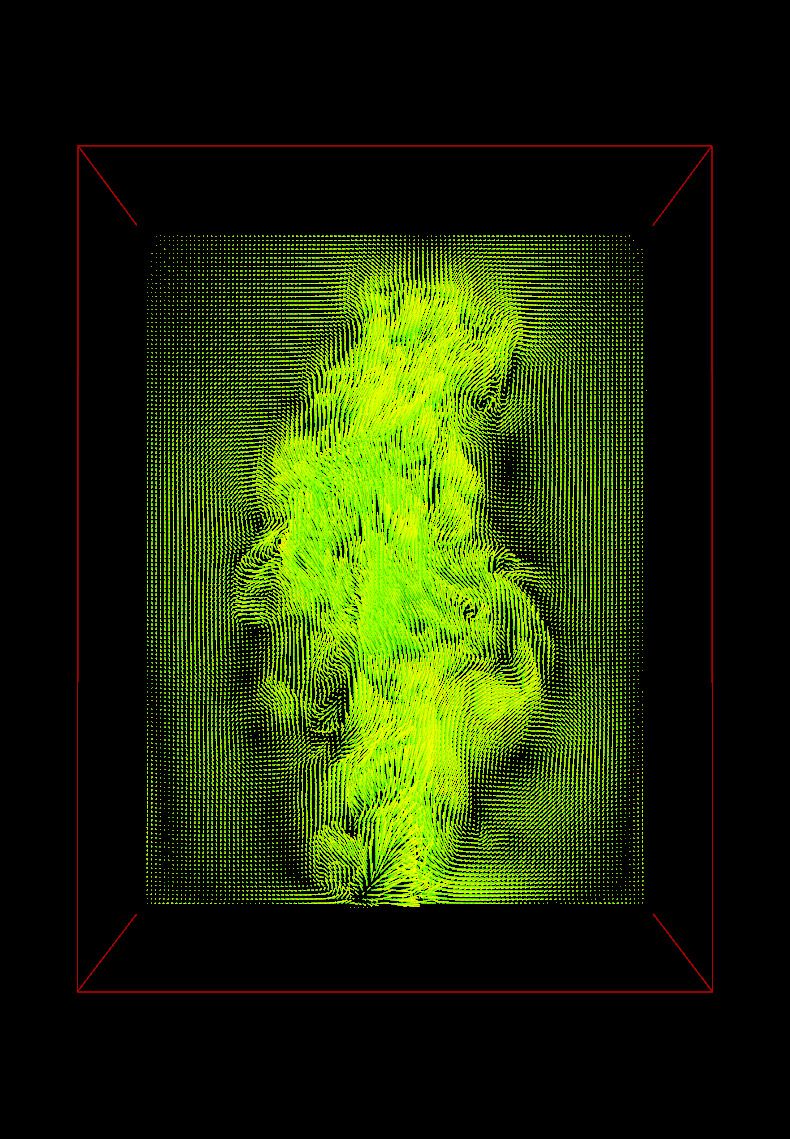}};
	\small \color{white} \draw (-0.65, 0.75) node {$h)$ Gregson et al., side}; 
	\end{tikzpicture}
	\caption{Center slice of ground truth, our single view, our double view and Gregson et al.'s reconstructed velocities at $t=30,60,96$.}	\label{fig:compareTomoOF_vel}
\end{figure}

\begin{figure*}[th!]
	\centering
	\begin{subfigure}{.165\textwidth}
		\centering
		\includegraphics[trim={1.9cm 0.4cm 2.5cm 2.5cm},clip,width=\linewidth]{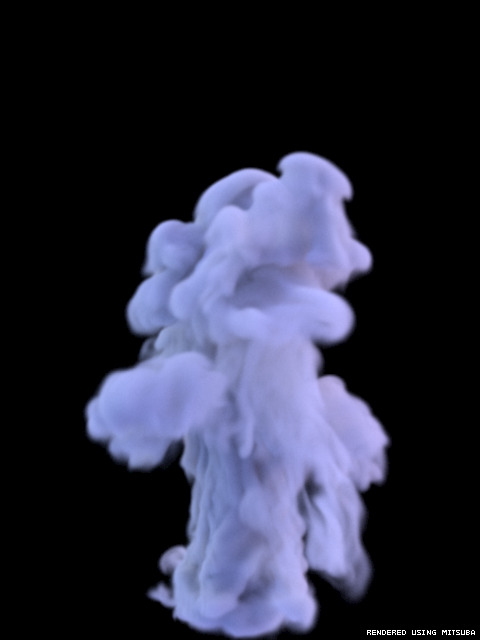}
	\end{subfigure}\hfill
	\begin{subfigure}{.165\textwidth}
		\centering
		\includegraphics[trim={1.9cm 0.4cm 2.5cm 2.5cm},clip,width=\linewidth]{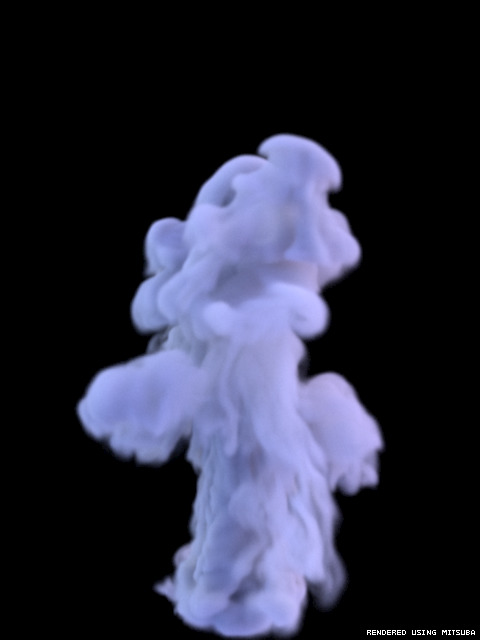}
	\end{subfigure}\hfill
	\begin{subfigure}{.165\textwidth}
		\centering
		\includegraphics[trim={1.9cm 0.4cm 2.5cm 2.5cm},clip,width=\linewidth]{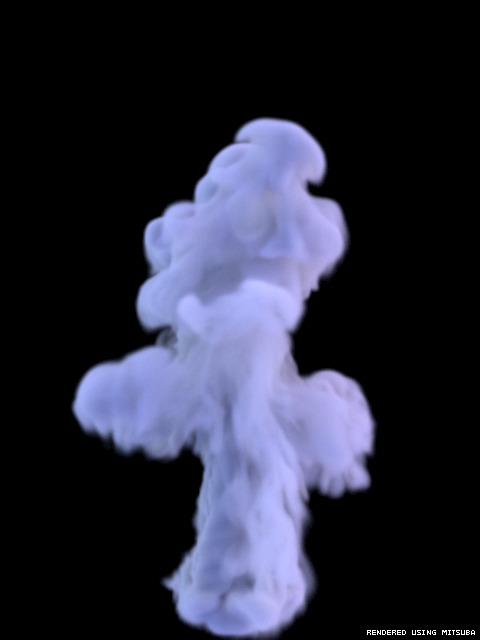}
	\end{subfigure}\hfill
	\begin{subfigure}{.165\textwidth}
		\centering
		\includegraphics[trim={1.9cm 0.4cm 2.5cm 2.5cm},clip,width=\linewidth]{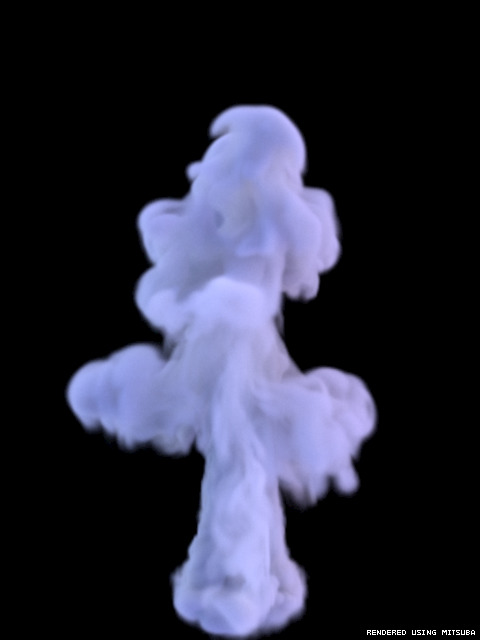}
	\end{subfigure}\hfill
	\begin{subfigure}{.165\textwidth}
		\centering
		\includegraphics[trim={1.9cm 0.4cm 2.5cm 2.5cm},clip,width=\linewidth]{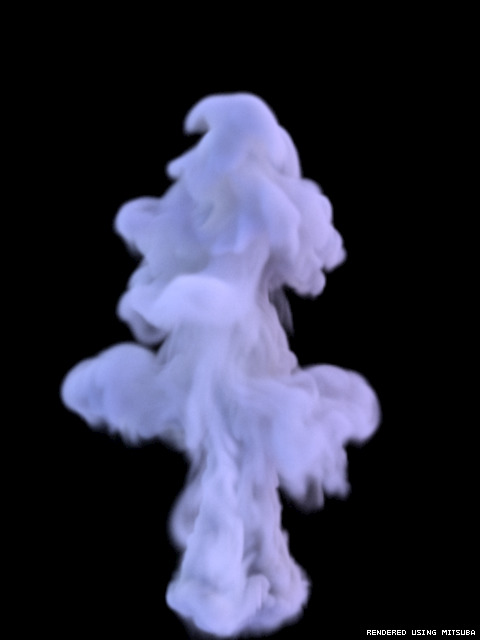}
	\end{subfigure}\hfill
	\begin{subfigure}{.165\textwidth}
		\centering
		\includegraphics[trim={1.9cm 0.4cm 2.5cm 2.5cm},clip,width=\linewidth]{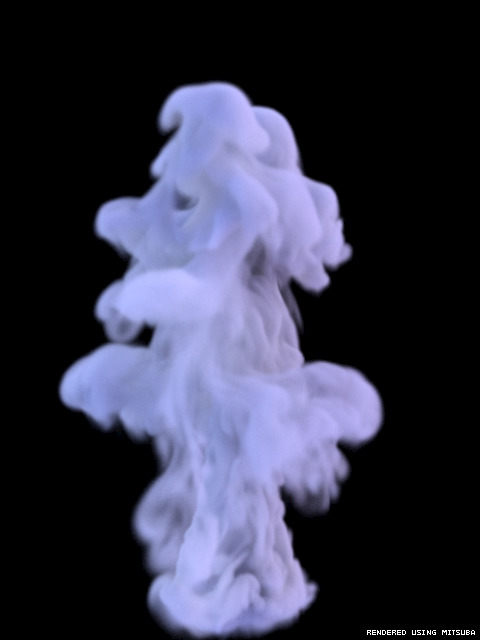}
	\end{subfigure}
	\caption{Re-simulated plume from a rotating viewpoint at $t=84,86,88,90,92,94$ with a $240\times320\times240$ domain.}	\label{fig:plume_resim}
\end{figure*}

While the advected densities already indicate motion, we additionally show the velocity's center slice from the front and side view for ground truth and reconstructed velocities in~\myreffig{fig:compareTomoOF_vel} a) - d).
Both center slices show a plausible motion, although velocity covers a larger depth than the ground truth.
\myreffig{fig:plume_resim} shows a re-simulation of the plume reconstruction in~\myreffig{fig:jetobs_mitsuba} c) - d), where we advect a two times finer smoke field 
in our reconstructed velocity, varying both in time and viewing angle.
These images indicate the amount of detail captured by our flow fields, which could be refined further
with procedural turbulence methods.

\begin{figure}[t]
	\centering
	\begin{overpic}[trim={0.4cm 0cm 9.0cm 1.7cm},clip,width=0.164\linewidth]{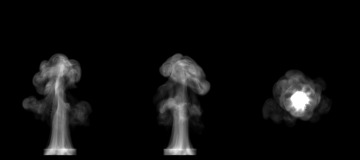}
	\put(4,85){\small \color{white}{$a)$}}\end{overpic}\hspace{0em}%
	\includegraphics[trim={4.7cm 0cm 4.7cm 1.7cm},clip,width=0.164\linewidth]{images/frontAs/input/density_076}\hfill
	\begin{overpic}[trim={0.4cm 0cm 9.0cm 1.7cm},clip,width=0.164\linewidth]{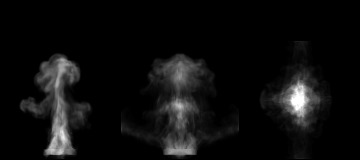}
	\put(4,85){\small \color{white}{$b)$}}\end{overpic}\hspace{0em}%
	\includegraphics[trim={4.7cm 0cm 4.7cm 1.7cm},clip,width=0.164\linewidth]{images/frontAs/back/volRecon_076}\hfill
	\begin{overpic}[trim={0.4cm 0cm 9.0cm 1.7cm},clip,width=0.164\linewidth]{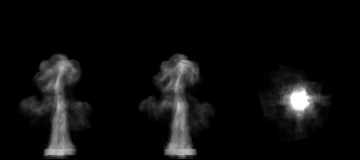}
	\put(4,85){\small \color{white}{$c)$}}\end{overpic}\hspace{0em}%
	\includegraphics[trim={4.7cm 0cm 4.7cm 1.7cm},clip,width=0.164\linewidth]{images/frontAs/side/volRecon_076}
	\caption{a) Input, b,c) reconstructions with front view as constraint for back and side view at $t=76$.}
	\label{fig:frontAs}
\end{figure}

\textbf{Multiple View Constraints:}
We can re-use a single input image to further constrain the reconstruction from other views.
One alternative is to use the mirrored input image from angle $0^{\circ}$ as input for $180^{\circ}$.
Additionally, we could assume rotational invariance and use the front view as a constraint from the side.
While this would be overly limiting in a general setting, we demonstrate the effect
of this additional constraint in \myreffig{fig:frontAs}.
We down-weight the side image by $10x$, such that the input front view is still matched with highest priority.
In \myreffig{fig:frontAs}, we observe that the reconstruction quality drastically increases for both additional input view cases. 
The input front view is matched closely, while the side view is either symmetric for front-as-back reconstructions or very similar to the front view for front-as-side reconstructions.
Especially the front-as-back variant is very useful, since it is close to the ground reference motion.
Both examples were produced without any depth regularizers.

\textbf{Jet Stream and Obstacle Scene:}
Our reconstruction method is also able to handle flows that are not rotationally symmetric, as well as flows with solid obstacles.
\myreffig{fig:jetobs_mitsuba} e) - h) show a jet stream flowing in from the left and rising up due to buoyancy. A rising smoke plume with a sphere in the center of the domain is shown in i) - l).
Our single-view jet reconstruction matches the input view very well and the depth motion is plausibly reconstructed.
For this setup, imposing artificial additional view constraints (as in the previous paragraph) is clearly unsuitable. 
It is a good example of a situation that particularly benefits from a single-view reconstruction.

Note that our solver has knowledge of the obstacle region for the plume with obstacle scene. For a generic reconstruction 
from videos, we could use techniques for single-view geometry estimation. However, 
we leave this as an extension for future work.
Our reconstruction of this smoke cloud from the front view is likewise very close to the input and our algorithm estimates a realistic 
behavior for the unconstrained motions throughout the volumetric density.

\begin{figure*}[tb!]
	\centering
	\begin{tikzpicture}
		\draw (0,0) node[inner sep=0] {
		\includegraphics[trim={3cm 1cm 3cm 5.8cm},clip,width=0.165\linewidth]{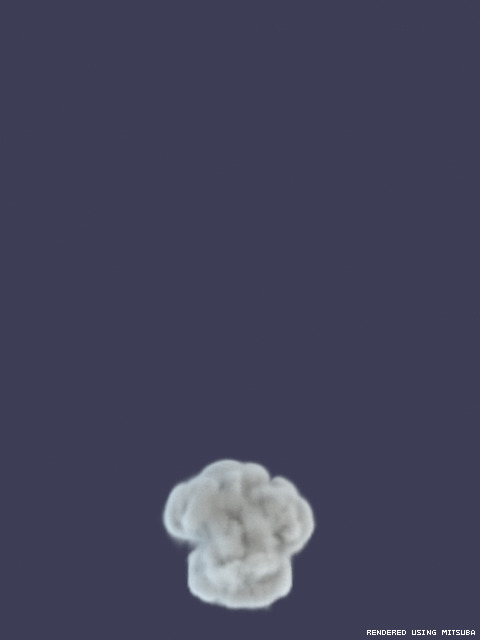}\hspace{0.0cm}%
		\includegraphics[trim={3cm 1cm 3cm 5.8cm},clip,width=0.165\linewidth]{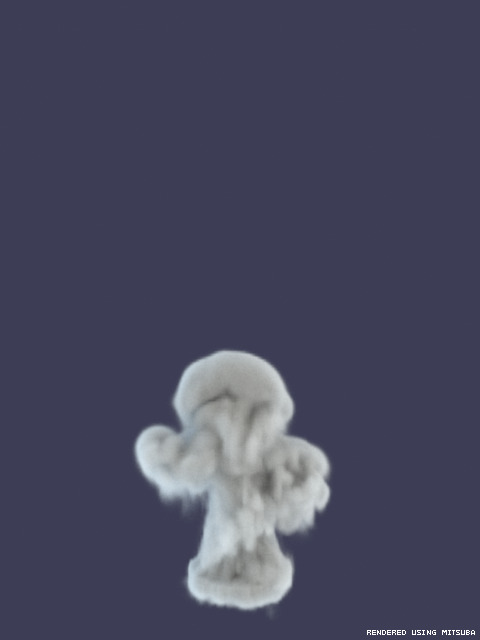}\hspace{0.0cm}%
		\includegraphics[trim={3cm 1cm 3cm 5.8cm},clip,width=0.165\linewidth]{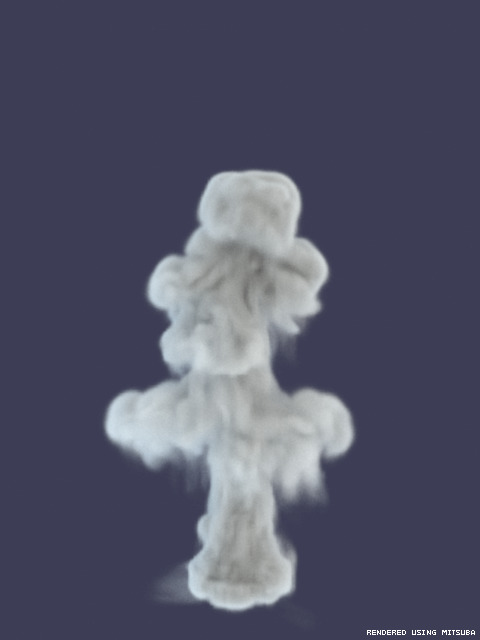}};
	\small \color{white} \draw (-3.5, 1.8) node {$a)$ ours, front};
	\end{tikzpicture}\hfill
	\begin{tikzpicture}
	\draw (0,0) node[inner sep=0] {
		\includegraphics[trim={3cm 1cm 3cm 5.8cm},clip,width=0.165\linewidth]{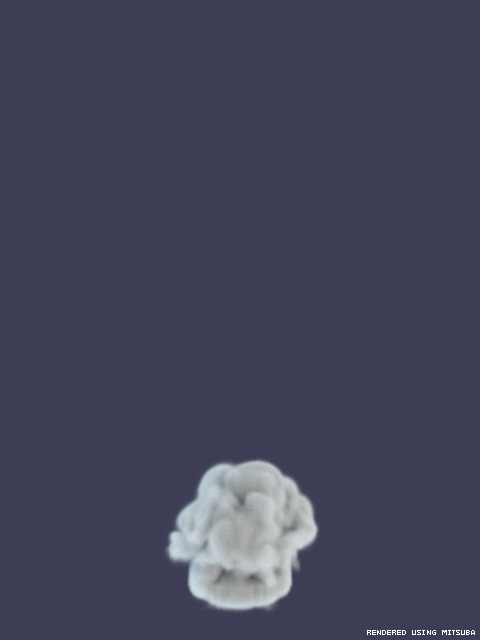}\hspace{0.0cm}%
		\includegraphics[trim={3cm 1cm 3cm 5.8cm},clip,width=0.165\linewidth]{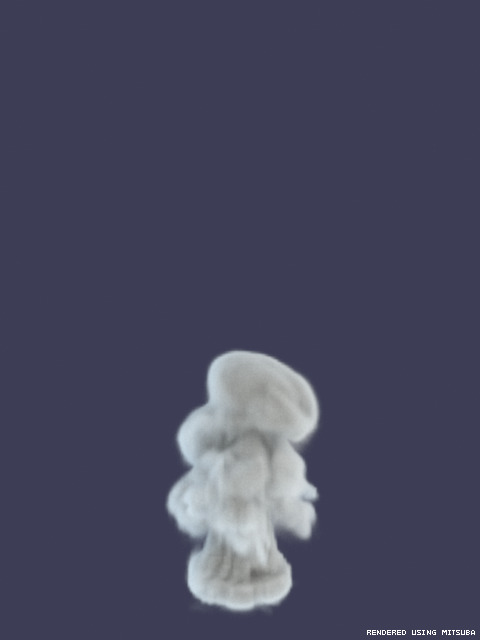}\hspace{0.0cm}%
		\includegraphics[trim={3cm 1cm 3cm 5.8cm},clip,width=0.165\linewidth]{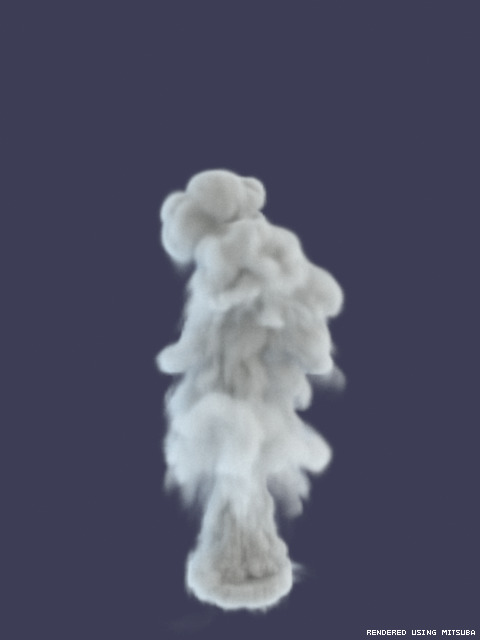}};
	\small \color{white} \draw (-3.5, 1.8) node {$b)$ ours, side};
	\end{tikzpicture}\\
	\vspace{0.5mm}
	\begin{tikzpicture}
	\draw (0,0) node[inner sep=0] {
		\includegraphics[trim={3cm 1cm 3cm 5.8cm},clip,width=0.165\linewidth]{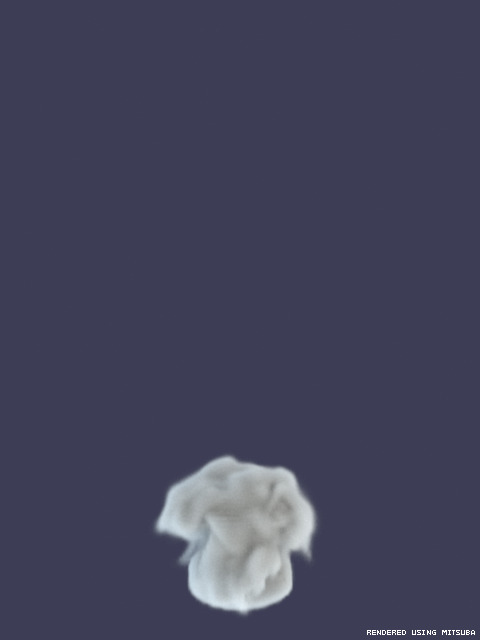}\hspace{0.0cm}%
		\includegraphics[trim={3cm 1cm 3cm 5.8cm},clip,width=0.165\linewidth]{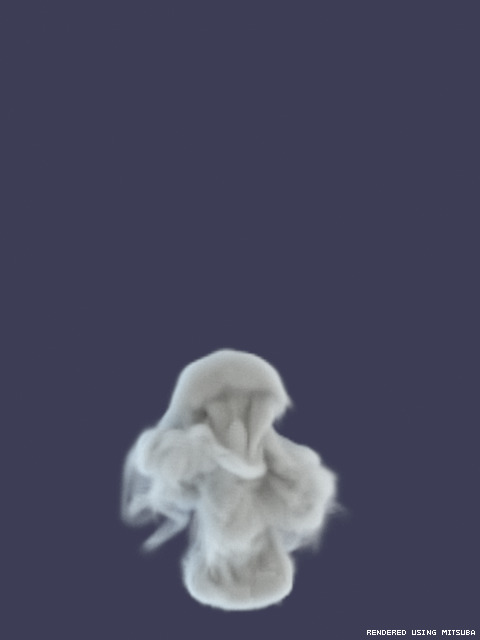}\hspace{0.0cm}%
		\includegraphics[trim={3cm 1cm 3cm 5.8cm},clip,width=0.165\linewidth]{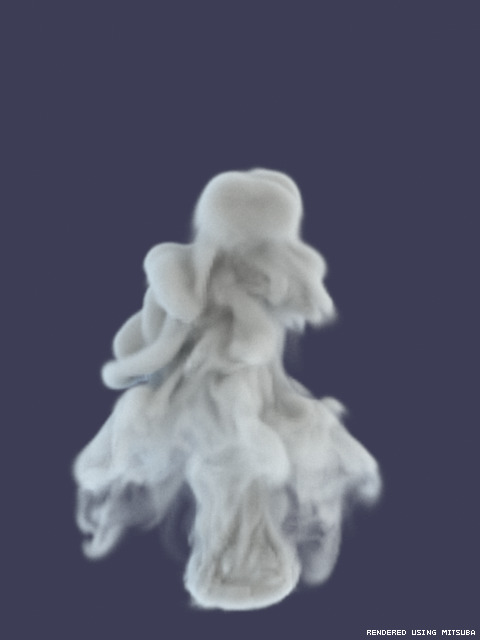}};
	\small \color{white} \draw (-2.95, 1.8) node {$c)$ Gregson et al., front};
	\end{tikzpicture}\hfill
	\begin{tikzpicture}
	\draw (0,0) node[inner sep=0] {
		\includegraphics[trim={3cm 1cm 3cm 5.8cm},clip,width=0.165\linewidth]{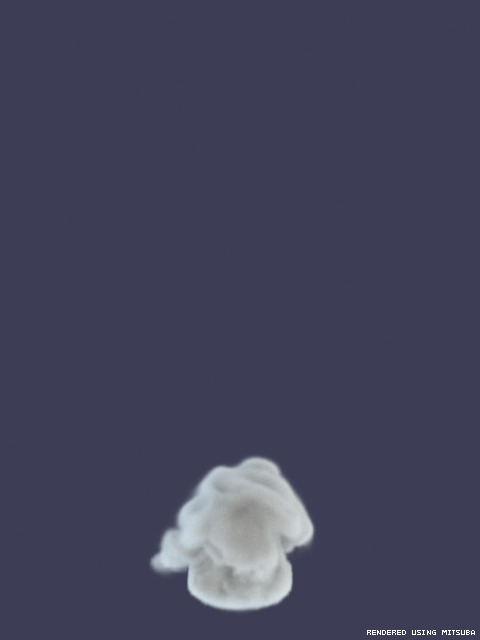}\hspace{0.0cm}%
		\includegraphics[trim={3cm 1cm 3cm 5.8cm},clip,width=0.165\linewidth]{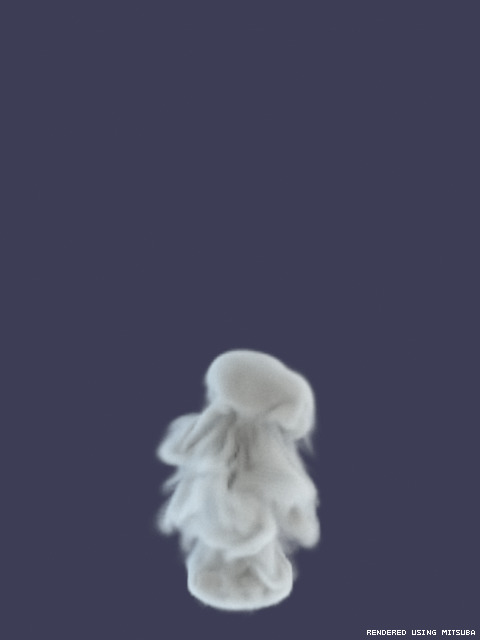}\hspace{0.0cm}%
		\includegraphics[trim={3cm 1cm 3cm 5.8cm},clip,width=0.165\linewidth]{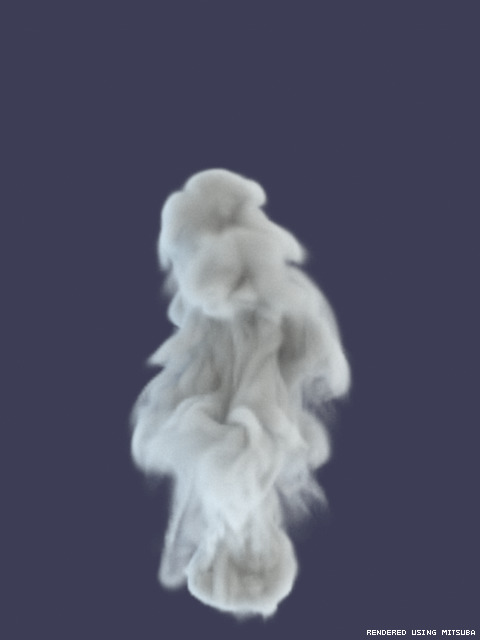}};
	\small \color{white} \draw (-2.95, 1.8) node {$d)$ Gregson et al., side};
	\end{tikzpicture}
	\caption{Our reconstruction a,b) and the approach by Gregson et al.~\cite{Gregson:2014} c,d) for a synthetic rising plume scene.}	\label{fig:compareTomoOF}
\end{figure*}

\textbf{Comparison to Previous Work:}
We compare our results with the tomography plus optical flow approach by Gregson et al.~\shortcite{Gregson:2014}
with the synthetic plume scene from~\myreffig{fig:jetobs_mitsuba} a) - b).
While our method works with a single view, we introduce a second view for the sake of comparing it to previous work here.
As shown in~\myreffig{fig:compareTomoOF}, our approach matches the ground truth motion significantly better. 
It also has less volume expansion and resembles the given views more closely.
Examples of the reconstructed velocities are shown in~\myreffig{fig:compareTomoOF_vel} e) - h).
While our reconstructed velocity matches the velocity field quite accurately for both views, the approach by Gregson et al.~\shortcite{Gregson:2014} works best for accurate density reconstructions, 
where the flow is corrected by targeting a reliable density at each time step.
Otherwise, decoupling the velocity and density estimation for very sparse views leaves 
too many degrees of freedom for the optical flow solve, degrading its quality. 

While Okabe et al. \shortcite{Okabe:2015} also target single view reconstructions, we believe it is not meaningful
to compare our algorithm directly. As their work focuses primarily on the reconstruction of the volumetric 
appearance, the reconstructed motions are naturally less reliable. Combining this method with ours, however,
could be an interesting future extension.

\begin{figure}[t]
	\centering
	\begin{overpic}[trim={0.5cm 0cm 9.1cm 1.7cm},clip,width=0.164\linewidth]{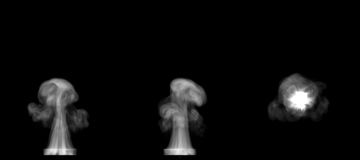}
		\put(4,85){\small \color{white}{$a)$}}\end{overpic}\hspace{0em}%
	\includegraphics[trim={4.8cm 0cm 4.8cm 1.7cm},clip,width=0.164\linewidth]{images/extrapolation/density_060}\hfill
	\begin{overpic}[trim={0.5cm 0cm 9.1cm 1.7cm},clip,width=0.164\linewidth]{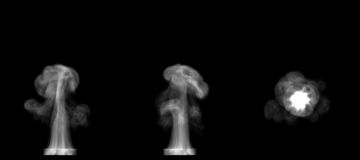}
		\put(4,85){\small \color{white}{$b)$}}\end{overpic}\hspace{0em}%
	\includegraphics[trim={4.8cm 0cm 4.8cm 1.7cm},clip,width=0.164\linewidth]{images/extrapolation/density_070}\hfill
	\begin{overpic}[trim={0.5cm 0cm 9.1cm 1.7cm},clip,width=0.164\linewidth]{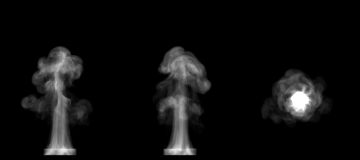}
		\put(4,85){\small \color{white}{$c)$}}\end{overpic}\hspace{0em}%
	\includegraphics[trim={4.8cm 0cm 4.8cm 1.7cm},clip,width=0.164\linewidth]{images/extrapolation/density_080}\\
	\begin{overpic}[trim={0.5cm 0cm 9.1cm 1.7cm},clip,width=0.164\linewidth]{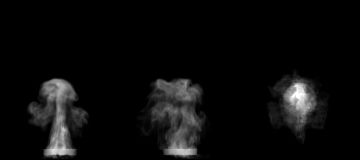}
		\put(4,85){\small \color{white}{$d)$}}\end{overpic}\hspace{0em}%
	\includegraphics[trim={4.8cm 0cm 4.8cm 1.7cm},clip,width=0.164\linewidth]{images/extrapolation/volRecon_060}\hfill
	\begin{overpic}[trim={0.5cm 0cm 9.1cm 1.7cm},clip,width=0.164\linewidth]{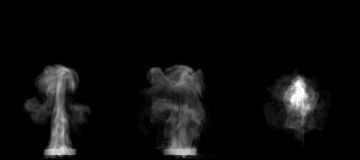}
		\put(4,85){\small \color{white}{$e)$}}\end{overpic}\hspace{0em}%
	\includegraphics[trim={4.8cm 0cm 4.8cm 1.7cm},clip,width=0.164\linewidth]{images/extrapolation/volRecon_070}\hfill
	\begin{overpic}[trim={0.5cm 0cm 9.1cm 1.7cm},clip,width=0.164\linewidth]{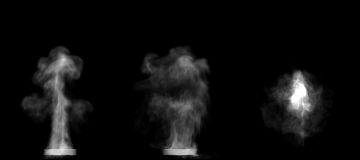}
		\put(4,85){\small \color{white}{$f)$}}\end{overpic}\hspace{0em}%
	\includegraphics[trim={4.8cm 0cm 4.8cm 1.7cm},clip,width=0.164\linewidth]{images/extrapolation/volRecon_080}\\
	\begin{overpic}[trim={0.5cm 0cm 9.1cm 1.7cm},clip,width=0.164\linewidth]{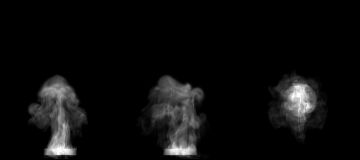}
		\put(4,85){\small \color{white}{$g)$}}\end{overpic}\hspace{0em}%
	\includegraphics[trim={4.8cm 0cm 4.8cm 1.7cm},clip,width=0.164\linewidth]{images/extrapolation/simulatedDensity_060}\hfill
	\begin{overpic}[trim={0.5cm 0cm 9.1cm 1.7cm},clip,width=0.164\linewidth]{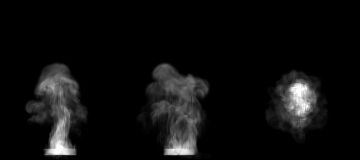}
		\put(4,85){\small \color{white}{$h)$}}\end{overpic}\hspace{0em}%
	\includegraphics[trim={4.8cm 0cm 4.8cm 1.7cm},clip,width=0.164\linewidth]{images/extrapolation/simulatedDensity_070}\hfill
	\begin{overpic}[trim={0.5cm 0cm 9.1cm 1.7cm},clip,width=0.164\linewidth]{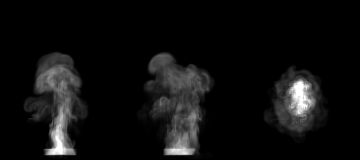}
		\put(4,85){\small \color{white}{$i)$}}\end{overpic}\hspace{0em}%
	\includegraphics[trim={4.8cm 0cm 4.8cm 1.7cm},clip,width=0.164\linewidth]{images/extrapolation/simulatedDensity_080}
	\caption{Ground truth input in a,b,c), reconstruction in d,e,f) and extrapolation by simulation in g,h,i) 
	after 12, 22, and 32 frames.}
	\label{fig:extrapolation}
\end{figure}

\textbf{Time Extrapolation by Simulation:}
We evaluate the robustness of our reconstruction by reconstructing only the first half of two scenes (smoke plume and jet stream) and compute the second half starting at $t=48$
with a regular forward simulation. 
A comparison for the plume scene is shown in~\myreffig{fig:extrapolation}. 
The extrapolation follows the reconstructed motion for the first \textasciitilde 25 frames and is very similar to the input simulation.
This indicates that our reconstructed density and velocity are indeed reliable.
As future work, it would be interesting to automatically tune simulation parameters like buoyancy in order to match the ground truth input longer and more closely.
The full sequences for this comparison can be found in the accompanying video.

\subsection{Reconstruction of Real Smoke Flow}
\begin{figure*}[tb!]
	\centering
	\begin{tikzpicture}
	\draw (0,0) node[inner sep=0] {
		\includegraphics[trim={0cm 0.7cm 0cm 0.7cm},clip,width=0.125\linewidth]{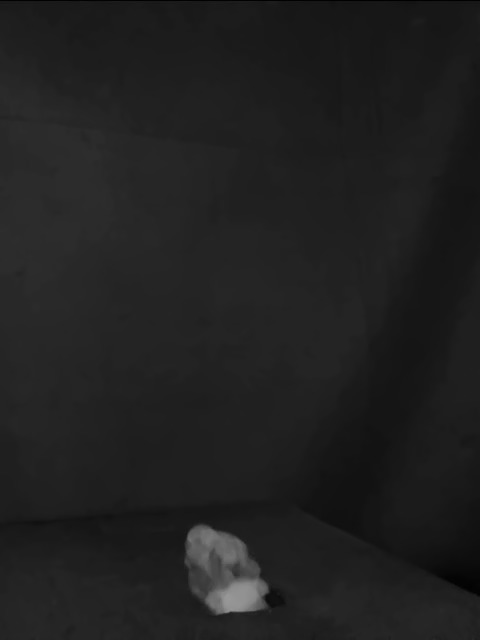}\hspace{0.0cm}%
		\includegraphics[trim={0cm 0.7cm 0cm 0.7cm},clip,width=0.125\linewidth]{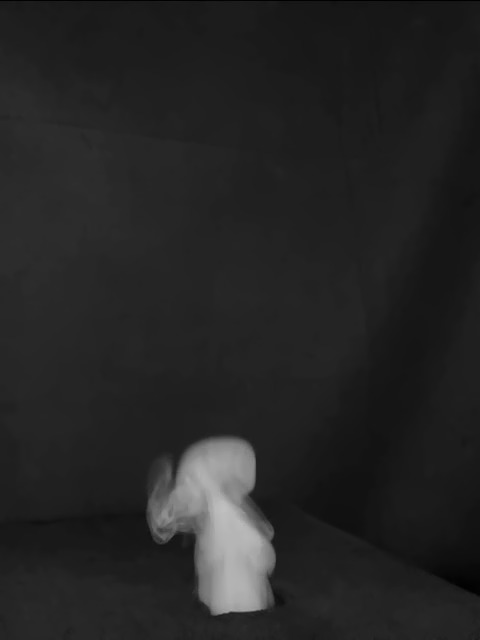}\hspace{0.0cm}%
		\includegraphics[trim={0cm 0.7cm 0cm 0.7cm},clip,width=0.125\linewidth]{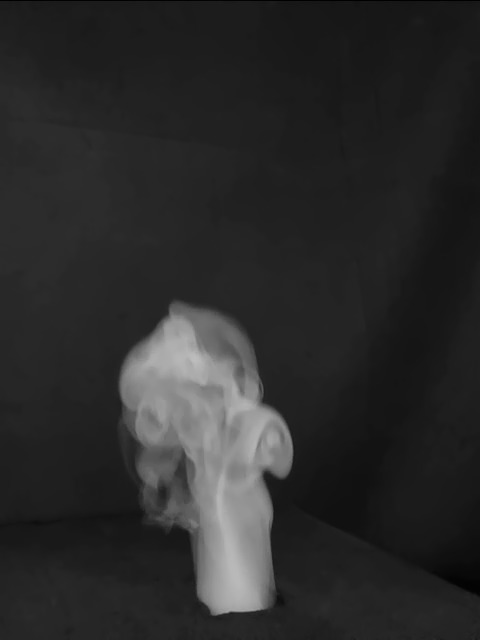}\hspace{0.0cm}%
		\includegraphics[trim={0cm 0.7cm 0cm 0.7cm},clip,width=0.125\linewidth]{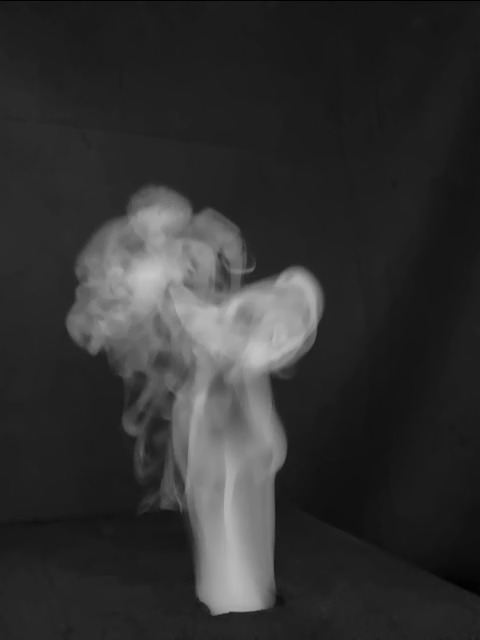}\hspace{0.0cm}%
		\includegraphics[trim={0cm 0.7cm 0cm 0.7cm},clip,width=0.125\linewidth]{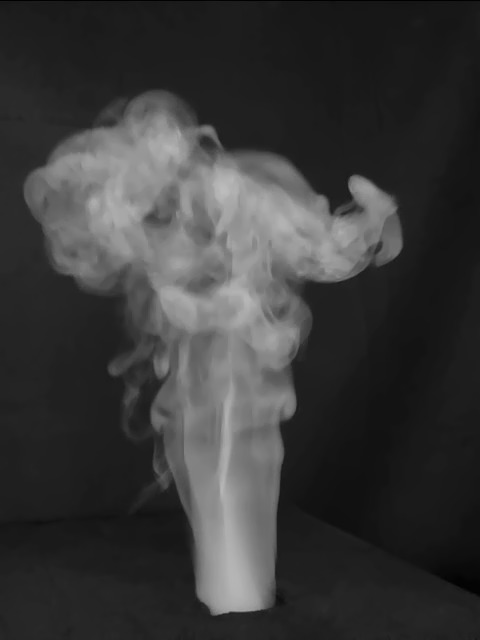}\hspace{0.0cm}%
		\includegraphics[trim={0cm 0.7cm 0cm 0.7cm},clip,width=0.125\linewidth]{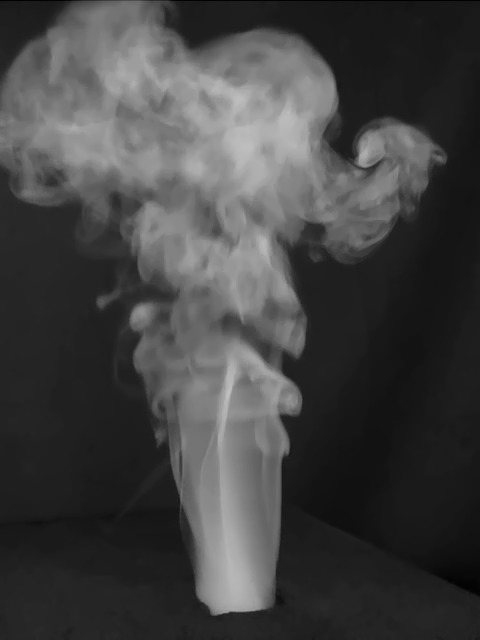}\hspace{0.0cm}%
		\includegraphics[trim={0cm 0.7cm 0cm 0.7cm},clip,width=0.125\linewidth]{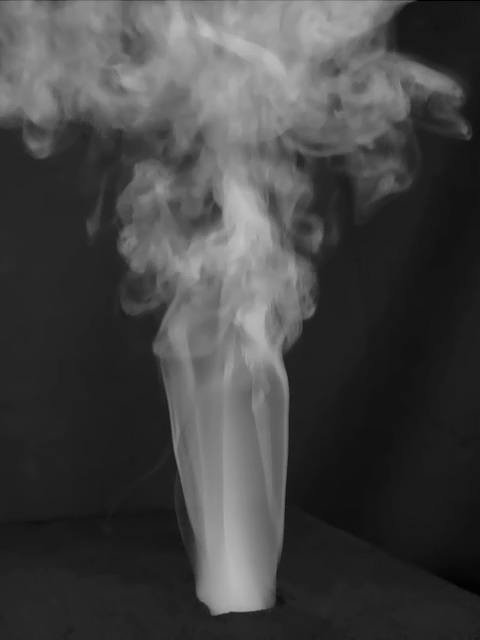}\hspace{0.0cm}%
		\includegraphics[trim={0cm 0.7cm 0cm 0.7cm},clip,width=0.125\linewidth]{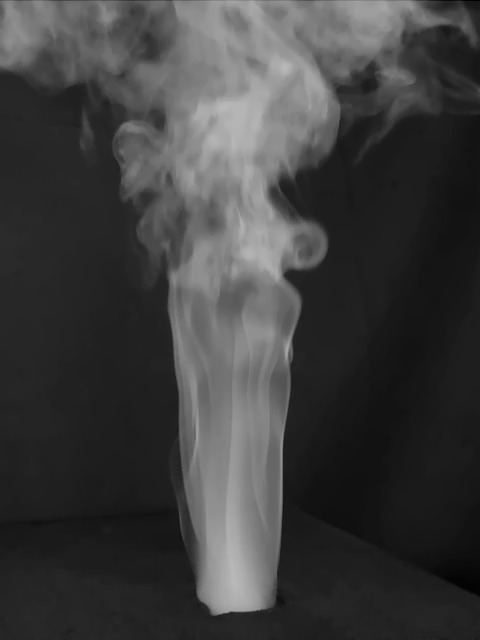}};
	\small \color{white} \draw (-7.8, 1.1) node {$a)$ real capture};
	\end{tikzpicture}\\
	\vspace{0.2mm}
	\begin{tikzpicture}
	\draw (0,0) node[inner sep=0] {
		\includegraphics[trim={0cm 0.7cm 0cm 0.7cm},clip,width=0.125\linewidth]{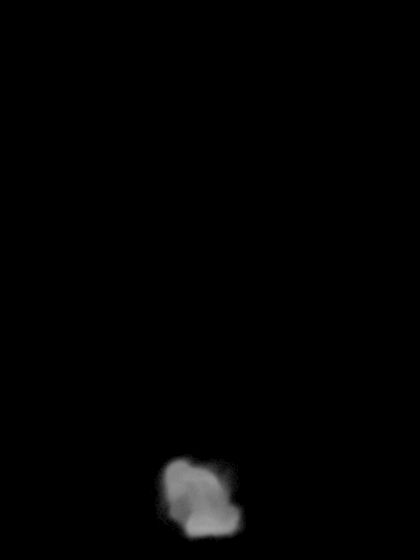}\hspace{0.0cm}%
		\includegraphics[trim={0cm 0.7cm 0cm 0.7cm},clip,width=0.125\linewidth]{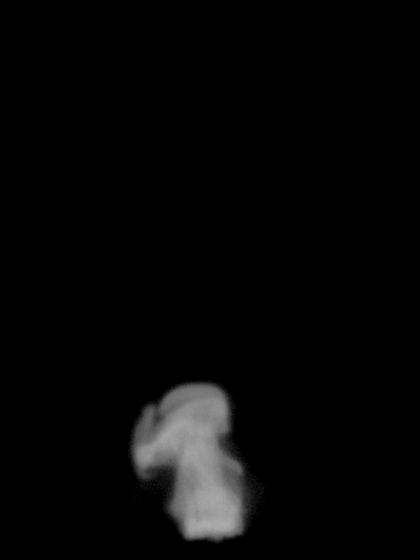}\hspace{0.0cm}%
		\includegraphics[trim={0cm 0.7cm 0cm 0.7cm},clip,width=0.125\linewidth]{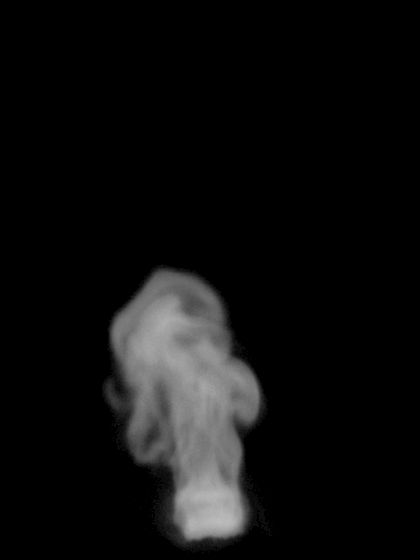}\hspace{0.0cm}%
		\includegraphics[trim={0cm 0.7cm 0cm 0.7cm},clip,width=0.125\linewidth]{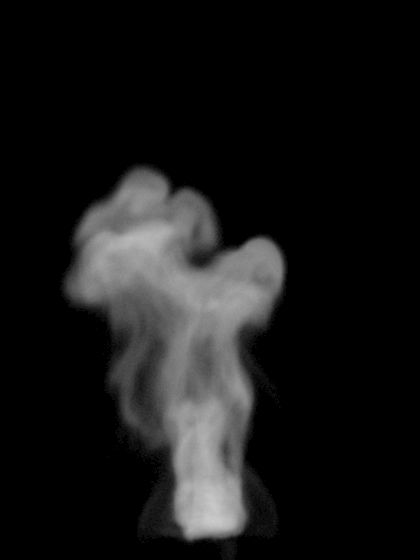}\hspace{0.0cm}%
		\includegraphics[trim={0cm 0.7cm 0cm 0.7cm},clip,width=0.125\linewidth]{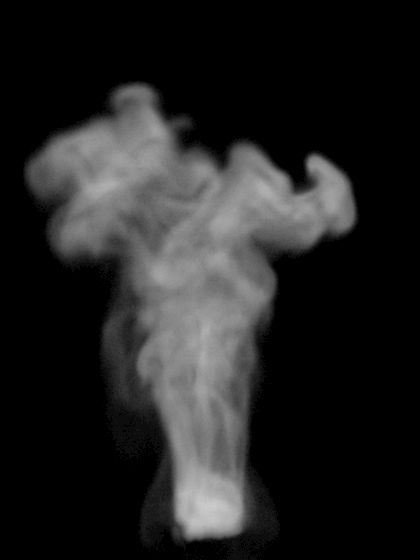}\hspace{0.0cm}%
		\includegraphics[trim={0cm 0.7cm 0cm 0.7cm},clip,width=0.125\linewidth]{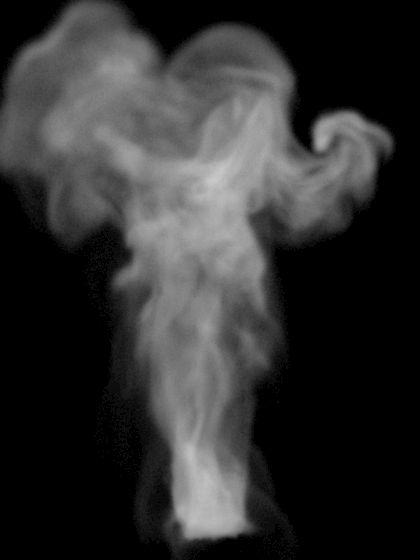}\hspace{0.0cm}%
		\includegraphics[trim={0cm 0.7cm 0cm 0.7cm},clip,width=0.125\linewidth]{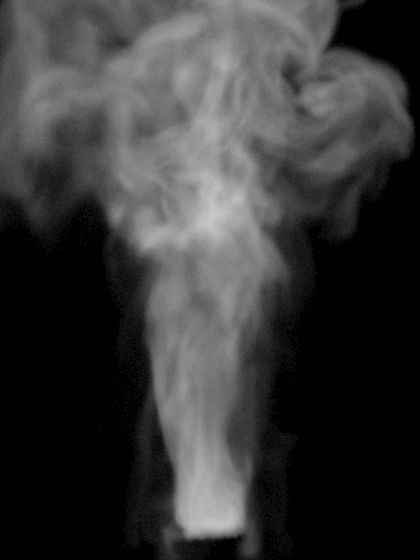}\hspace{0.0cm}%
		\includegraphics[trim={0cm 0.7cm 0cm 0.7cm},clip,width=0.125\linewidth]{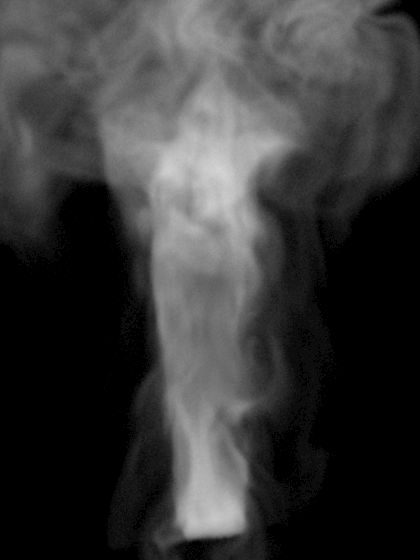}};
	\small \color{white} \draw (-7.35, 1.1) node {$b)$ reconstruction, front};
	\end{tikzpicture}\\
	\vspace{0.2mm}
	\begin{tikzpicture}
	\draw (0,0) node[inner sep=0] {
		\includegraphics[trim={0cm 0.7cm 0cm 0.7cm},clip,width=0.125\linewidth]{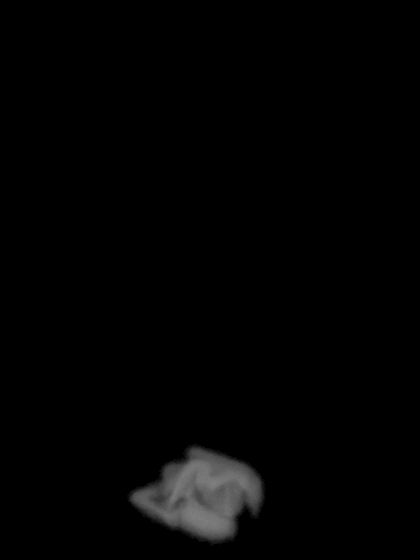}\hspace{0.0cm}%
		\includegraphics[trim={0cm 0.7cm 0cm 0.7cm},clip,width=0.125\linewidth]{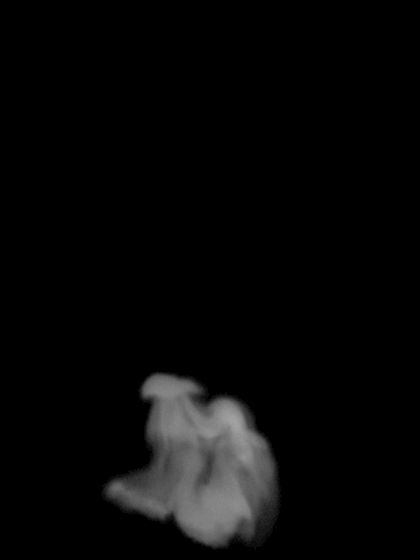}\hspace{0.0cm}%
		\includegraphics[trim={0cm 0.7cm 0cm 0.7cm},clip,width=0.125\linewidth]{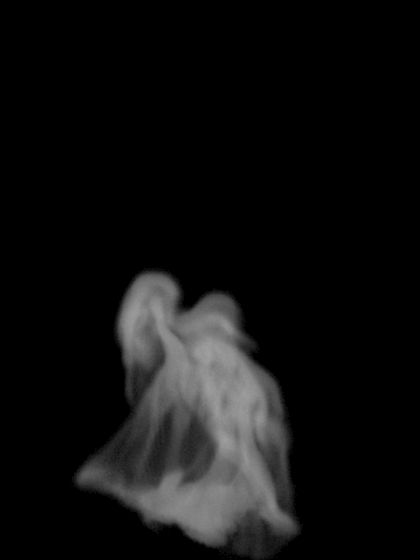}\hspace{0.0cm}%
		\includegraphics[trim={0cm 0.7cm 0cm 0.7cm},clip,width=0.125\linewidth]{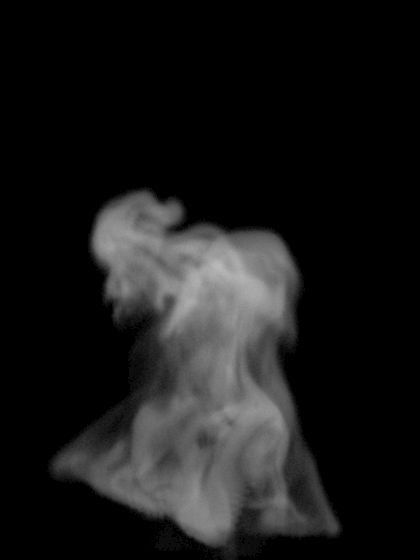}\hspace{0.0cm}%
		\includegraphics[trim={0cm 0.7cm 0cm 0.7cm},clip,width=0.125\linewidth]{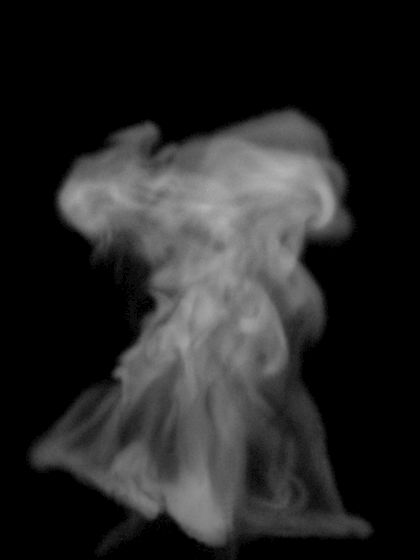}\hspace{0.0cm}%
		\includegraphics[trim={0cm 0.7cm 0cm 0.7cm},clip,width=0.125\linewidth]{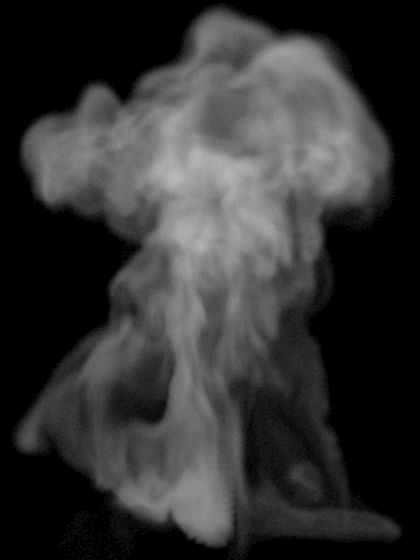}\hspace{0.0cm}%
		\includegraphics[trim={0cm 0.7cm 0cm 0.7cm},clip,width=0.125\linewidth]{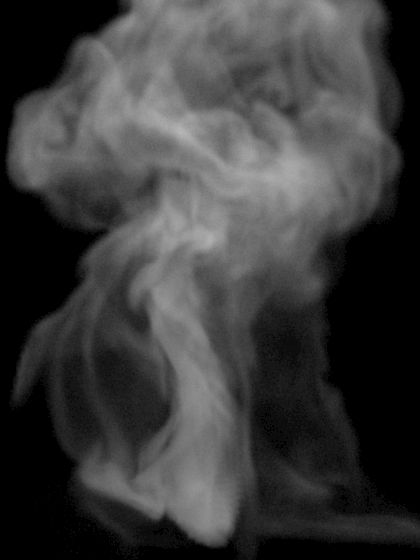}\hspace{0.0cm}%
		\includegraphics[trim={0cm 0.7cm 0cm 0.7cm},clip,width=0.125\linewidth]{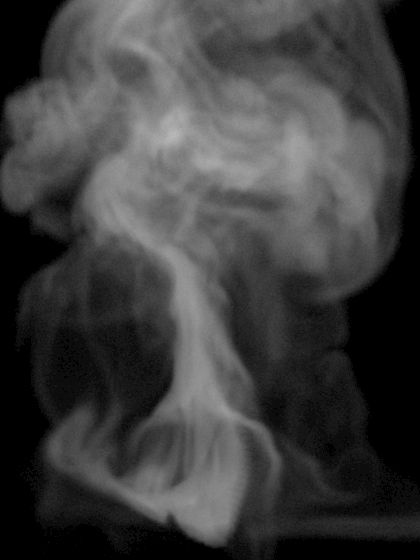}};
	\small \color{white} \draw (-7.4, 1.1) node {$c)$ reconstruction, side};
	\end{tikzpicture}\\
	\vspace{1mm}  % -------------------------------------- part 2
	\begin{tikzpicture}
	\draw (0,0) node[inner sep=0] {
		\includegraphics[trim={0cm 0.7cm 0cm 0.7cm},clip,width=0.125\linewidth]{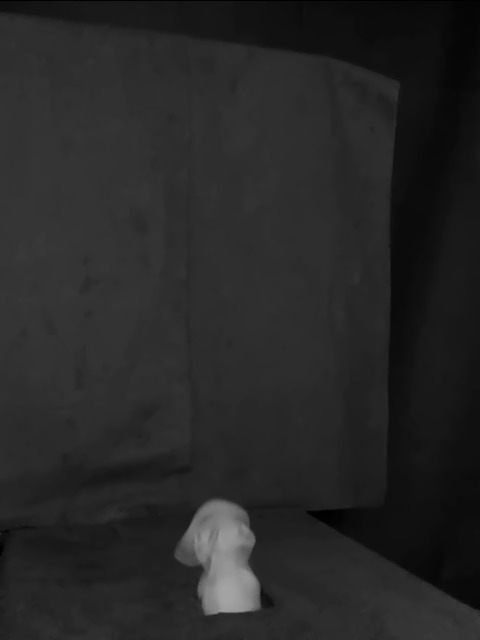}\hspace{0.0cm}%
		\includegraphics[trim={0cm 0.7cm 0cm 0.7cm},clip,width=0.125\linewidth]{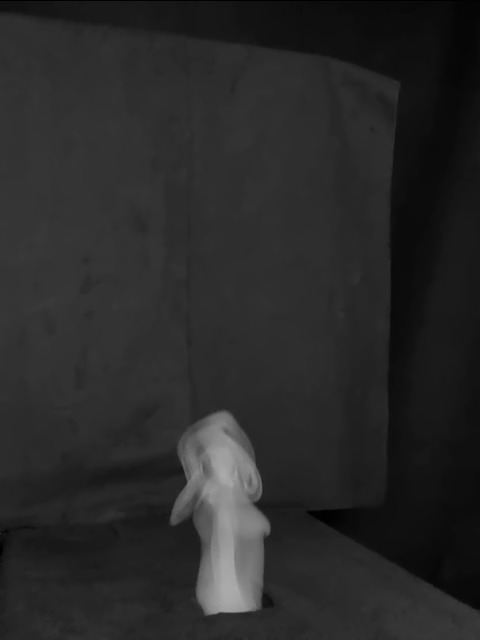}\hspace{0.0cm}%
		\includegraphics[trim={0cm 0.7cm 0cm 0.7cm},clip,width=0.125\linewidth]{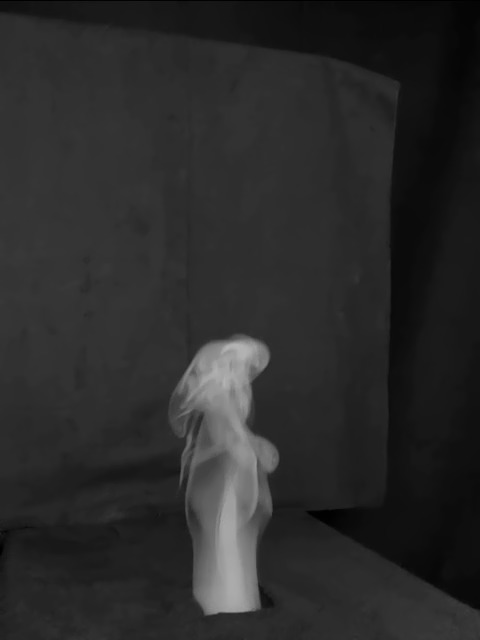}\hspace{0.0cm}%
		\includegraphics[trim={0cm 0.7cm 0cm 0.7cm},clip,width=0.125\linewidth]{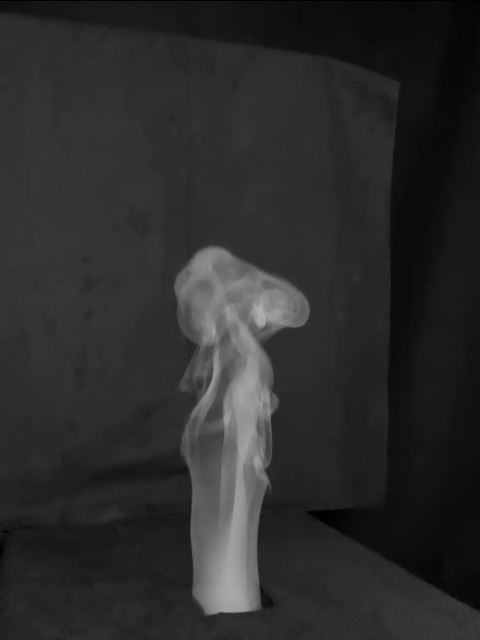}\hspace{0.0cm}%
		\includegraphics[trim={0cm 0.7cm 0cm 0.7cm},clip,width=0.125\linewidth]{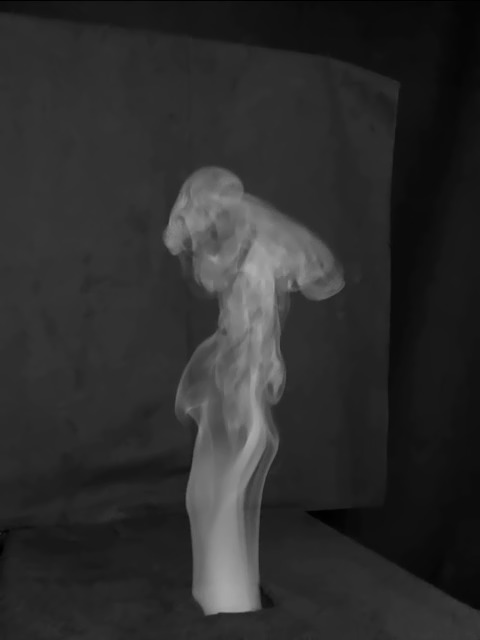}\hspace{0.0cm}%
		\includegraphics[trim={0cm 0.7cm 0cm 0.7cm},clip,width=0.125\linewidth]{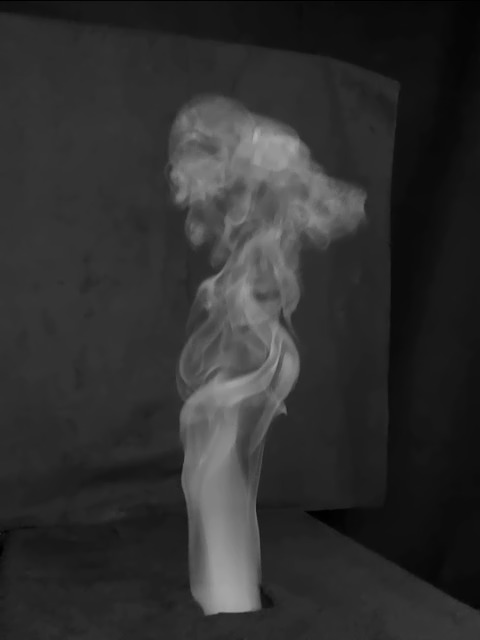}\hspace{0.0cm}%
		\includegraphics[trim={0cm 0.7cm 0cm 0.7cm},clip,width=0.125\linewidth]{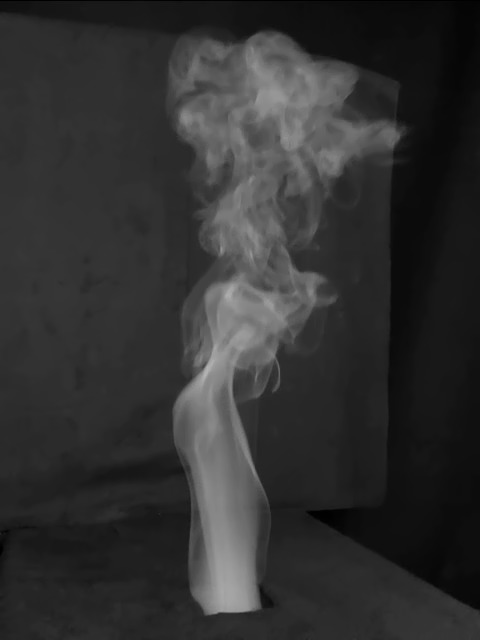}\hspace{0.0cm}%
		\includegraphics[trim={0cm 0.7cm 0cm 0.7cm},clip,width=0.125\linewidth]{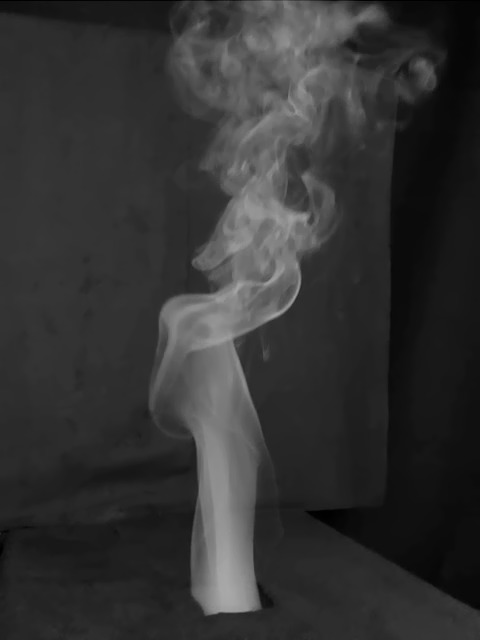}};
	\small \color{white} \draw (-7.8, 1.1) node {$d)$ real capture};
	\end{tikzpicture}\\
	\vspace{0.2mm}
	\begin{tikzpicture}
	\draw (0,0) node[inner sep=0] {
		\includegraphics[trim={0cm 0.7cm 0cm 0.7cm},clip,width=0.125\linewidth]{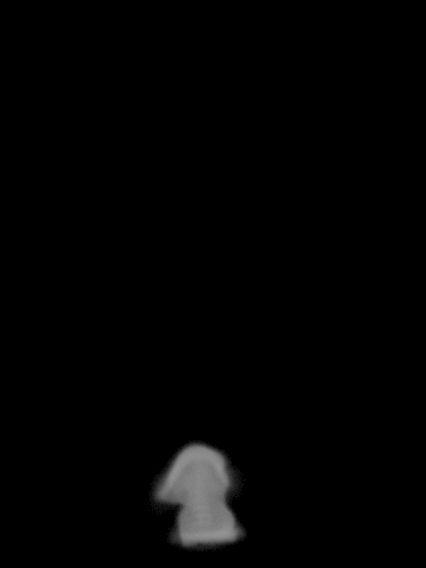}\hspace{0.0cm}%
		\includegraphics[trim={0cm 0.7cm 0cm 0.7cm},clip,width=0.125\linewidth]{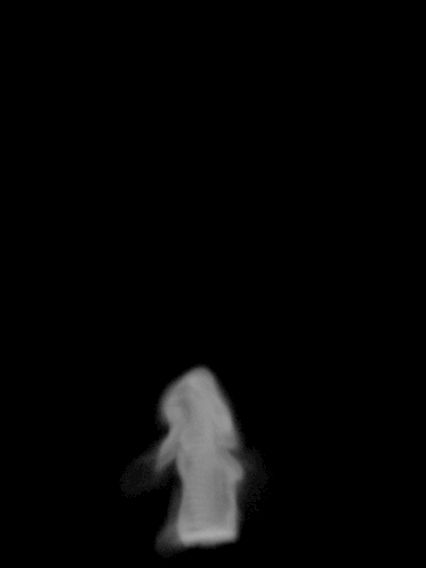}\hspace{0.0cm}%
		\includegraphics[trim={0cm 0.7cm 0cm 0.7cm},clip,width=0.125\linewidth]{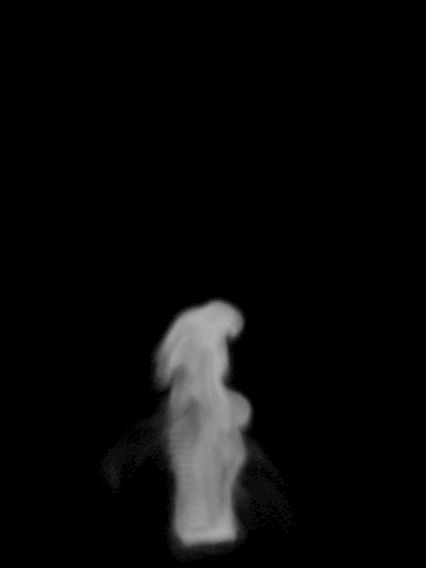}\hspace{0.0cm}%
		\includegraphics[trim={0cm 0.7cm 0cm 0.7cm},clip,width=0.125\linewidth]{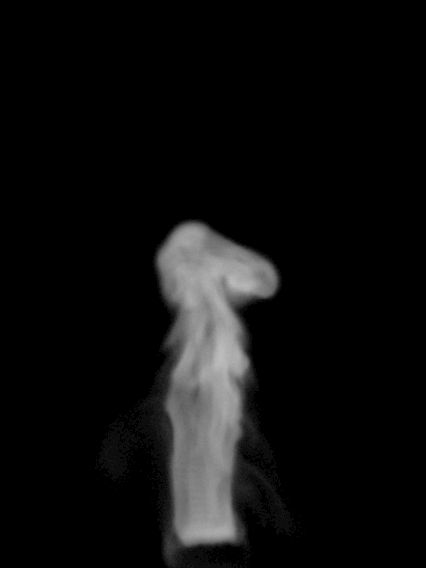}\hspace{0.0cm}%
		\includegraphics[trim={0cm 0.7cm 0cm 0.7cm},clip,width=0.125\linewidth]{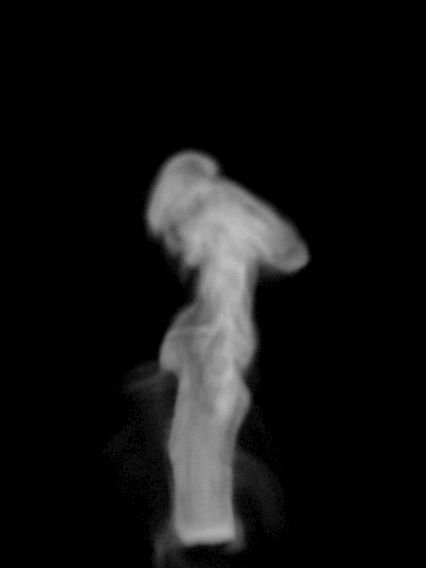}\hspace{0.0cm}%
		\includegraphics[trim={0cm 0.7cm 0cm 0.7cm},clip,width=0.125\linewidth]{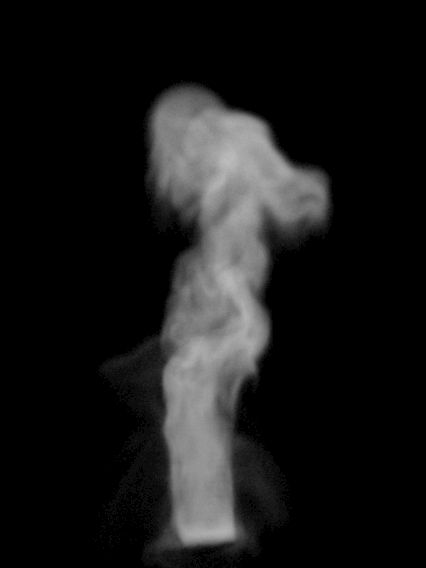}\hspace{0.0cm}%
		\includegraphics[trim={0cm 0.7cm 0cm 0.7cm},clip,width=0.125\linewidth]{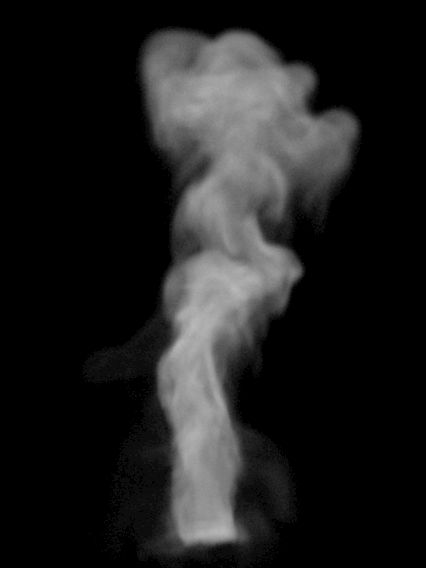}\hspace{0.0cm}%
		\includegraphics[trim={0cm 0.7cm 0cm 0.7cm},clip,width=0.125\linewidth]{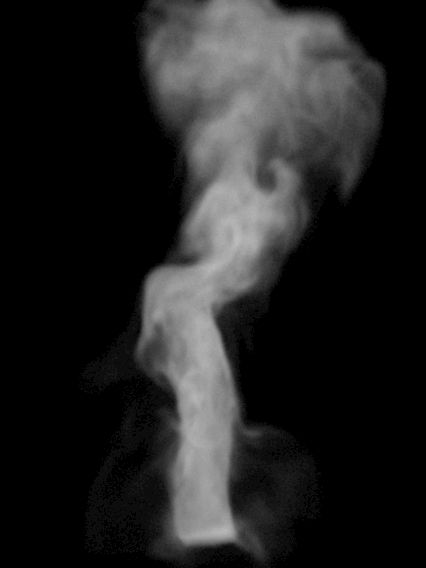}};
	\small \color{white} \draw (-7.35, 1.1) node {$e)$ reconstruction, front};
	\end{tikzpicture}\\
	\vspace{0.2mm}
	\begin{tikzpicture}
	\draw (0,0) node[inner sep=0] {
		\includegraphics[trim={0cm 0.7cm 0cm 0.7cm},clip,width=0.125\linewidth]{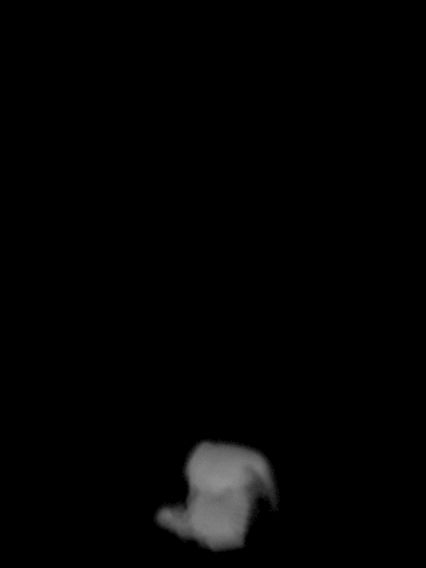}\hspace{0.0cm}%
		\includegraphics[trim={0cm 0.7cm 0cm 0.7cm},clip,width=0.125\linewidth]{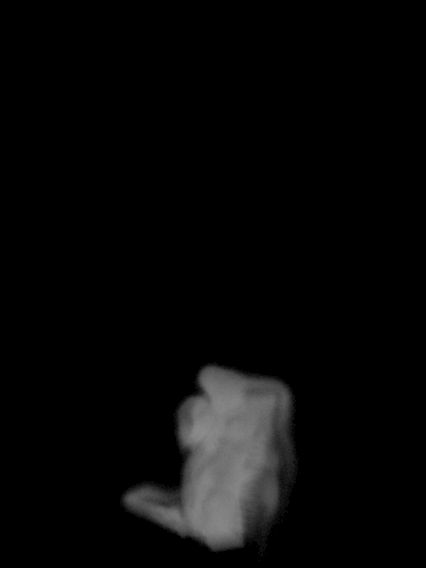}\hspace{0.0cm}%
		\includegraphics[trim={0cm 0.7cm 0cm 0.7cm},clip,width=0.125\linewidth]{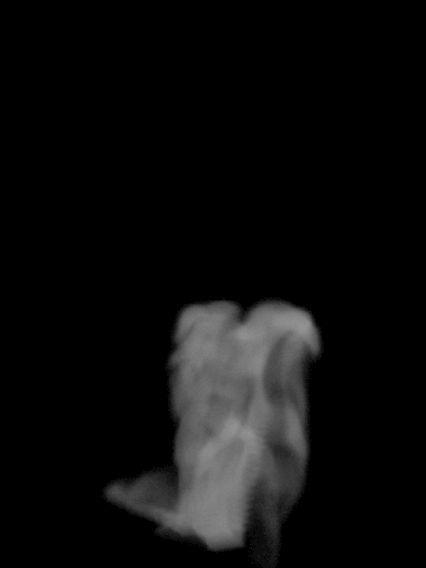}\hspace{0.0cm}%
		\includegraphics[trim={0cm 0.7cm 0cm 0.7cm},clip,width=0.125\linewidth]{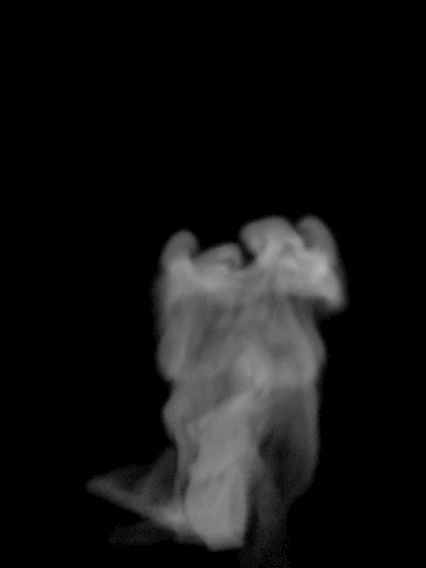}\hspace{0.0cm}%
		\includegraphics[trim={0cm 0.7cm 0cm 0.7cm},clip,width=0.125\linewidth]{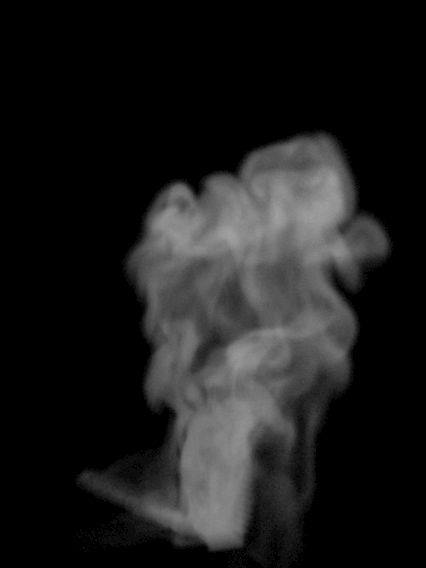}\hspace{0.0cm}%
		\includegraphics[trim={0cm 0.7cm 0cm 0.7cm},clip,width=0.125\linewidth]{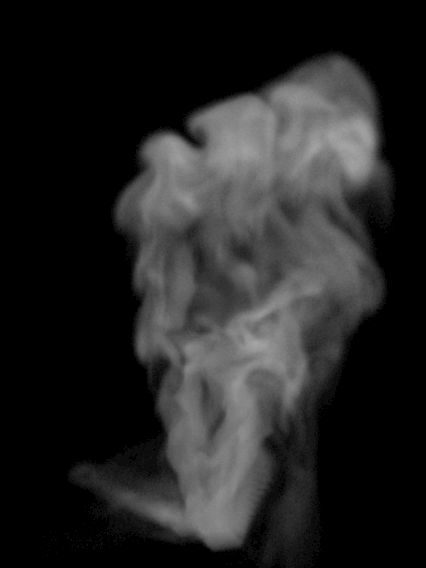}\hspace{0.0cm}%
		\includegraphics[trim={0cm 0.7cm 0cm 0.7cm},clip,width=0.125\linewidth]{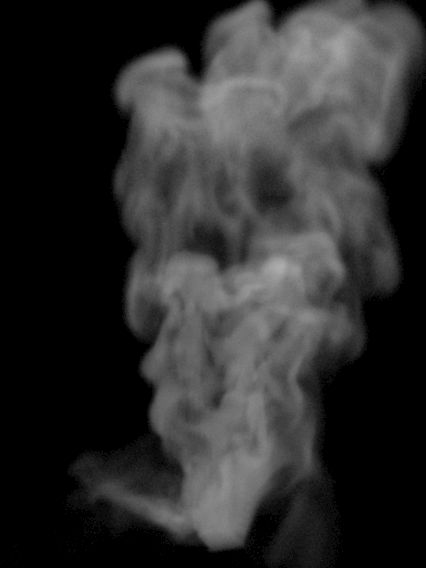}\hspace{0.0cm}%
		\includegraphics[trim={0cm 0.7cm 0cm 0.7cm},clip,width=0.125\linewidth]{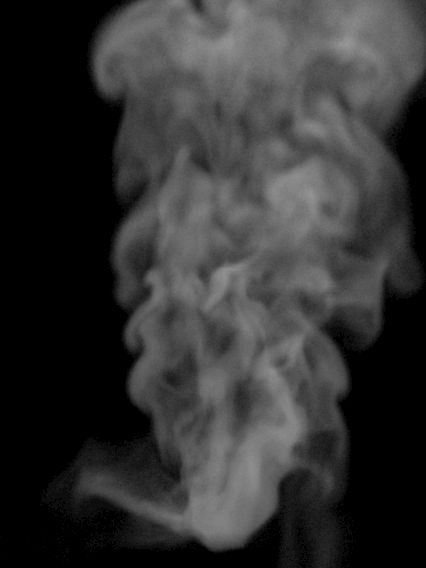}};
	\small \color{white} \draw (-7.4, 1.1) node {$f)$ reconstruction, side};
	\end{tikzpicture}
	\caption{Two captures of real fluids a,d) and reconstructions front d,e) and side views c,f). Frames are shown in 15 frame intervals.
	}	\label{fig:real2}
\end{figure*}

We additionally recorded rising plumes of real smoke from a fog machine with a Raspberry Pi camera.
This hardware represents an inexpensive capturing setup that is easy to use.
Two of our raw input recordings are shown in~\myreffig{fig:real2} a) and d). 
We post-process these input videos with a gray-scale mapping, noise reduction and background separation. 
If images are captured with more complex lighting and background, we 
could potentially include more advanced methods from the image processing area in the post-processing step.
Regarding the real capturing setup, the unknown and temporally changing smoke inflow at the bottom of the images presents an
additional challenge.
Therefore, we impose a velocity in the inflow region for both real reconstructions.
The velocity is the same for both captures and is roughly estimated from the videos.
Our algorithm still reconstructs a realistic volumetric motion that matches the input very well 
and produces realistic swirls from the side view.
Renderings with a similar visual style but slightly different camera properties are shown below each raw input row
of~\myreffig{fig:real2}.
\subsection{Performance}
\begin{table}
	\footnotesize
	\begin{tabular}{ | l | l | l |l |}
		\hline
		Scene & Grid Size & \#Frames & Avg. T \\ \hline\hline
		Plume, ~\myreffig{fig:jetobs_mitsuba} & $120\times160\times120$ & 96 & 61m\\ \hline
		Re-sim., ~\myreffig{fig:plume_resim} & $240\times320\times240$ & 96 & 08m\\ \hline
		Front = Side, ~\myreffig{fig:frontAs} & $120\times160\times120$ & 76 & 52m\\ \hline
		Jet, ~\myreffig{fig:jetobs_mitsuba}& $120\times160\times120$ & 96 & 68m\\ \hline
		Obstacle, ~\myreffig{fig:jetobs_mitsuba}& $120\times160\times120$ & 112 & 45m\\ \hline
		2 views, ~\myreffig{fig:compareTomoOF}& $120\times160\times120$ & 107 & 44m\\ \hline
		Real, ~\myreffig{fig:real2} a)& $120\times200\times120$ & 120  & 62m\\  \hline
		Real, ~\myreffig{fig:real2} d)& $120\times200\times120$ & 120  & 54m\\ 
		\hline
		\hline
	\end{tabular}
	\caption{Performance details for all reconstructed scenes. The time is given as an average over the whole sequence.}
	\label{tab:perf}
\end{table}
We show an overview of the grid sizes and reconstruction times in~\myreftab{tab:perf}.
Note that later time steps with larger visual hulls take more time than earlier time steps.
This especially increases the run time for the jet, the plume and the first real capture reconstructions, where the density fills most of the domain in the end.
We have been using one CPU for synthetic data and two for real data reconstructions, usually Intel(R) Xeon(R) CPU E5-1650 v3 @ 3.50GHz.
Our implementation computes and stores the tomography matrix $P^2$ explicitly and employs a CG solver, which simplifies the implementation, but is not optimal regarding run time or memory. Thus, this step of our pipeline could be optimized in the future.

\section{Conclusion}
%summary
We have demonstrated a first reconstruction algorithm for tomographic density and velocity from just one single input view.
With a novel density update that is tied to the estimated velocity, we are able to produce volumetric reconstructions without the need for complex capturing setups.
Our depth regularizers additionally increase control over the motion in depth. 
This way and only from monocular videos, our algorithm calculates dense and realistic volumetric flow fields that can be flexibly used for re-simulations or further editing and guiding operations.
Our method outperforms previous work in terms of reconstruction quality.

\vspace{-0.01cm}
%limitations
By construction, one inherent limitation of our algorithm is the lack of reliable information about motion in depth. 
Improving the estimation of depth motion with extensions such
as data-driven regularizers would be an interesting avenue for future work.
Additionally, we would like to improve the inflow handling in terms of both shape and flux, since 
the inflow has substantial influence on the reconstruction quality.
One could optimize for best initial conditions, e.g. source position, size, shape, and amount of inflow velocity over time.
Designing the inflow in a more flexible way could also lead to better control for the end users of our algorithm.

\vspace{-0.01cm}
Incorporating a matrix-free or SART approach as tomography part could reduce run time 
and memory consumption significantly.
Also, our current linear image formation model is potentially limiting. 
We have found it to work well for cases like the filmed smoke in \myrefsec{sec:results}, 
but very dense clouds will require more complex image formation models.
Despite these open topics, our algorithm demonstrates the usefulness of physics simulations
for under-constrained reconstructions and represents a first step towards capturing
more general 3D motions from simple, single view video streams.

\section{Acknowledgments}
This work was funded by the ERC Starting Grant {\em realFlow} (StG-2015-637014) and supported by King Abdullah University of Science and Technology under Individual Baseline Funding.

\bibliographystyle{eg-alpha-doi}
\bibliography{references,fluids} 

\end{document}